\documentclass[A4paper,useAMS,usenatbib]{mnras}

\usepackage[T1]{fontenc}
\usepackage{ae,aecompl}
\usepackage{graphicx}
\usepackage{rotating}
\usepackage{float}
\usepackage{array}
\usepackage{mathenv}
\usepackage{multirow}
\usepackage{url}
\usepackage{subcaption}
\usepackage{times,txfonts}

\begin{document}

 \title[The MIXR sample]{The MIXR sample: AGN activity versus star formation across the cross-correlation of WISE, 3XMM, and FIRST/NVSS.}


\author[B. Mingo et al.]{B. Mingo$^{1}$\thanks{E-mail:bmingo@extragalactic.info}, M. G. Watson$^{1}$, S. R. Rosen$^{1}$, M. J. Hardcastle$^{2}$, A. Ruiz$^{3}$, A. Blain$^{1}$, \newauthor F. J. Carrera$^{3}$, S. Mateos$^{3}$, F.-X. Pineau$^{4}$, and G. C. Stewart$^{1}$\\
$^{1}$Department of Physics and Astronomy, University of Leicester, University Road, Leicester LE1 7RH, UK\\
$^{2}$School of Physics, Astronomy \& Mathematics, University of Hertfordshire, College Lane, Hatfield AL10 9AB, UK\\
$^{3}$Instituto de F\'isica de Cantabria (CSIC-UC), Avenida de los Castros, 39005 Santander, Spain\\
$^{4}$CNRS, Universit\`e de Strasbourg, Observatoire Astronomique, 11 rue de l'Universit\`e, 67000 Strasbourg, France\\}

   \date{Received ; accepted}

\maketitle

\begin{abstract}
We cross-correlate the largest available Mid-Infrared (WISE), X-ray (3XMM) and Radio (FIRST+NVSS) catalogues to define the MIXR sample of AGN and star-forming galaxies. We pre-classify the sources based on their positions on the WISE colour/colour plot, showing that the MIXR triple selection is extremely effective to diagnose the star formation and AGN activity of individual populations, even on a flux/magnitude basis, extending the diagnostics to objects with luminosities and redshifts from SDSS DR12. We recover the radio/mid-IR star formation correlation with great accuracy, and use it to classify our sources, based on their activity, as radio-loud and radio-quiet AGN, LERGs/LINERs, and non-AGN galaxies. These diagnostics can prove extremely useful for large AGN and galaxy samples, and help develop ways to efficiently triage sources when data from the next generation of instruments becomes available. We study bias in detail, and show that while the widely-used WISE colour selections for AGN are very successful at cleanly selecting samples of luminous AGN, they miss or misclassify a substantial fraction of AGN at lower luminosities and/or higher redshifts. MIXR also allows us to test the relation between radiative and kinetic (jet) power in radio-loud AGN, for which a tight correlation is expected due to a mutual dependence on accretion. Our results highlight that long-term AGN variability, jet regulation, and other factors affecting the $Q/L$$_{bol}$ relation, are introducing a vast amount of scatter in this relation, with dramatic potential consequences on our current understanding of AGN feedback and its effect on star formation.
\end{abstract}

   \begin{keywords}
   		galaxies: active --
   		galaxies: starburst --
   		infrared: galaxies --
		X-rays: galaxies --
		radio continuum: galaxies --
   \end{keywords}

%

\section{Introduction}\label{Intro}

Over the last few decades, and in particular in the last 10 to 15 years, our understanding of active galactic nuclei (AGN), their underlying physical mechanisms, their environments, and their observational properties, has greatly increased. Although the unification model proposed by \citet{Antonucci1993} still holds true in many aspects, subsequent revisions \citep[see e.g.][]{Netzer2015} illustrate what we have learned about the structure of the obscuring torus, the mechanisms that provide feedback, the variability timescales involved, and where the radio-loud sources fit (or do not fit) in the grand AGN unification scheme. We are living in what could be considered a golden era of surveys, which allow us, for the first time, to construct large, consistent, multiwavelength samples of AGN with the potential to push our understanding of these objects even further. 

Although only $\sim$10--20 per cent of the AGN we observe are classified as radio-loud, recent evidence shows that jets and lobes could be far more ubiquitous than we previously thought. There is an increasingly large number of Seyfert galaxies, and even QSOs, where jets and lobes, or excess radio emission, have been detected \citep[e.g.][]{Hota2006,Gallimore2006,DelMoro2013,Singh2015,Harrison2015}, throwing into question the radio-loud/quiet classification, which, being based on optical (B band) to radio (5 GHz) flux ratios \citep[e.g.][]{Kellerman1989}, classifies most of these objects as radio-quiet. This `jet mode' or `radio mode' is fundamental to our understanding of the AGN/host relationship, not only for very powerful sources in clusters, where the jet-driven shocks can offset radiative cooling of the gas \citep[e.g.][]{McNamara2012,HLarrondo2015}, but especially for low power sources ($L_{1.4 GHz} \le 10^{23}$ W Hz$^{-1}$ sr$^{-1}$), because it is in these systems that the effect of the AGN on the surrounding interstellar gas (on 10--100 kpc scales) can have the largest potential impact on the evolution and star formation history of the host galaxy \citep[e.g.][]{Cattaneo2009,Croston2011,Mingo2011,Mingo2012}. 

Radio-loud sources are also useful in that they allow us to unequivocally identify sources in the radiatively inefficient accretion regime \citep{Narayan1995}. This population, originally identified as low excitation radio galaxies (LERGs) by \citet{Hine1979}, lacks the `traditional' AGN disc and torus, shows very low Eddington rates \citep{Hardcastle2007b,Hardcastle2009,DeGasperin2011,Best2012,Mingo2014,Paggi2016}, and seems to be channeling most of the gravitational energy into jets rather than radiative output, in a similar manner to the low/hard state of low mass X-ray binaries \citep[see e.g. the review by][]{Fender2014}. In the optical, radiatively inefficient sources are typically classified as LINERs (low ionisation nuclear emission line regions) as initially proposed by \citet{Heckman1980}, or even appear as fully `quiescent' (i.e. not containing an AGN) galaxies \citep[see e.g.][]{Kimball2008}. This classification, however, is misleading, in the sense that other processes such as shocks or emission from an old stellar population can also produce low ionisation spectra \citep[see e.g.][and references therein]{Balmaverde2015}. Therefore, finding low ionisation optical emission lines does not guarantee the presence of a radiatively inefficient AGN, while finding active radio jets in an otherwise `quiescent' looking galaxy does. As radiatively inefficient AGN only produce soft X-rays related to the jet \citep[e.g.][]{Hardcastle1999}, their typical X-ray luminosity is $10^{39}-10^{41}$ erg/s, which precludes them from being included in most X-ray selected AGN surveys.

Recent results show that the interplay between AGN activity, outflows, and star formation may be more complex than we previously thought, and fundamental to understanding galaxy evolution and black hole growth \citep[e.g.][]{Alexander2012,Magliocchetti2014,Davies2014}. Although we are beginning to better understand the transition between the regimes in which AGN and star formation activity dominate, and how radio AGN activity, in particular, affects star formation \citep[e.g.][]{Smolcic2009,Dicken2012,DelMoro2013,Hardcastle2013b,Kalfountzou2014,Villarroel2014,Gurkan2015,Rawlings2015,Hardcastle2016,Drouart2016,Tadhunter2016}, there is still a distinct lack of agreement on how and when AGN activity influences star formation \citep{Harrison2012,Ishibashi2012,Symeonidis2013,Symeonidis2014,AlonsoHerrero2013,Heckman2014,Balmaverde2016,Brusa2015,Rosario2013,Rosario2015,Stanley2015,Bernhard2016,Alberts2016}. Although the large timescales involved probably cause part of this confusion \citep{Georgakakis2008,Wild2010,RamosAlmeida2013,Best2014}, and it is clear that we still do not fully understand long AGN variability timescales \citep[see e.g.][]{Hickox2014}, it is also true that dedicated samples that encompass sources in both regimes, as well as the transition, still tend to be limited either in wavelength, scope, redshift, or size. 

Obtaining large multiwavelength samples of radio-loud AGN is challenging for several reasons, the main two being the extended nature of radio emission and the low sky density of radio-loud AGN, and the number of sources decreases rapidly if selections in more than two bands are required. These surveys also tend to focus on particular populations of radio-loud AGN (or star-forming galaxies). There is a wealth of on-going and upcoming instruments and surveys that will open a wide field of potential exploration in both fields: LOFAR, SKA, e-MERLIN, JVLA, in the radio; e-Rosita, and Athena in the X-rays, CTA in the gamma-ray band, LSST, and JWST and ALMA at infrared and sub-mm wavelengths, respectively. Now is the perfect time to assess which questions our current data can and cannot answer, to set a framework and potential diagnostic tools for the next generation of results.

The ARCHES FP7 collaboration\footnote{\url{http://www.arches-fp7.eu/}} is a project dedicated to fully exploiting the capabilities of the 3XMM catalogue of X-ray sources, by creating multiwavelength products (cross-correlated catalogues and tools, spectral energy distributions, and a cluster catalogue and finder tool). As part of this collaboration, we have built and describe in this paper the MIXR sample: a systematic, large sample of sources detected in the Mid-IR (WISE all-sky survey), X-rays (3XMM DR5) and Radio (FIRST/NVSS). By requiring a detection in all three bands, we find a wide range of populations: from radiatively inefficient (LERG/LINER) systems in otherwise quiescent galaxies, to low luminosity Seyfert-like sources where the host emission dominates in some bands, to nearby starburst objects, to high luminosity radio-loud and radio-quiet Seyferts and QSOs. The MIXR sample allows us to derive efficient diagnostics for star formation and AGN activity (both radiatively efficient, as seen in `traditional' AGN, and radiatively inefficient, as seen in LERG/LINER), even in host-dominated sources that are normally considered quiescent and discarded from most mid-IR and X-ray AGN samples. We also test the  radiative (luminosity) versus kinetic (jet) output in our AGN, to explore the extent and possible causes for the scatter we observed in \citet{Mingo2014}, in contradiction with the well-known correlation of \citet{Rawlings1991}. Our analysis also helps us pinpoint several sources of bias that affect selections performed in one or more of the bands we use, helping us better understand what AGN populations are included and excluded in each selection. 

In section \ref{Data} we discuss in detail the MIXR sample construction. In section \ref{Diagnostics} we use WISE colours to pre-classify the sources, and carry out a series of early diagnostics to test these classifications, using hardness ratios, radio versus X-ray `loudness', and flux/magnitude diagrams. In section \ref{z} we add redshift information from SDSS, which we use in section \ref{LDiag} to derive luminosities for the MIXR sources, and extend our diagnostics to verify the underlying type of activity for the MIXR sources. In section \ref{RL_RQ_section} we re-classify the sources based on their activity (radio-quiet and radio-loud AGN, including LERGs/LINERs, and galaxies). For sections \ref{Eddington} and \ref{Power} we focus on the AGN, assessing their Eddington rates and their radiative versus kinetic (jet) output, to highlight the strengths and limitations of current surveys, and address some of the open questions on AGN variability and its impact on the AGN/host relationship.

For this work we have used the latest cosmological values released by the Planck collaboration \citep{Planck2015}: $H_{0} = 67.74$ km/s/Mpc, $\Omega_{m} = 0.3089$, and $\Omega_{\Lambda} = 0.6911$. The catalogue we describe in this paper is available on-line for download at \url{http://www.arches-fp7.eu/index.php/tools-data/downloads/mixr-catalogue} and will be made available on VizieR.

%

\section{Data and Sample Construction}\label{Data}

Our aim is to select a large, clean (i.e. avoiding mis-classifications, but also contaminants, such as stars) sample of sources, with data that will allow us to characterise the accretion properties of the AGN population, as well as to explore the extent of star formation present; we need large, uniform surveys at wavelengths where AGN and star formation activity can be detected unequivocally: Mid-Infrared (3.4--12 $\mu$m), X-rays (0.2--10 keV) and Radio (1.4 GHz) (MIXR).

X-ray and mid-IR emission are very good probes of accretion in AGN, the former being produced in the accretion disc and hot corona in the inner regions of the AGN, and the latter being the region of the spectrum where the bulk of the thermal (blackbody) emission from the dusty torus peaks \citep[see e.g.][]{Horst2008}. Although it is possible to obtain clean selections of samples using only X-ray and mid-IR data, some caution must be applied to eliminate X-ray binaries and galaxies with X-ray and infrared emission associated with star formation, rather than an AGN. The process typically involves cuts in mid-IR colours and X-ray hardness ratio \citep[e.g.][]{Assef2010,Assef2013,Stern2012,Mateos2012,Rovilos2014}. While these studies are extremely successful in characterising the properties of QSO-like and bright Seyfert-type sources, they cannot include fainter galaxies, where AGN emission cannot be detected unequivocally, as well as radiatively inefficient (LINER or LERG) sources where most of the energy is channelled through a jet, rather than as radiative output \citep[see e.g][]{Best2012,Hardcastle2009,Mingo2014}. It is also important to keep in mind that a strict hardness ratio cut can eliminate sources with a soft excess, most relevantly radio-loud AGN, where jet-related emission produces soft X-rays. 

An additional radio selection could prove very advantageous in this context. There are two mechanisms that can produce bulk radio emission: star formation \citep[free-free emission from HII regions, some synchrotron radiation from particle acceleration in winds and supernova explosions, and some thermal emission from cold gas and dust, plus HI at 21cm, see e.g.][]{Harwit1975,Condon1992} and AGN activity (synchrotron radiation from jets, hotspots and lobes). While thermal emission from dust becomes very relevant at higher frequencies ($>$10 GHz), low to intermediate frequency radio production from star formation is remarkably inefficient \citep[e.g.][]{Bell2003} and is generally detected only for very nearby starburst galaxies. It is now known that star formation-related radio emission may be more significant at sub-mJy level \citep{Padovani2011A,Bonzini2013} \citep[see also the recent LOFAR results of][]{Williams2016}, but even at low fluxes AGN processes may still dominate the emission in systems with moderate star formation, rather than powerful starbursts \citep{White2015}. Recent evidence also shows that the fraction of radiatively efficient (`traditional', radiative mode, IR and X-ray bright) AGN that show accretion-related radio emission inversely correlates with radio power \citep{Padovani2015}. AGN radio emission is also unaffected by obscuration, and thus relatively unbiased with respect to orientation \citep[there is a slight bias towards favouring core-dominated, face-on sources, see e.g. the discussion by][]{Mingo2014}.

Aside from the obvious bias introduced by selecting only sources that produce radio emission, the main downside of requiring radio detections is a substantial reduction in the number of sources in the final sample, given the low sky density of the radio sky. This disadvantage, however, is more than made up for by the fact that, without any additional filtering, a combination of radio, mid-IR and X-rays can produce a clean, uniform selection of AGN across all luminosities, host types and accretion modes, as well as identifying nearby starburst galaxies. As such, our study focuses mostly on AGN, but the star-forming galaxies provide the necessary framework to quantify star formation and AGN activity in sources where both contributions are hard to disentangle.  

Although using radio data would guarantee a very clean selection of extragalactic sources \citep[the number of individual stars identified at 1.4 GHz is very low, see e.g.][]{McMahon2002,FIRST2015} we decided to minimise the incidence of Galactic sources across all catalogues by imposing a high Galactic latitude cut ($|b_{II}| \ge 20^{\circ}$).

Most multi-survey samples use positional matching techniques to cross-correlate the sources across the different catalogues. For MIXR we have used the statistical \textsc{xmatch} cross-correlation tool developed for the ARCHES collaboration, described in more detail in section \ref{Sample}, which allowed us to quantitatively, efficiently and simultaneously establish the source associations across the three catalogues we used to create MIXR.

%

\subsection{X-rays: 3XMM}\label{3XMM}

For our X-ray data we have used the DR5 release of the 3XMM catalogue \citep{Rosen2016}. This catalogue comprises results from 7781 individual pointings taken between February 2000 and the end of 2013, resulting in 565962 individual detections and 396910 unique sources covered by XMM-Newton's EPIC cameras (pn, MOS1, MOS2) in the 0.2--12 keV band, making it the largest X-ray catalogue ever produced. The sources in 3XMM are resolved on scales of $\sim6$ arcsec, and the typical positional error is $\sim1.5$ arcsec.

The fluxes in the catalogue are calculated for 5 bands (0.2--0.5, 0.5--1, 1--2, 2--4.5, and 4.5--12 keV) from the count rate of each instrument in each individual observation \citep[see][]{Mateos2009,Watson2009,Rosen2016}, and, when more than one observation per source exists, combined for each source using weighted average based on the flux errors. For the diagnostic plots in section \ref{HR} we have combined the fluxes in the first three bands (0.2--2 keV) to obtain the soft X-ray flux, and the fluxes of the fourth and fifth band (2--12 keV) for the hard X-ray flux.

As 3XMM has the smallest sky area of all our catalogues ($\sim800$ square degrees), it is the limiting factor in this respect. However, X-ray observations are essential to diagnose AGN activity, particularly in complex samples such as ours. Sources that appear extended in the 3XMM catalogue, at high Galactic latitudes, typically fall under two categories: very nearby galaxies or relatively nearby galaxy clusters \citep[the intracluster gas is very hot and typically emits soft X-rays, with a spectral shape that is a combination of bremmstrahlung, recombination and 2-photon radiation, and peaks around 2--5 keV, see e.g.][]{Boehringer2010,Ineson2013,Ineson2015}. By eliminating extended X-ray sources from our sample we can avoid some of the potentially problematic sources in our final sample. Given the method we used to combine FIRST sources (see section \ref{Radio}), this also minimises any cases in which we might have combined radio components from distinct sources in the same cluster. 

After eliminating extended sources, and those with very low detection probabilities (by imposing SC\_EXTENT $\le0.0$, and SC\_DET\_ML $\ge10$), as well as applying the high Galactic latitude cut ($|b_{II}| \ge 20^{\circ}$), we are left with $\sim150000$ 3XMM sources.

It is worth mentioning that 3XMM is not corrected for pile-up effects. Pile-up is only relevant for fluxes above $10^{-11}$ erg cm$^{-2}$ s$^{-1}$. Given the flux distribution of our sources, and the fact that we are using catalogue fluxes, rather than performing full spectroscopic fits, we do not expect pile-up issues to affect our results.

%

\subsection{Mid-IR: WISE all-sky catalogue}\label{WISE}

The WISE catalogue covers the entire sky in four mid-IR bands, 3.4, 4.6, 12, and 22 $\mu$m (W1 to W4, respectively), with a spatial resolution of 6.1--6.5 arcsec for the first three bands, and 12 arcsec for the fourth band \citep{WISE2010,WISE2011}. As such, it is ideally suited to probe AGN activity, and to characterise the host galaxies of systems where AGN activity is not the dominant source of emission in one or more of the bands selected in our catalogue. Given the lower sensitivity and larger pass band of W4, we have focused our analysis on the first three WISE bands, imposing a signal/noise cut of 5 on W1, W2, and of 3 for W3.

For our work we have used the allWISE IPAC release from November 2013 \citep{Cutri2014}, adding up to a total of nearly 750 million sources. As the entire WISE catalogue is very large, and given that our statistical cross-matching tool requires matching sky areas between all catalogues (see section \ref{Sample}), we worked with a subset of WISE sources, obtained by uploading the list of $\sim150000$ pre-selected (see section \ref{3XMM} for the selection criteria) 3XMM source positions to the IPAC allWISE query form\footnote{\url{http://irsa.ipac.caltech.edu/cgi-bin/Gator/nph-scan?submit=Select&projshort=WISE}}, and searching for WISE sources within 60 arcsec of each 3XMM source. The average separation between 3XMM sources (in our sample) is of the order of twice that value; hence such a selection radius guaranteed that the WISE cutout would cover roughly the same sky area as our 3XMM pre-selection. The resulting WISE subset has $\sim$ 1.8 million sources, a small fraction of the original number. We carried out the WISE signal/noise cuts after cross-correlating the catalogues, to keep as many sources for as long as possible. The S/N cuts in W1 and W2 do not reduce the number of sources by a large amount ($\sim8$ per cent), but W3 is far less sensitive, and even a required detection on a 3$\sigma$ level, rather than 5$\sigma$, cuts our sample size by half. While the 12$\mu$m band is essential to characterise the AGN emission related to the torus, there is a large amount of diagnostics and science that can be carried out simply with W1 and W2, particularly for sources not dominated by the AGN.

For the following sections we have considered, separately, those sources that pass the signal to noise cut in W1 and W2, but not W3 (Full Sample, Full Redshift Sample, see section \ref{z}), and those that pass also the W3 signal to noise cut (W3 Sample, W3 Redshift Sample).

The vast majority of the WISE sources in our samples are classified as point-like in the catalogue (ext\_flg=0), and in those cases we have used the standard apertures provided (w1mpro, w2mpro, w3mpro). For sources labelled as extended (ext\_flg=3 and ext\_flg=5, $\sim0.6$ per cent and $\sim17$ per cent of our sources, respectively) we used the provided 2MASS corrected elliptical apertures (w1gmag, w2gmag, w3gmag) instead of the standard apertures, as suggested in the on-line documentation, as they are likely to give more accurate results. We also checked the quality flags for potential problems, and found them to be good after the S/N cuts were implemented.

%

\subsection{Radio: combining FIRST and NVSS}\label{Radio}

At low to intermediate radio frequencies, specifically at 1.4 GHz, there are two large radio surveys that would be ideally suited for our purposes: FIRST \citep[Faint Images of the Radio Sky at Twenty centimetres][]{FIRST1995,FIRST2015} and NVSS \citep[NRAO VLA Sky Survey][]{NVSS1998}. FIRST has higher spatial resolution ($\sim5$ arcsec, versus $\sim45$ arcsec for NVSS) and thus better positional accuracy, and it is deeper (1 mJy detection level, versus 2 mJy for NVSS) but it covers a smaller area of the sky, as it was designed to coincide with the Sloan Digital Sky survey ($\sim10500$ square degrees, while NVSS covers 82 per cent of the sky, all the area north of $\delta=-40^{\circ}$). Most importantly, given the frequently extended and multi-component nature of radio sources, FIRST can split sub-components from resolved FRI and FRII \citep{FR1974} galaxies into several catalogue entries, which could create false matches in a cross-correlation with a higher sky density catalogue (e.g. independent optical sources matched to the core and lobes of the same radio galaxy) and, in some correct matches, yield only partial integrated fluxes (e.g. the radio core is matched to a counterpart at other wavelengths, but the lobes are not). The lower resolution of NVSS can avoid these problems, but when combined with its lower positional accuracy, especially at low fluxes, it can result in off-centre positions and inaccurate errors, which in turn increase the risk of missed matches. We also want to reach the lowest possible fluxes, to include objects with small jets and lobes that would normally be classified as radio-quiet, as well as objects where the radio emission is produced by star formation, rather than AGN activity. 

There have been some notable attempts to combine these two radio surveys before. The Unified Radio Catalogue of \citet{Kimball2008,Kimball2014} also includes sources from the Green Bank 6 cm (GB6) and Westerbork Northern Sky (WENSS) 92 cm survey, which could potentially be useful to estimate the spectral indices of some sources, but it only provides lists of possible counterparts for each entry in those catalogues, leaving to the user how to group and use them. The catalogue of \citet{Best2005}, and its later improved version by \citet{Donoso2009}, uses a very reliable method to group multi-component sources, but relies on prior assumptions by first cross-matching the NVSS sources with Sloan (SDSS) detected optical sources. As a general rule, we prefer to avoid imposing prior cuts on the data, as it is possible to assess the nature of the matches at a later stage and minimise the bias. We therefore decided to combine NVSS and FIRST using criteria that would suit our specific purposes.

We used the latest version of NVSS, which contains 1773484 sources with integrated fluxes and positional errors, and is available through the NVSS public ftp server\footnote{\url{ftp://nvss.cv.nrao.edu/pub/nvss/CATALOG/FullNVSSCat.text}}, and through VizieR. For FIRST we used the March 2014 release, which includes 946432 sources\footnote{\url{http://sundog.stsci.edu/first/catalogs/readme_14mar04.html}}. The catalogue does not provide specific position errors, so we used uncertainties of 0.5 arcsec for fluxes larger than 3mJy, and 1 arcsec at lower fluxes, as suggested by the on-line documentation. While this is likely an overestimation, the positional errors are still much smaller than those in NVSS, and comparable to those in the X-ray survey we will be using (see section \ref{3XMM}). 

The first step was to choose a suitable way to combine potential FIRST sub-components, in cases with and without an NVSS potential match. Although the clustering of the radio sky is low \citep[see e.g.][]{Magliocchetti2016b}, and the likelihood of finding more than one (relatively powerful) radio galaxy in a cluster is also low, there is always a risk of grouping distinct sources into one. We chose a conservative grouping radius, 30 arcsec, which is compatible with separations found in earlier studies based on FIRST \citep[e.g.][]{Cress1996,Magliocchetti1998,Gubanov2003}. These studies also explored the clustering of the radio sky at larger radii, finding typical scales of a few arcmin. 

For very nearby sources this 30 arcsec radius might prove too small to include all the subcomponents of an extended radio source. Our selection criteria for the X-ray sources, however, minimise this problem (we excluded extended X-ray sources, see section \ref{3XMM}). The X-ray selection excludes not only nearby clusters, but also nearby galaxies, which may appear extended in 3XMM, thus we minimise the presence of very nearby radio sources that might have bright radio components with very large separations on the sky. We decided to use a single collapsing radius, rather than the flux-dependent approach used by \citet{Magliocchetti1998}, for simplicity, as using larger radii for bright FIRST sources would have required us to group NVSS sources as well (the beam used for NVSS had a 45 arcsec FWHM, so for separations around and beyond that scale the possibility of NVSS groups would have to be considered), and would have badly impacted the positional accuracy of the resulting combined sources.

\begin{figure}
\centering
\includegraphics[width=0.46\textwidth]{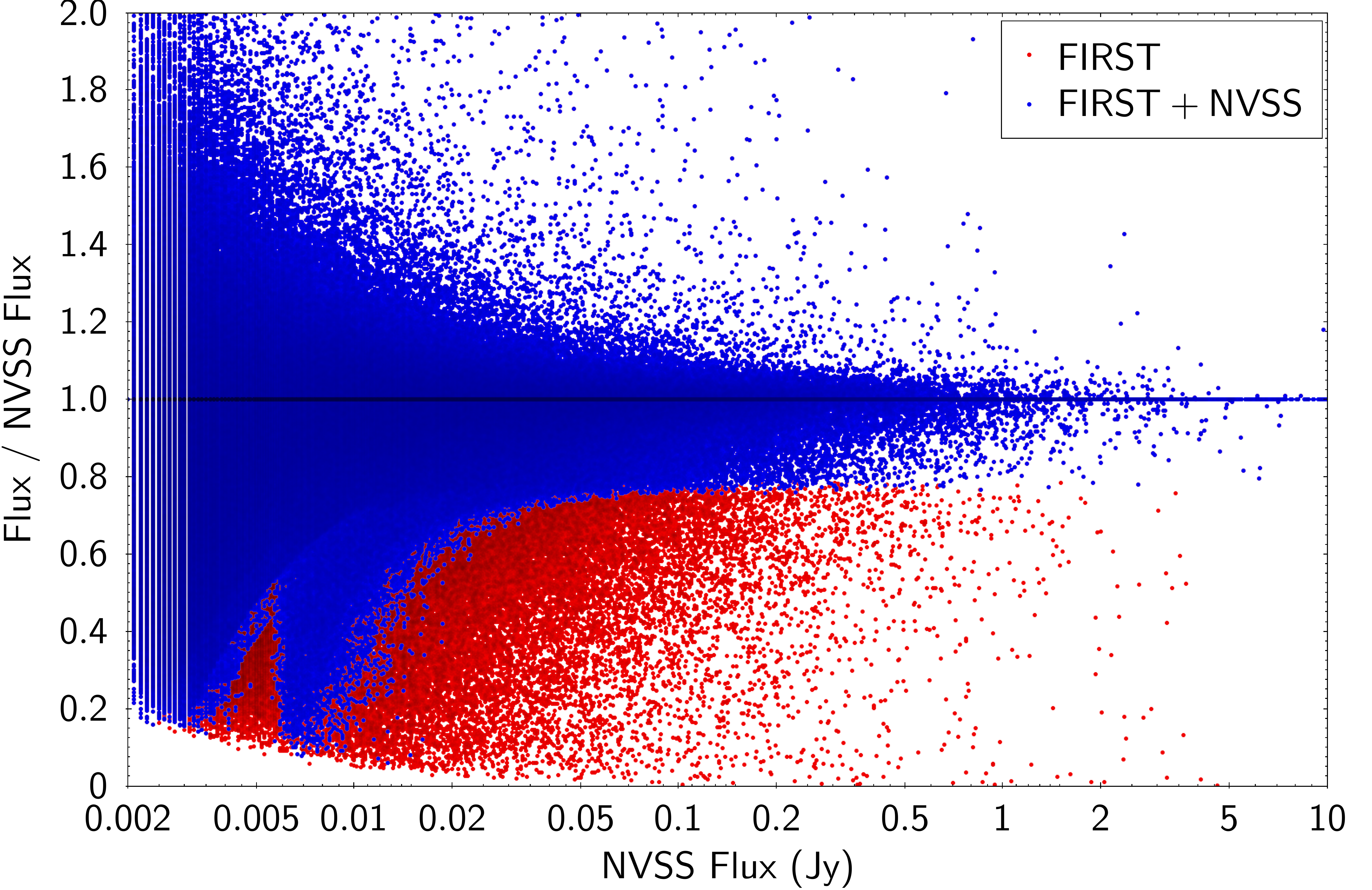}
\caption{Ratio between the integrated flux from FIRST (red) or the combined catalogue (blue, overlaid) over the NVSS integrated flux, as a function of the NVSS flux. This plot illustrates the result of applying our 5$\sigma$ criterion when selecting the combined flux, and serves as a direct comparison to Fig. 11 in \citet{FIRST2015}.}\label{RadioFlux}
\end{figure}

\begin{table}\small
\caption{Number of FIRST subcomponents within 30 arcsec, collapsed into a single source. Only $\sim10$ per cent of the sources have multiple components.}\label{nComp}
\centering
\begin{tabular}{ccc}\hline
Subcomponents&Number of cases&Percentage of total groups\\\hline
2&80105&81.2\\
3&14659&14.9\\
4&3145&3.2\\
$\geq 5$&738&0.7\\\hline
\end{tabular}
\end{table}

We therefore grouped FIRST sources within 30 arcsec of an NVSS source, or, where no NVSS counterpart was present, within 30 arcsec of another FIRST source. For each combination we co-added the fluxes, and combined the positions using a flux-weighted average, which is likely to give the best estimate for the core position in all cases \citep[the brightest lobe is closer to the core in most FRI and FRII sources, see][]{Magliocchetti1998}. The positional errors were assumed to be the larger between a flux-weighted sum of the individual positional errors and the standard deviation of the individual positions, for both of which we assessed RA and DEC separately. To minimise issues with the inaccurate positions in NVSS at low fluxes, for the combined entries in the catalogue we used FIRST fluxes and positions unless an NVSS match was present and had a flux larger by at least 5$\sigma$. Fig. \ref{RadioFlux} illustrates the effect of this selection. In this Figure we have plotted $F_{FIRST}/F_{NVSS}$ versus $F_{NVSS}$ in red, and overlaid $F_{Combined}/F_{NVSS}$ versus $F_{NVSS}$ in blue. For the combined (blue) distribution, any sources where we have used the NVSS value follow the 1:1 horizontal line, and any sources where we used a FIRST flux (different from the NVSS flux) will deviate from the 1:1 line. So any areas of the plot where the red FIRST distribution appears represent sources where the FIRST fluxes were smaller by at least 5$\sigma$, and NVSS fluxes were used. The plot shows a very large scatter at low NVSS flux values, a consequence of the large uncertainties in the NVSS fluxes near the detection limit, thus our method selected the (grouped) FIRST values for the majority of these cases. At fluxes around 0.05--0.1 Jy, some FIRST fluxes are noticeably smaller than their NVSS counterparts, perhaps due to our conservative collapsing radius, so the NVSS fluxes were used. For larger fluxes FIRST and NVSS tend to agree, and the distribution becomes narrower around the 1:1 line. While the effect is subtle, our plot shows some differences from Fig. 11 in \citet{FIRST2015}. It is difficult to determine how much of this difference is caused by the much larger number of sources we are using, and how much it is due to the fact that we are grouping FIRST subcomponents (this should narrow the distribution). 

Using these criteria, we found 98647 groups ($\sim 4.6$ per cent of the total number of sources, 2129340, $\sim 10$ per cent of the number of FIRST sources). The number of subcomponents per collapsed source is given in Table \ref{nComp}. These numbers give us a rough idea of the number of resolved FRI and FRII galaxies in FIRST, but they are not representative of the entire population, as many FRI and some distant FRII are not split into separate objects in FIRST, even if they show resolved structures ($\sim35$ per cent of the sources have resolved structures on scales of 2--30 arcsec). The number of subcomponents is also not a reliable diagnostic for the radio morphology, as for many FRI sources the core and one of the lobes might be grouped due to orientation effects or Doppler suppression, making them appear as doubles in Table \ref{nComp}, rather than triples.

As for WISE (section \ref{WISE}), only a small fraction of the radio sources fall within 60 arcsec of a 3XMM source, $\sim17500$ (see the details of the area of overlap at the end of the next section). When we compare this number to those in 3XMM and the equivalent fraction of the WISE catalogue, the low radio sky density becomes immediately apparent.

As our combined FIRST+NVSS catalogue might prove useful to other researchers, and only a subset of its sources are used in the final MIXR sample, we have made it available on-line as a stand-alone file at \url{http://www.arches-fp7.eu/index.php/tools-data/downloads/combined-radio-catalogue} and will also upload it to VizieR. As with any catalogue, it may not suit every purpose: please consider carefully the caveats described in this section.

%

\subsection{MIXR Sample Construction}\label{Sample}

For our sample construction we used a cross-correlation tool developed as part of the ARCHES collaboration products \citep[Pineau et al. 2016, subm.; ][]{Pineau2015}. This tool is based on an earlier version, tested on the 2XMMi and SDSS DR7 catalogues \citep{Pineau2011}. Our version of the tool also uses a chi-square ($\chi^{2}$) statistical hypothesis test to select probable candidates, but applied on n-catalogues instead of only two. The likelihoods we use to compute Bayesian probabilities are $\chi$ distributions of various degrees of freedom and multidimensional Poisson distributions. These distributions are normalised so their integration over the $\chi^{2}$-test region of acceptance equals 1. Priors are derived both from local densities of sources in each catalogue and from the results of the cross-correlation of each possible subset of catalogues. Currently, the tool is able to provide probabilities for up to 8 catalogues. For more information see Pineau et al. (2016, subm.), and the ARCHES website\footnote{\url{http://www.arches-fp7.eu/index.php/tools-data/online-tools/cross-match-service}}. The \textsc{xmatch} tool is available as a web service through the ARCHES website, and will eventually supersede the two-catalogue tool currently available at the CDS cross-match service\footnote{\url{http://cdsxmatch.u-strasbg.fr/xmatch}}.

The association probabilities for any given tuple of sources depend on the normalised distances of the individual sources from the averaged position (up to the equivalent of the 1-dimension 3$\sigma$ level, i.e. 99.7 per cent completeness), as well as the sky density for each given catalogue (which is why it is fundamental that the sky coverage of all the cross-matched catalogues coincide and are accurate). 

Although WISE covers the entire sky, FIRST and NVSS only cover latitudes north of $-40^{\circ}$, and 3XMM covers patches throughout the entire sky. The area of overlap between the three individual catalogues is roughly 135 square degrees. We carried out a simultaneous match of the overlapping sections of the combined FIRST/NVSS catalogue ($\sim17500$ sources), WISE ($\sim1.8$ million sources) and the cleaned-up 3XMM ($\sim150000$ sources), using two inner joins (i.e. keeping only the tuples that had a candidate in each catalogue). For the sources resulting from the three-catalogue cross-correlation, we aimed for maximum completeness, requiring an association probability greater than 1$\sigma$ ($\sim70$ per cent), and thus obtaining a sample of 2753 sources (with a reliability of 90.15 per cent), reduced to 2529 in the full sample, and 1575 in the W3 sample.

%

\section{Activity diagnostics}\label{Diagnostics}

To accurately characterise the multiwavelength behaviour of an extragalactic source, we need to know its distance (redshift), from which we can derive its luminosity in each band. However, redshifts are not always available, consistent, or accurate. In this section we demonstrate how it is possible to pre-emptively diagnose the type of activity present in a sample of sources, using the three bands in MIXR. These diagnostics are very useful to test the accuracy of single-band activity markers commonly employed in other surveys, especially those that do not require a radio selection, as well as to better constrain an a priori range of models for systematic spectral or SED fitting, which can yield more accurate redshift values than those obtained by cross-correlation with an extra catalogue. In sections \ref{LDiag} and \ref{RL_RQ_section} we will verify the accuracy of these preliminary diagnostics.

Please note that, because we want these diagnostic plots to be as straightforward as possible for the end user, we have not converted between flux and magnitude systems: for WISE we plot aperture-corrected magnitudes; for the X-rays we plot fluxes in cgs (erg cm$^{-2}$ s$^{-1}$); for the radio, we plot fluxes in mJy.

%

\subsection{The WISE colour/colour plot}\label{C_C}

\begin{figure*}
\centering
\includegraphics[width=0.92\textwidth]{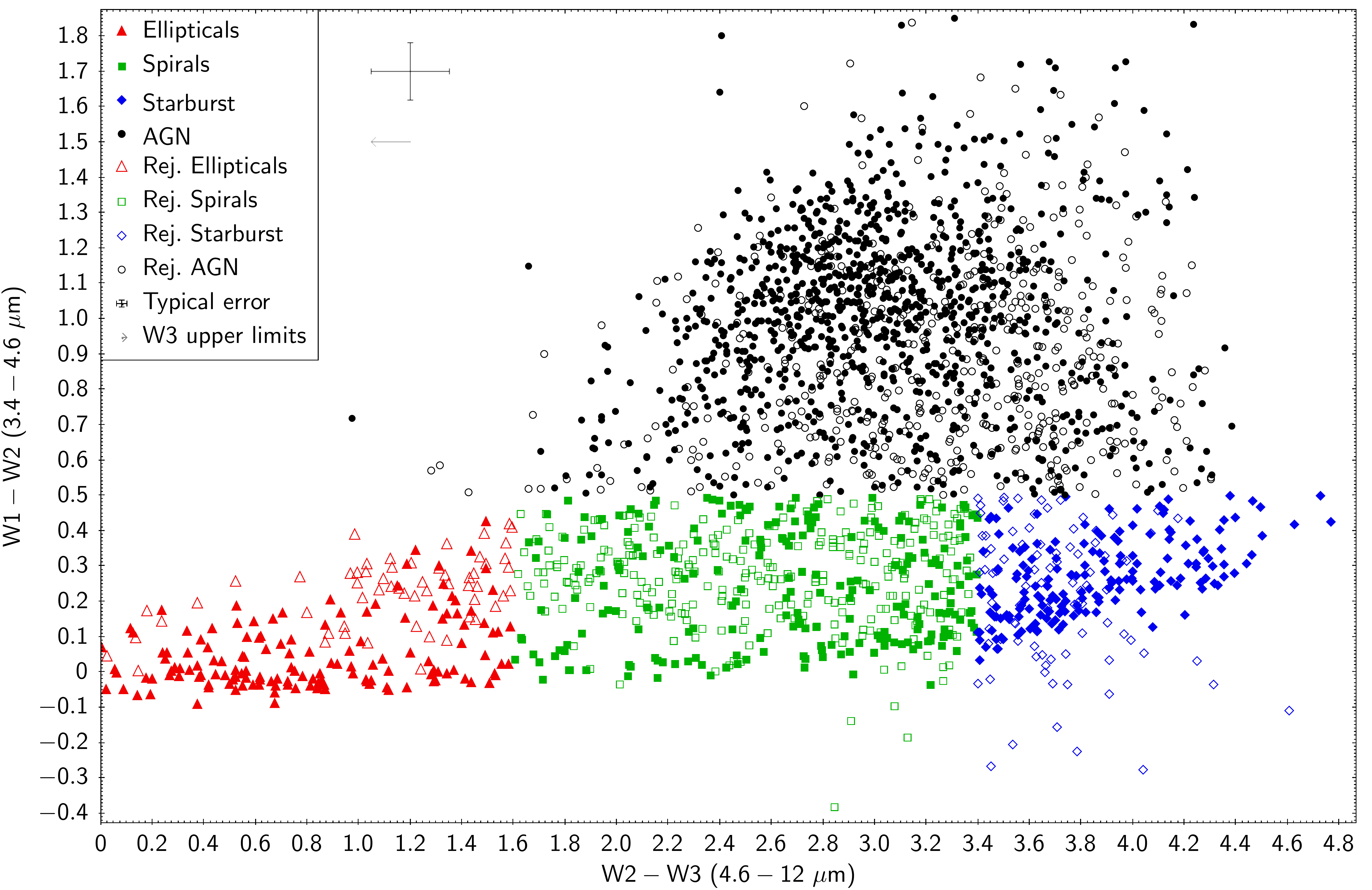}
\caption{Colour/colour diagram for the sources in our Full Sample (see Table \ref{nSources} for details on sample and source type statistics). We have plotted with empty symbols the W3 rejected sources, and with full symbols the sources in the W3 Sample. The cross and the arrow by the legend indicate the typical size of the errors and the direction of the W3 upper limits, respectively.}\label{C_C_all}
\end{figure*}

The first diagnostic we tested involves mid-IR colours, and is based on the work by \citet{Lake2012}. For their work, Lake et al. generated a series of synthetic SEDs for a wide range of astronomical populations, and plotted them on the WISE colour/colour diagram (W1-W2 versus W2-W3 magnitudes, see their Fig. 1) to test which regions of the parameter space they occupied. While this diagnostic is extremely useful for a first approach to study what the AGN and the host galaxy are doing, it is important to keep in mind that there is overlap between populations, even more so when obscuration and redshift evolution are taken into account \citep[see e.g. Fig. 1 of][]{Hainline2014}, and that several colour cuts have been proposed to identify the AGN population in particular \citep[e.g.][]{Ashby2009,Assef2010,Assef2013,Stern2012,Mateos2012}.

\begin{table*}\small
\caption{Activity table. For each of our source types, selected on the WISE colour/colour plot, this table shows the types of activity most likely to be found at each wavelength. Please note that, for each colour category, several combinations of the elements in columns 2--4 may be possible, e.g. in the first group, an elliptical galaxy in a cluster, with a radiatively inefficient AGN in X-rays, and a LERG in radio.  LINER stands for low-ionisation nuclear emission-line region. ULIRG stands for ultraluminous infrared galaxy. LERG stands for low excitation radio galaxy; high excitation sources (HERG) include NLRG (narrow line radio galaxies) and BLRG (broad line radio galaxies). Please see also Table \ref{nSources} for the statistics of each subset.}\label{Activity}
\centering
\begin{tabular}{lllll}\hline
Label&WISE colour selection&Mid-IR/Optical&X-rays&Radio\\\hline
\multirow{3}{*}{Elliptical}&\multirow{3}{*}{$W1-W2<0.5$;  $0<W2-W3<1.6$}&Elliptical galaxy (isolated)&\multirow{2}{*}{Rad. inefficient AGN}&\multirow{3}{*}{LERG}\\
&&Elliptical galaxy (cluster)&\multirow{2}{*}{Hot ICM gas}&\\
&&LINER&\multirow{-1}{*}{}&\\\hline
\multirow{3}{*}{Spiral}&\multirow{3}{*}{$W1-W2<0.5$;  $1.6\leq W2-W3<3.4$}&\multirow{2}{*}{Star-forming galaxy}&\multirow{2}{*}{Star formation}&Star formation\\
&&\multirow{2}{*}{Star-forming galaxy + AGN}&\multirow{2}{*}{Seyfert galaxy}&Low-L NLRG\\
&&&&LERG\\\hline
\multirow{2}{*}{Starburst}&\multirow{2}{*}{$W1-W2<0.5$; $W2-W3 \geq 3.4$}&Starburst galaxy&Star formation&Star formation\\
&&ULIRG&Seyfert galaxy&Low-L NLRG\\\hline
\multirow{3}{*}{AGN/QSO}&\multirow{3}{*}{$W1-W2 \geq 0.5$; $W2-W3<4.4$}&\multirow{3}{*}{AGN}&Luminous Seyfert galaxy&NLRG\\
&&&BL-Lac&BLRG\\
&&&QSO&QSO\\\hline
\end{tabular}
\end{table*}

At our flux and magnitude limits, and the high Galactic latitude we are working with, we do not expect to see the stellar objects that appear in the diagram of \citet{Lake2012}, but we pre-emptively excluded 15 sources with W2-W3 values smaller than zero. We also excluded 24 sources in the ULIRG/obscured AGN locus (W1-W2$>0.5$, W2-W3$\geq 4.4$), as this area of the diagram shows severe contamination from resolved star formation regions in extremely nearby galaxies, and it is unclear, for the extragalactic sources in this area, whether they could be treated systematically as AGN or starburst galaxies (including obscured AGN would require us to increase the range of $N_H$ values we use to calculate the X-ray luminosities in section \ref{LDiag}, skewing the results for the entire sample). We give the rest of the sources a rough characterisation based on the labels in the work by Lake et al.; the resulting colour/colour diagram is shown in Fig. \ref{C_C_all}. Table \ref{Activity} describes in detail the boundaries we imposed, and what type of activity we expect to find in each population \citep[see e.g. the source distributions on the equivalent WISE colour-colour plots of][]{Gurkan2014,Yang2015}. Table \ref{nSources} (in section \ref{z}) shows the statistics for each source type. Please note that, until we know more about the underlying properties of our sources, the categories in Table \ref{Activity} are only meant as a rough guide. Galaxies, like almost everything in the Universe, do not fall into neatly cut categories \citep[note how the classifications of][overlap on their colour/colour plot, as well as the size of the errors in Fig. \ref{C_C_all}]{Lake2012}. Thus, some sources may not fall into the general behaviour expected for their assigned categories (e.g. a source with the `elliptical' classification that shows signs of star formation or a radiatively efficient, bright AGN). Also note that, while radio sources are traditionally organised based on the Fanaroff-Riley classification \citep{FR1974}, we have not used this classification for Table \ref{Activity}, as the FRI/FRII divide is based both on morphology and on radio power, both of which depend heavily on the environment through which the jet and lobes propagate \citep[see e.g.][]{Hardcastle2013,Hardcastle2014,English2016}.

We have also pre-emptively labelled the sources that lack a reliable detection in the W3 band, represented by empty symbols in Fig. \ref{C_C_all} to signal that they are upper limits (the arrow next to the legend on the plot indicates the direction of the upper limits). We will show in the next Subsections that, overall, these sources behave very similarly to those in the same region of the colour/colour plot that do have a W3 S/N>3, demonstrating that they belong to the same populations, and that their faintness in the 12$\mu$m band is due to their larger distance or lower luminosity, rather than a mis-classification. We checked the W4 results, for consistency, and found even fewer detections than for W3, as expected.

In Table \ref{Activity} we see that the expected AGN classifications, both for the radio and the X-rays, are not clear-cut for each mid-IR population. Throughout this work we aim to study how accurate our mid-IR labels are with respect to the underlying activity (e.g. what fraction of moderately star-forming galaxies and what fraction of AGN we find among the `spiral' sources in Fig. \ref{C_C_all}). The diagnostic plots in the following Subsections aim to shed some light on this topic, but to get a clearer idea of what type of AGN each host harbours we need X-ray, bolometric and jet kinetic luminosities, which we will study in sections \ref{z} to \ref{Power}.

%

\subsection{X-ray hardness ratios and radio versus X-ray `loudness'}\label{HR}

\begin{figure*}
\centering
\begin{subfigure}{0.48\textwidth}
  \centering
  \includegraphics[width=.98\linewidth]{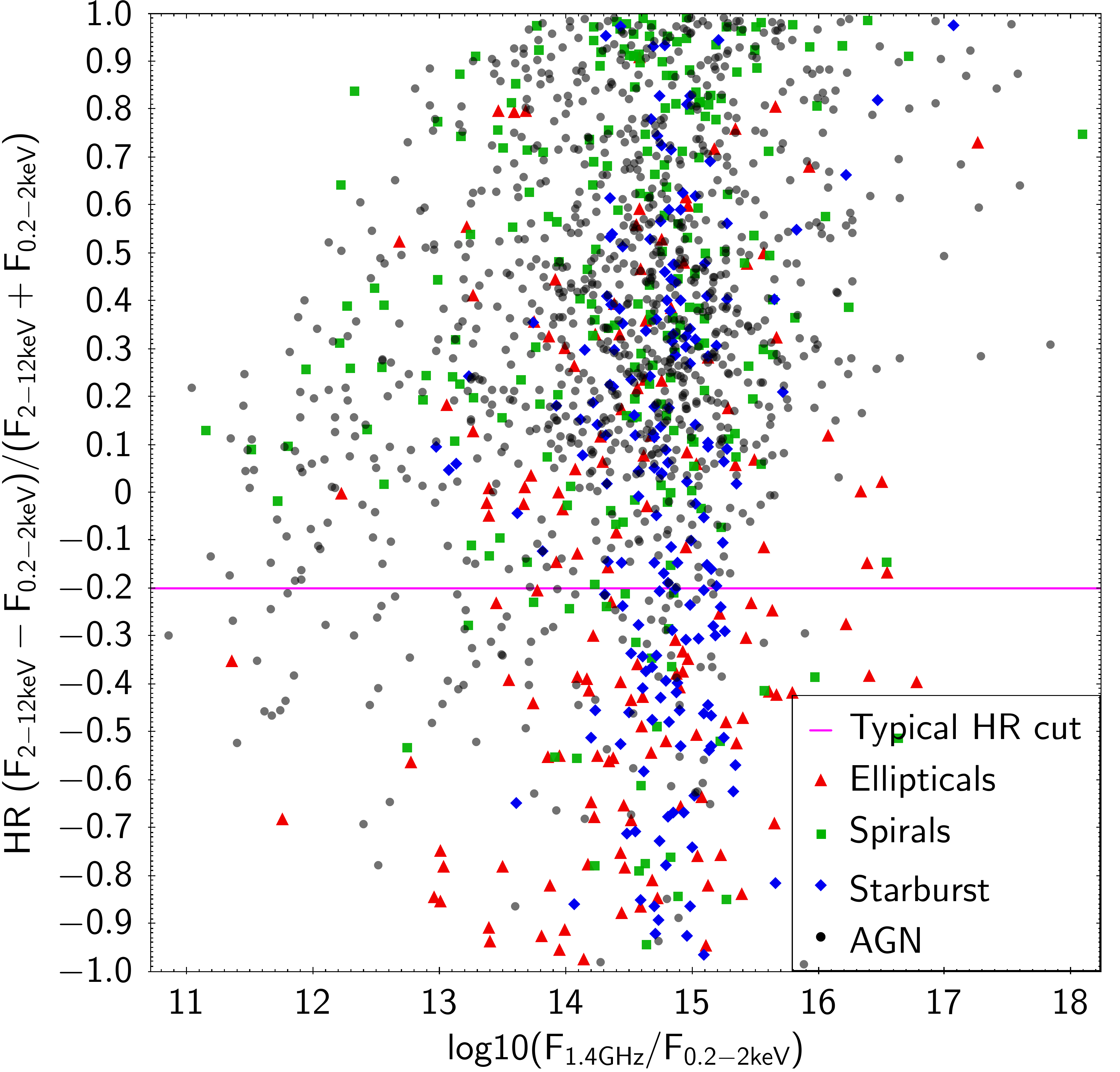}
  \caption{Hardness ratio - radio/soft X-ray flux}
  \label{HR_all}
\end{subfigure}
\begin{subfigure}{0.48\textwidth}
  \centering
  \includegraphics[width=.98\linewidth]{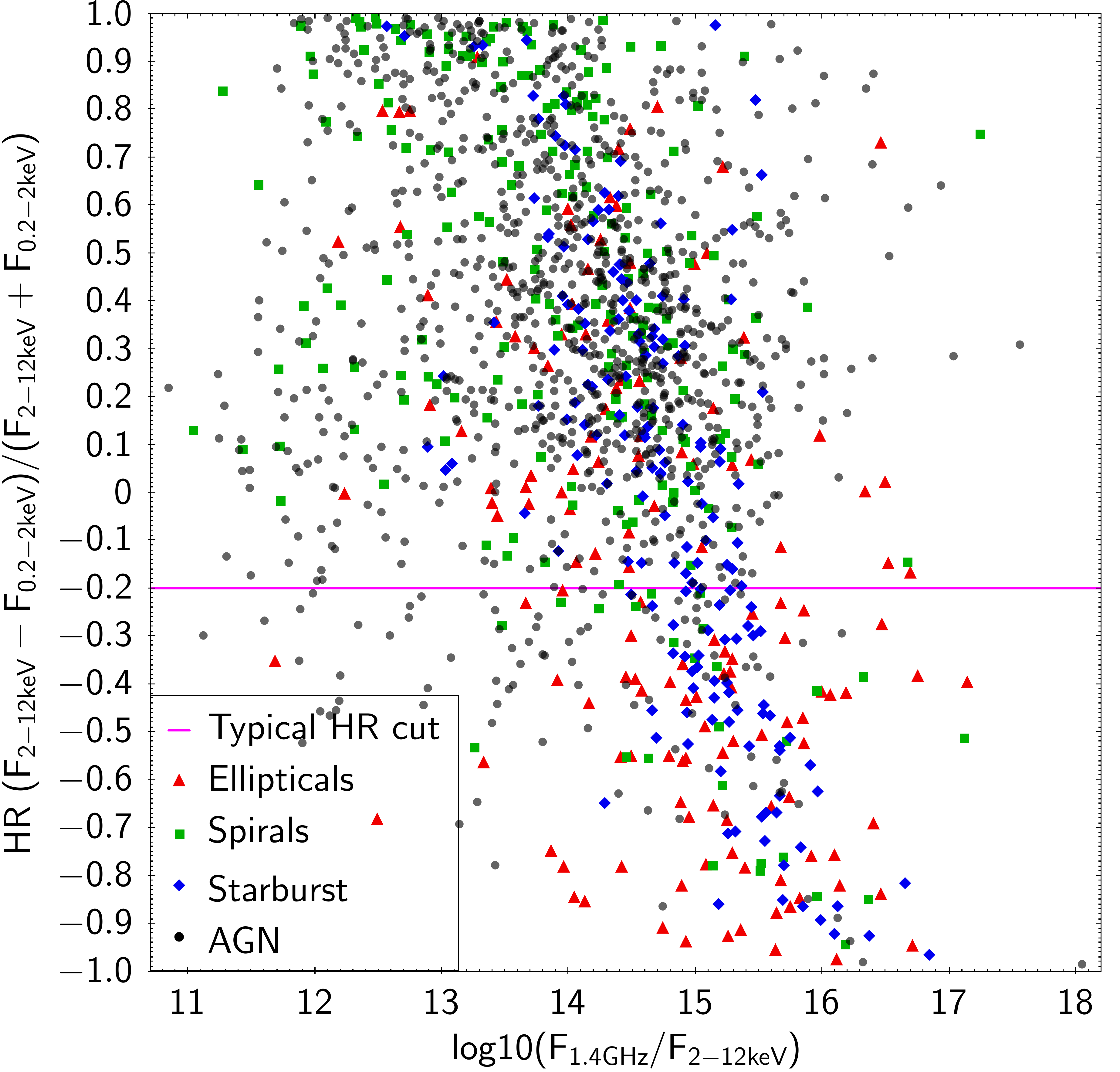}
  \caption{Hardness ratio - radio/hard X-ray flux}
  \label{HR_all_2}
\end{subfigure}
\caption{X-ray hardness ratio versus ratio of the 1.4 GHz radio flux (in mJy) to the X-ray flux (left: soft, 0.2--2 keV; right: hard, 2--10 keV, in erg cm$^{-2}$ s$^{-1}$) for all the sources. Colours and symbols are the same as in Fig. \ref{C_C_all}. Only sources in the W3 sample are plotted, for clarity.}
\label{HR_all_comb}
\end{figure*}

Many X-ray selected samples rely on hardness ratio cuts to eliminate non-AGN sources, as well as to determine the obscuration and distance of a given set of AGN. Normally it is preferable to use net (background-subtracted) counts, rather than fluxes, to estimate the hardness ratio. However, due to the nature of our X-ray catalogue, where more than one observation with multiple instruments can be present for any given source, we decided to use the averaged (over all the observations for each source), net fluxes for our analysis. The hardness ratio we use is defined as:
\begin{equation}\label{HR_eq}
HR=(F_{2-12 keV}-F_{0.2-2 keV})/(F_{2-12 keV}+F_{0.2-2 keV})
\end{equation}
The fluxes in our catalogue are biased by the model assumed to derive them from the raw counts \citep[see section \ref{LDiag} and][for details]{Mateos2009}, and thus the following plots should be considered carefully, particularly for sources with X-ray spectral shapes very different from the assumed spectral shape used to represent AGN emission ($\Gamma = 1.7$, $n_H\sim 10^{21}$ cm$^{-2}$). Reassuringly, star-forming sources do not greatly deviate from this approximation on average \citep{Ranalli2012}, but radiatively inefficient and Compton-thick AGN may be represented less accurately in these hardness ratio plots, the first due to their being dominated by the soft, jet-related component \citep[e.g.][]{Hardcastle1999}, the latter, which are not common in our sample (see section \ref{RL_RQ_section}), because of the heavy absorption and Compton reflection. Please note that we have used 2--12 keV fluxes, as we were constrained to the bands defined in the catalogue. At this point of the analysis our catalogue fluxes are also not corrected for foreground absorption, but given that we are working at high Galactic latitudes, the effect of Galactic obscuration should be unimportant in this band.

Fig. \ref{HR_all_comb} shows the distribution of hardness ratios for all the sources on the Y axis, and the ratio of the radio to the relative X-ray (soft and hard, Figs. \ref{HR_all} and \ref{HR_all_2}, respectively) flux on the X axis. This is a very good way to quickly assess the `radio loudness' and `X-ray loudness' of the sources, as well as to establish whether a soft excess or deficit may be related to the same processes that produce the radio emission. 

\begin{figure}
\centering
\includegraphics[width=0.46\textwidth]{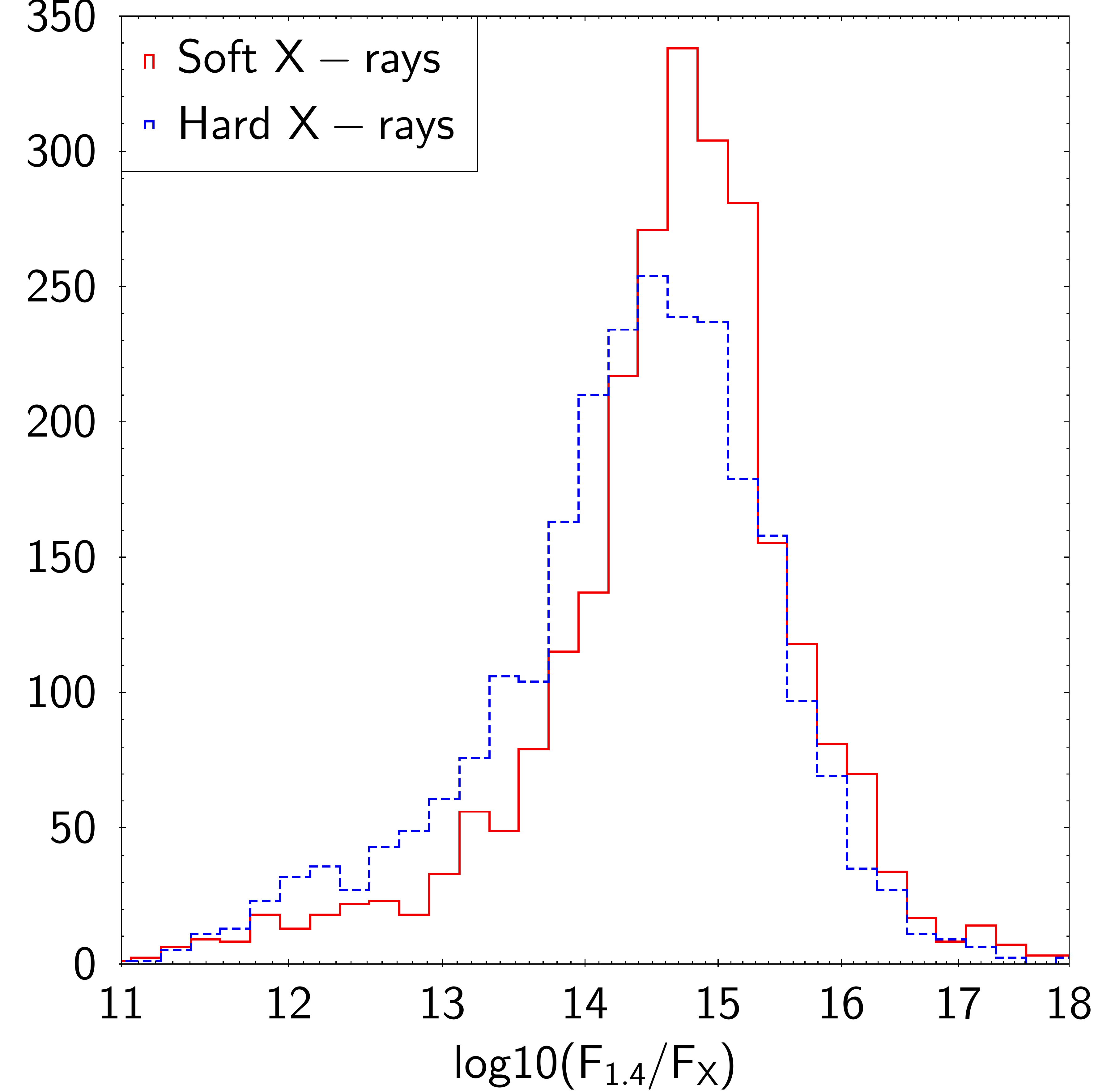}
\caption{Histogram of the radio to soft (red) and hard (blue, dashed) X-ray flux, for all the sources regardless of the W3 S/N cut (full sample).}\label{Fr_Fx_histo}
\end{figure}

\begin{figure}
\centering
\begin{subfigure}{0.238\textwidth}
  \centering
  \includegraphics[width=.95\linewidth]{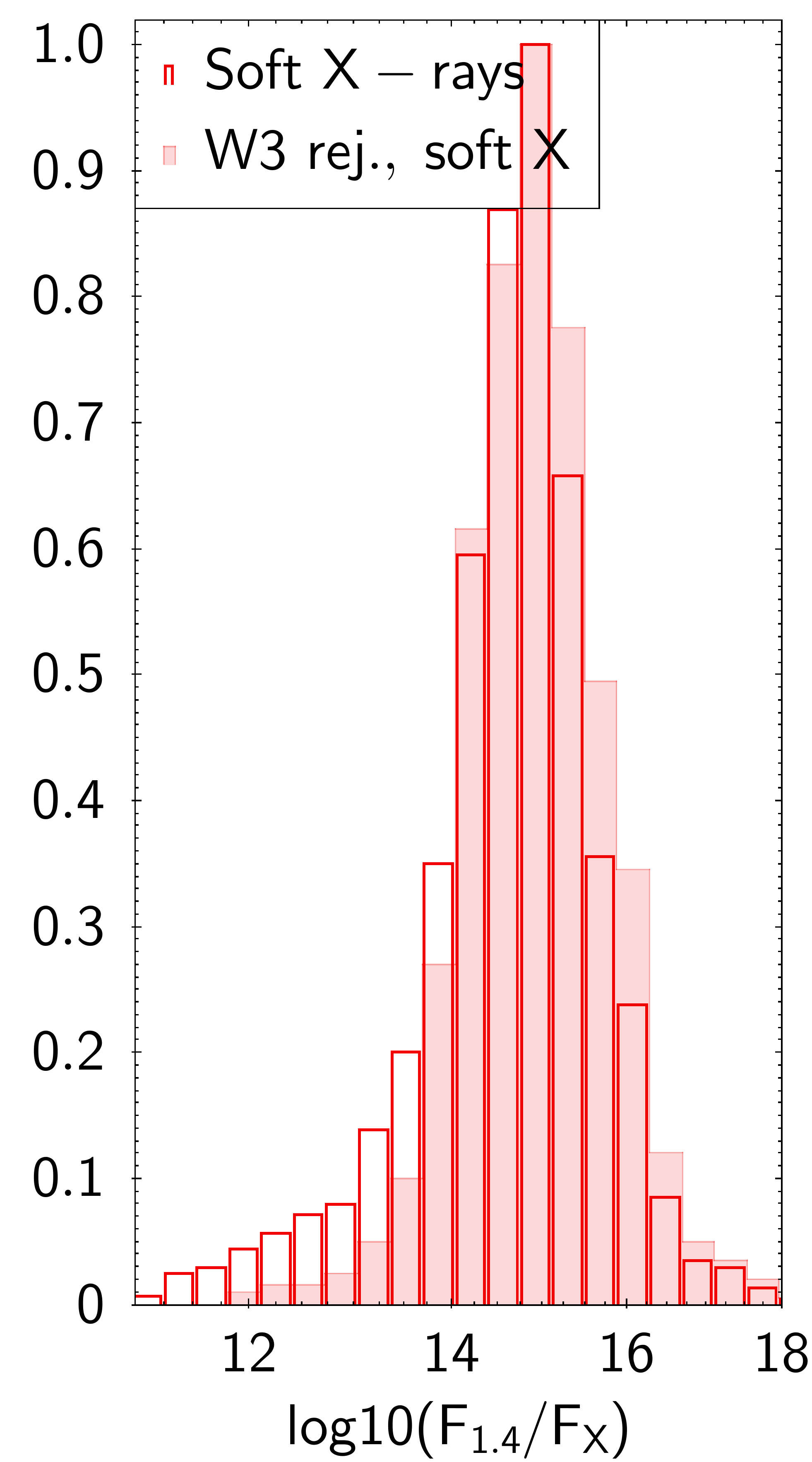}
  \caption{}
\end{subfigure}
\begin{subfigure}{0.238\textwidth}
  \centering
  \includegraphics[width=.95\linewidth]{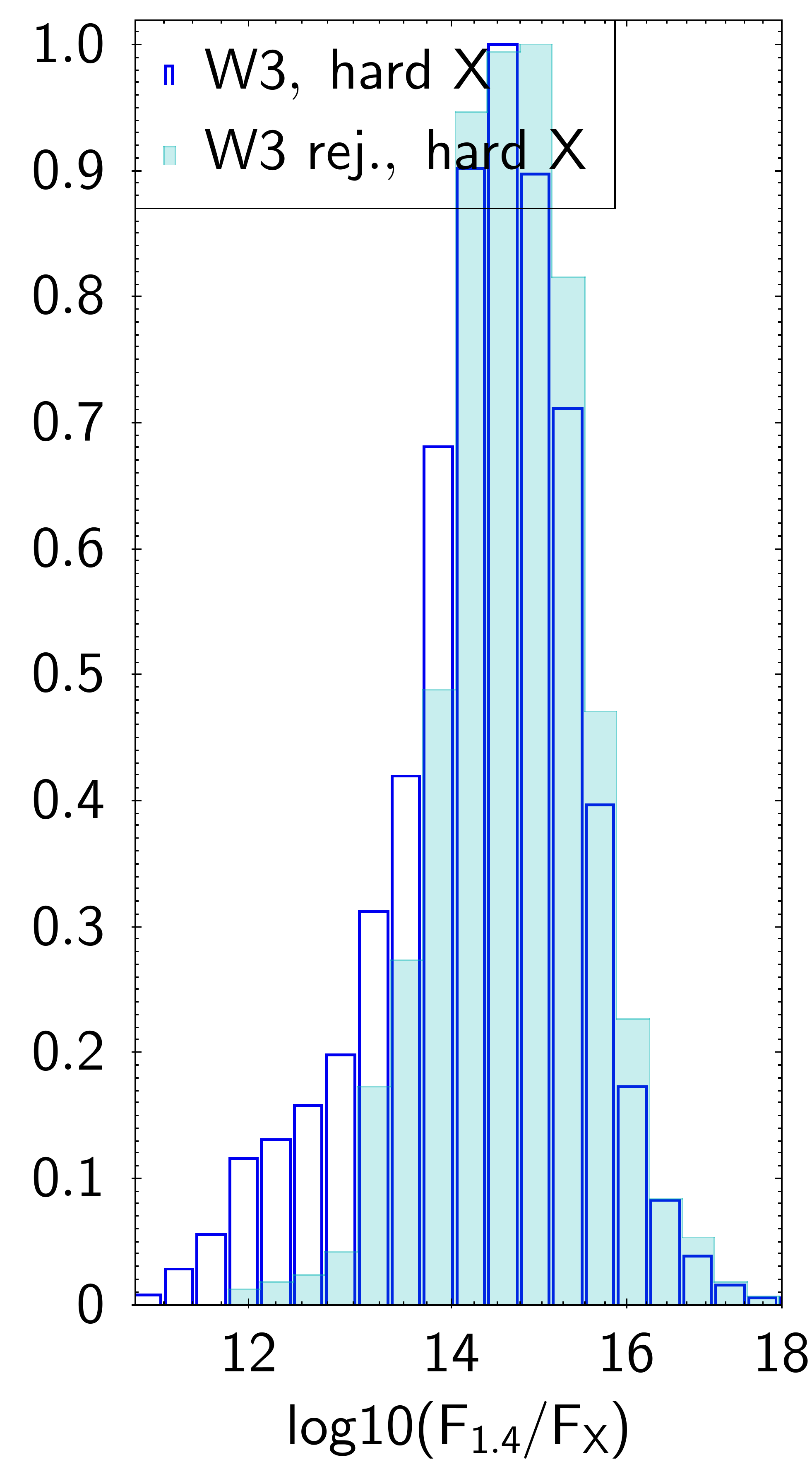}
  \caption{}
\end{subfigure}
\caption{Normalised histogram of the radio to soft (left, red) and hard (right, blue) X-ray flux, for the sources that pass (empty bars) and do not pass (full bars) the W3 S/N cut.}
\label{Fr_Fx_W3}
\end{figure}

Despite the large number of points, it is quite clear in these plots that most of the sources have rather high ($>0$) hardness ratios (as a guide: the `typical HR cut' barrier in our diagrams corresponds to an unabsorbed spectrum with $\Gamma=2$; an unabsorbed AGN with $\Gamma=1.7$ would have a HR of 0.17; an AGN with $N_{H}=10^{22}$ cm$^{-2}$, $\Gamma=1.7$, $z=1$, would have a HR of 0.53). The overall trend also varies depending on whether the hard or the soft X-ray flux is used for the ratio on the x axes: for soft X-rays (Fig. \ref{HR_all}) the ratio seems fairly constant, as evidenced by the aggregation of sources around a vertical line at $F_{1.4 GHz}/F_{0.2-2 keV}\sim10^{14}-10^{15}$, with some outliers, especially on the left side of the plot (larger relative X-ray fluxes). For the hard X-rays (Fig. \ref{HR_all_2}), the overall behaviour is slightly different, and there seems to be a slight negative trend between the radio/hard X-ray flux and the hardness ratio, meaning that more `hard X-ray loud' (less `radio-loud') sources have harder spectra. These behaviours become more evident when we plot the flux ratio histograms for both distributions (Fig. \ref{Fr_Fx_histo}). This negative trend probably arises from a combination of factors: the known radio-soft X-ray correlation of \citet{Hardcastle1999}, which will push radio-louder (softer) sources to the right of the plot, and radio-quieter (harder) sources slightly to the left; a higher intrinsic absorption for the hardest sources, which would hide a similar negative trend (pushing the hard sources to the right) in the soft X-rays; uncertainties derived from the underlying 3XMM flux derivation; higher intrinsic hard X-ray fluxes for the harder sources (e.g. from higher Eddington rates). Given that we are working with flux-limited samples, it is not possible to analyse the strength of a possible anticorrelation between the hardness ratio and the radio/hard X-ray flux ratio, as we do not know what sources may be missing from these plots beyond the flux limits.

\begin{figure*} 
  \begin{subfigure}[b]{0.5\linewidth}
    \centering
    \includegraphics[width=0.95\linewidth]{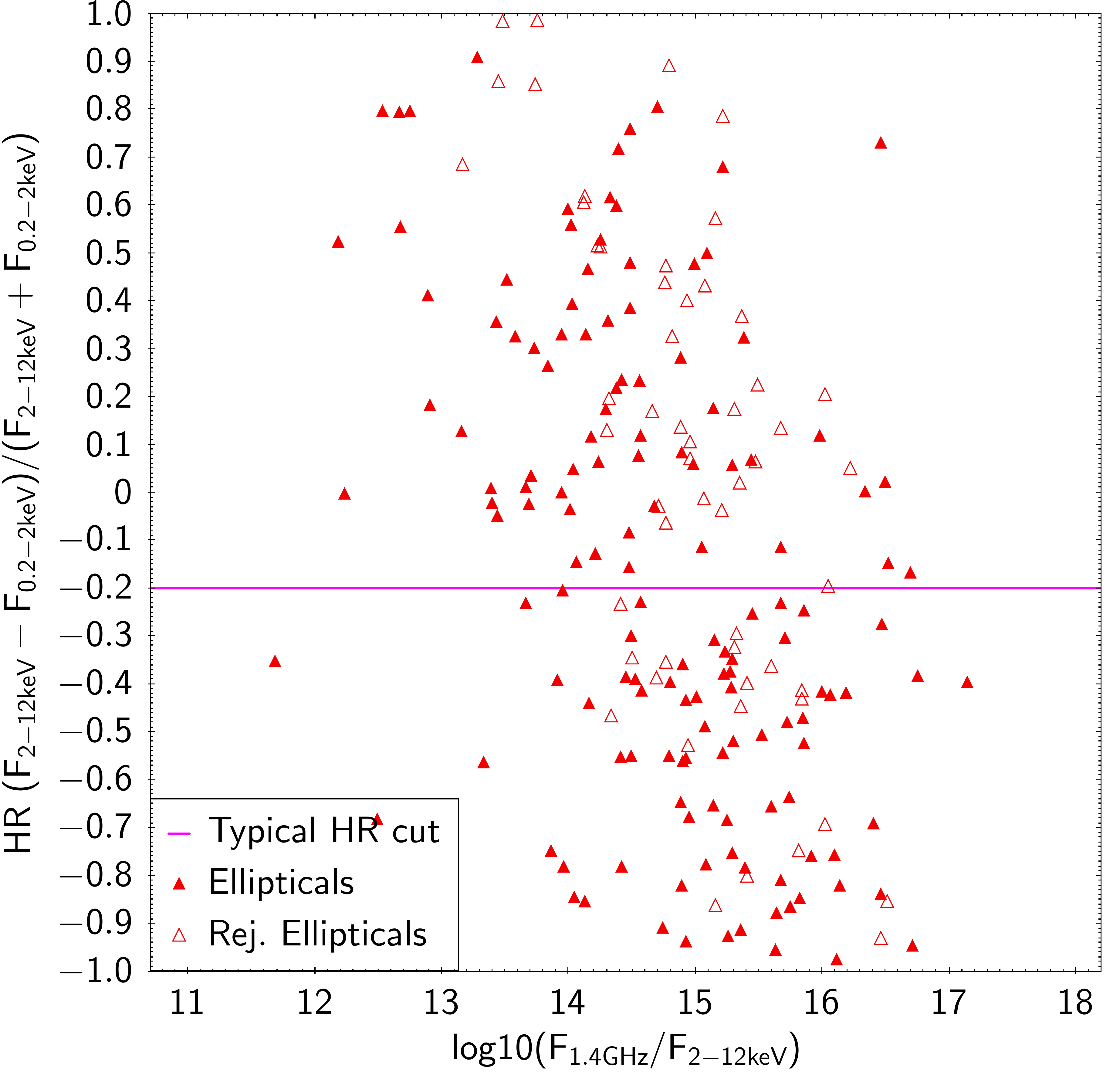} 
    \caption{Ellipticals} 
    \label{HR_ell_2} 
    \vspace{2ex}
  \end{subfigure}
  \begin{subfigure}[b]{0.5\linewidth}
    \centering
    \includegraphics[width=0.95\linewidth]{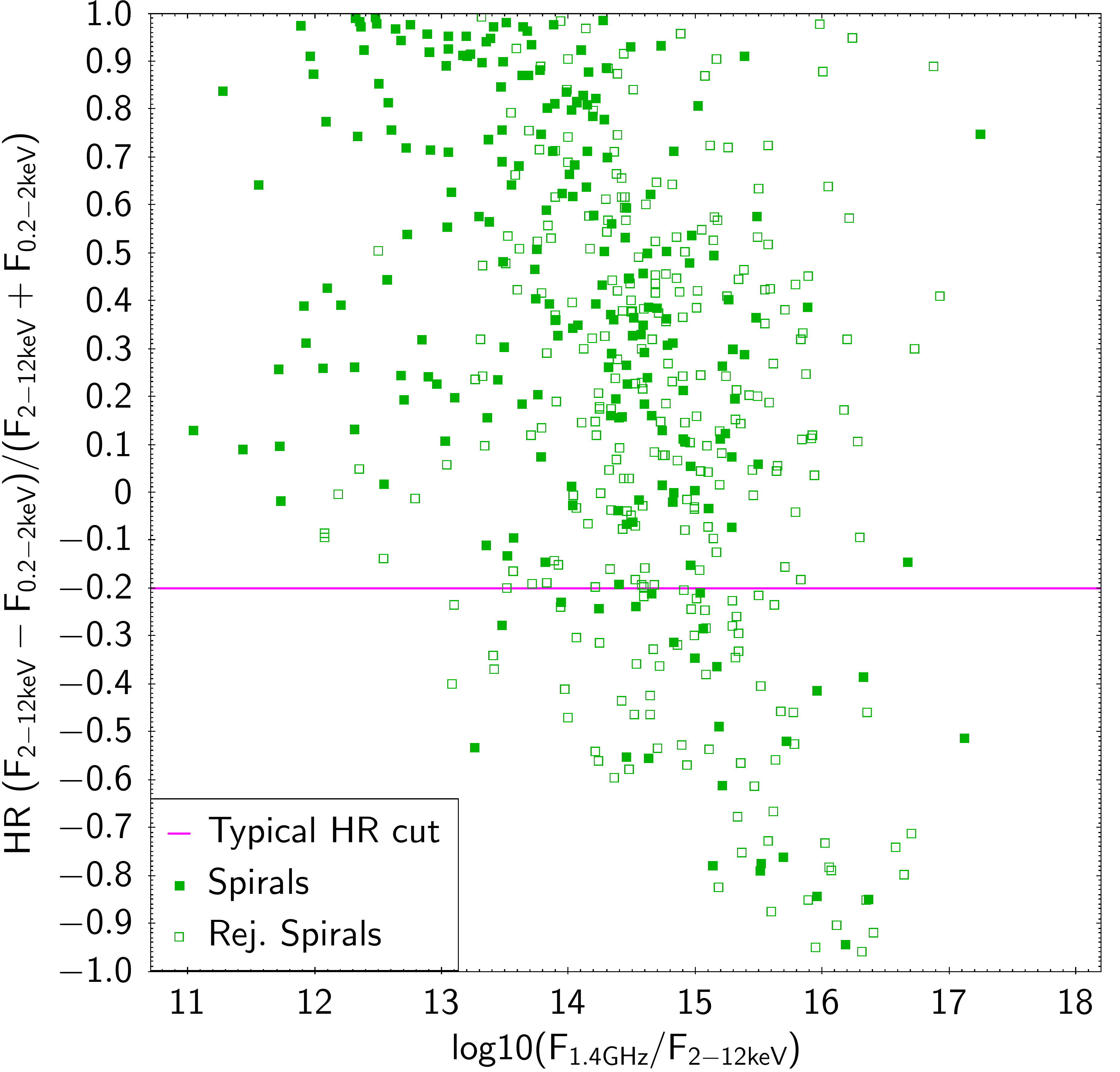} 
    \caption{Spirals} 
    \label{HR_spi_2} 
    \vspace{2ex}
  \end{subfigure}
  \begin{subfigure}[b]{0.5\linewidth}
    \centering
    \includegraphics[width=0.95\linewidth]{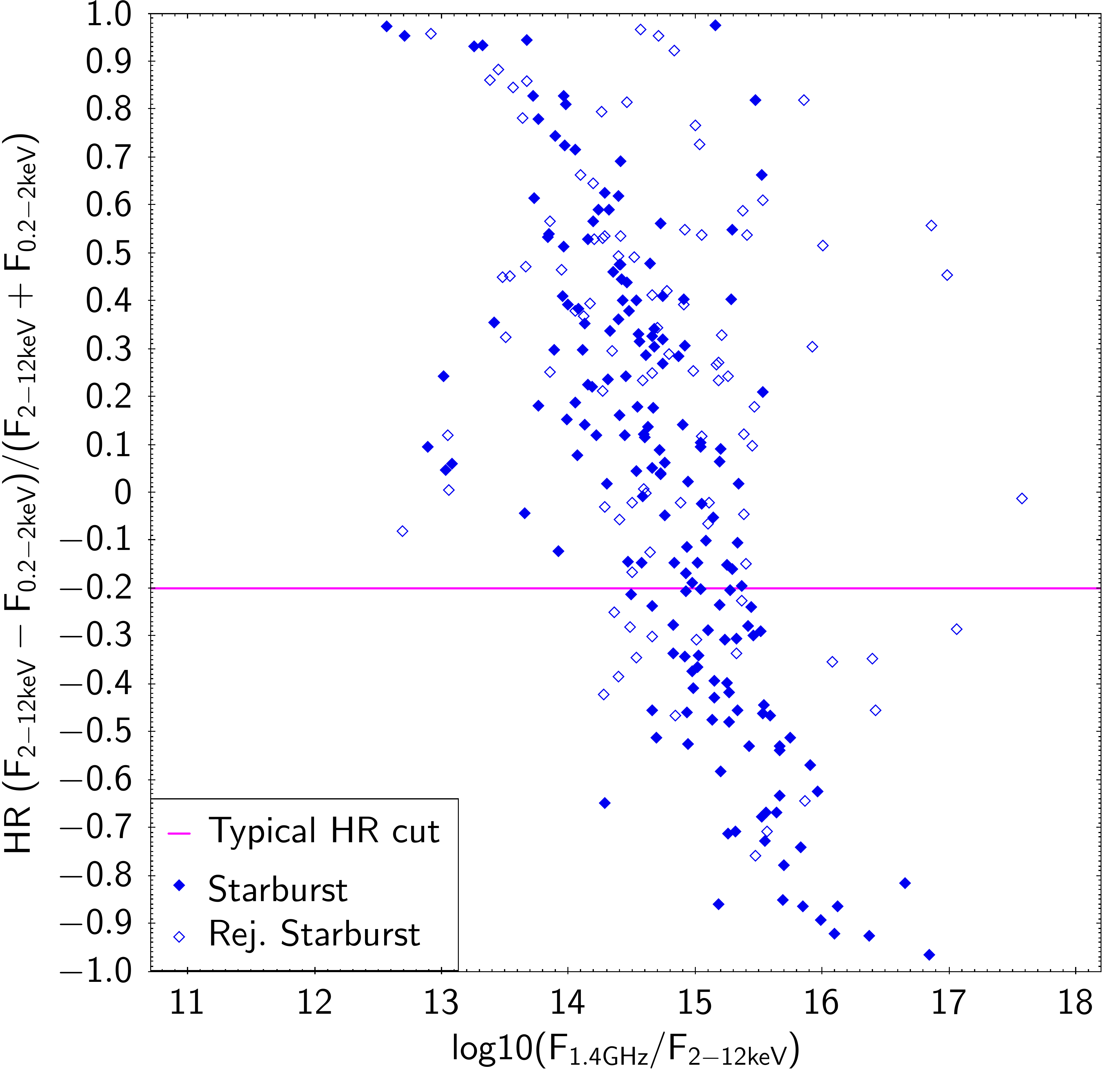} 
    \caption{Starburst} 
    \label{HR_sb_2} 
  \end{subfigure}
  \begin{subfigure}[b]{0.5\linewidth}
    \centering
    \includegraphics[width=0.95\linewidth]{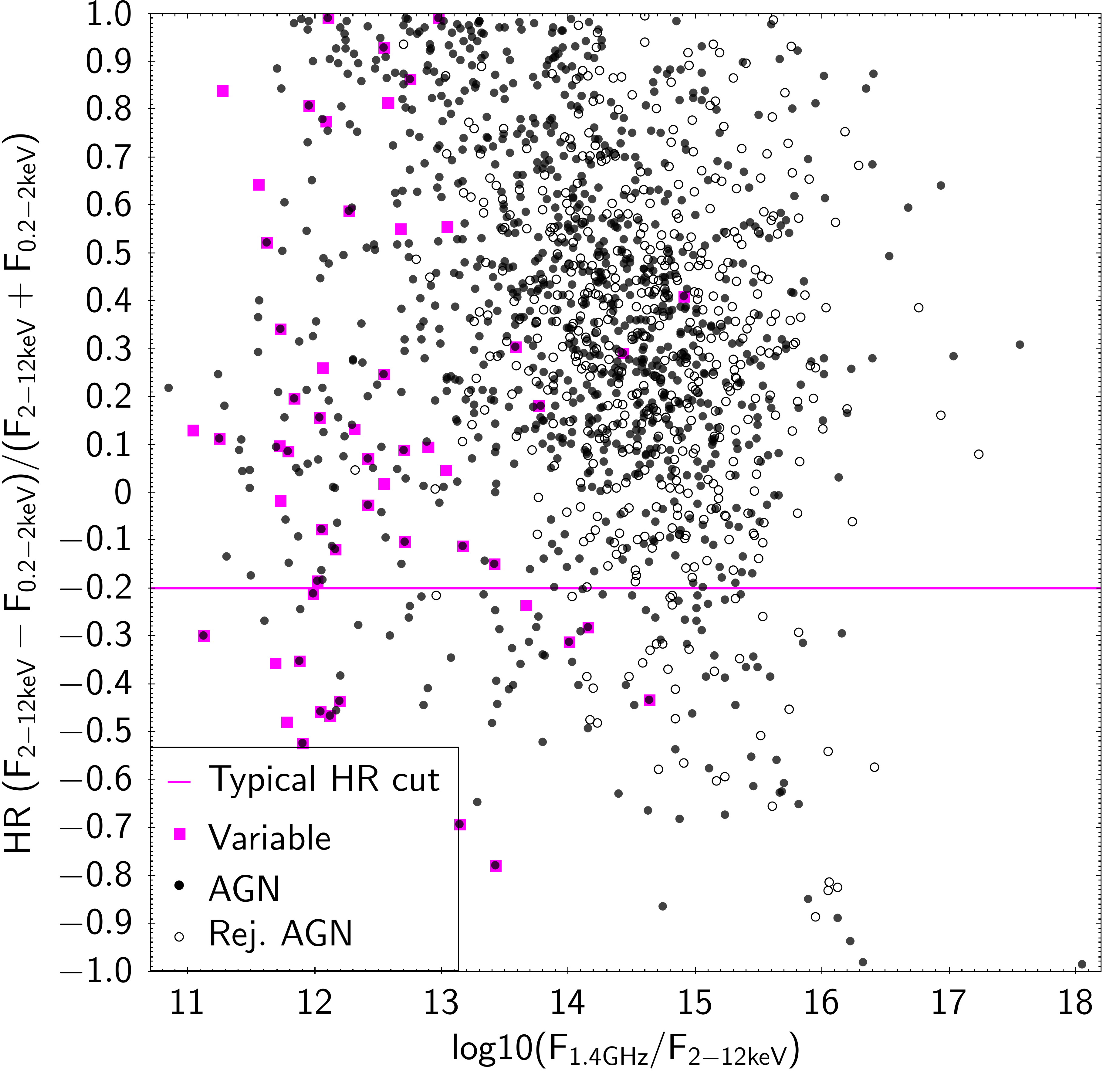} 
    \caption{AGN} 
    \label{HR_agn_2} 
  \end{subfigure} 
  \caption{Hardness ratio versus ratio of the 1.4 GHz radio flux (in mJy) to the hard X-ray flux (2--12 keV, in erg cm$^{-2}$ s$^{-1}$) for the various source populations. The colours and symbols follow the same scheme as those in Fig. \ref{C_C_all}, with full symbols representing the sources that pass the W3 S/N cut, and empty symbols those that do not. Sources labelled as variable on the 3XMM catalogue are plotted with larger, magenta squares, only in Fig. \ref{HR_agn_2}.}
  \label{HR_subsets_hard} 
\end{figure*}

Another effect that becomes more apparent when plotting some of the individual populations, and that we have displayed in more detail in Fig. \ref{Fr_Fx_W3}, is that the W3 S/N cut skews the sample slightly towards `radio-quieter' sources. This is expected, as the W3 cut essentially imposes a distance limit, and we know that, overall, AGN were radio-louder in the past \citep[see e.g.][and references therein]{Best2014,Williams2015}, and a majority of our sources are AGN. However, when studying the individual populations (Figs. \ref{HR_ell_2} to \ref{HR_agn_2}) we see that the distributions for the sources above and below the W3 S/N cut are similar enough that it is clear that we are essentially sampling the same types of sources, just at slightly different redshifts. We will study the effect of the W3 S/N cut and the redshift selection in more detail in section \ref{z}.

For brevity, we have only included the hard X-ray (2--12 keV) plots in Fig. \ref{HR_subsets_hard} for the individual populations. Overall, the various populations seem to behave as expected. The elliptical galaxies (Fig. \ref{HR_ell_2}) are `radio-loudest' (largest radio/X-ray flux ratios) and have the largest number of soft sources of all the groups, which is consistent with the idea that many of them host radiatively inefficient AGN. The spirals (Fig. \ref{HR_spi_2}) show quite a lot of scatter, which is consistent with the heterogeneous population of star-forming galaxies, low-luminosity Seyferts and LERG with spiral hosts, as we introduced in Table \ref{Activity}. All these populations are expected to quickly become undetectable in W3 as redshift increases, which is why our spirals are hit the hardest by the W3 cut. The starburst sources (Fig. \ref{HR_sb_2}) seem to have the narrowest distribution in the radio/X-ray flux ratio of all the populations, and the best consistency between the W3 accepted/rejected sources. This is reassuring, as we would expect a fairly clean selection for these sources, and a rather tight correlation between X-rays and radio where only star formation processes are responsible for both types of emission. The large range of hardness ratios covered by the starburst sources may seem surprising, but it is consistent with the picture presented by e.g. \citet{Ranalli2012}. The AGN (Fig. \ref{HR_agn_2}) show the largest scatter of all populations, and also seem to, overall, have the largest HR values, which is consistent with their expected spectral shape, $z$ evolution, and varying degrees of nuclear obscuration. There are probably several factors introducing scatter in the AGN plot, but along the X axis probably the most relevant one is the known scatter between jet output and radiative output, which we discuss in more detail in section \ref{Power}. The AGN sources are by far the most numerous at this point of the analysis, but they are also the ones most reduced by the introduction of redshifts (see Table \ref{nSources}), which is part of the motivation behind performing these diagnostics prior to carrying out the additional cross-correlation with SDSS.

The 3XMM catalogue includes a label for variable sources. These are sources that show X-ray variability within a single observation, thus in timescales of minutes to hours \citep[substantial variability on longer timescales is probably present for a large number of sources, but it is not described in the catalogue - see e.g.][]{Strotjohann2016} for examples of long-term variability from the XMM Slew Survey, and the EXTraS collaboration results\footnote{\url{http://www.extras-fp7.eu/index.php}} for examples across all the available XMM EPIC data). We have indicated with larger, magenta squares the sources with this classification that are retrieved in our sample, in Fig. \ref{HR_agn_2}, as most of them coincide with sources we have classified as AGN. Interestingly, the vast majority of these sources lie on the `X-ray louder' side of both the soft and hard X-ray HR plots. The fact that we do not find rapid X-ray variability for the radio-louder sources might be explained by a combination of factors. The variability timescales of the jet tend to be longer than those of the corona, where variations in the accretion flow are reflected quickly and abruptly. It is also possible that some of the radio flux is self-absorbed, that the relativistic boosting of the jet affects the radio and X-rays differently, or that the jet contribution is diluted in the X-rays.

\subsection{Flux correlations}\label{FCorr}

\begin{figure*}
\centering
\includegraphics[width=0.95\textwidth]{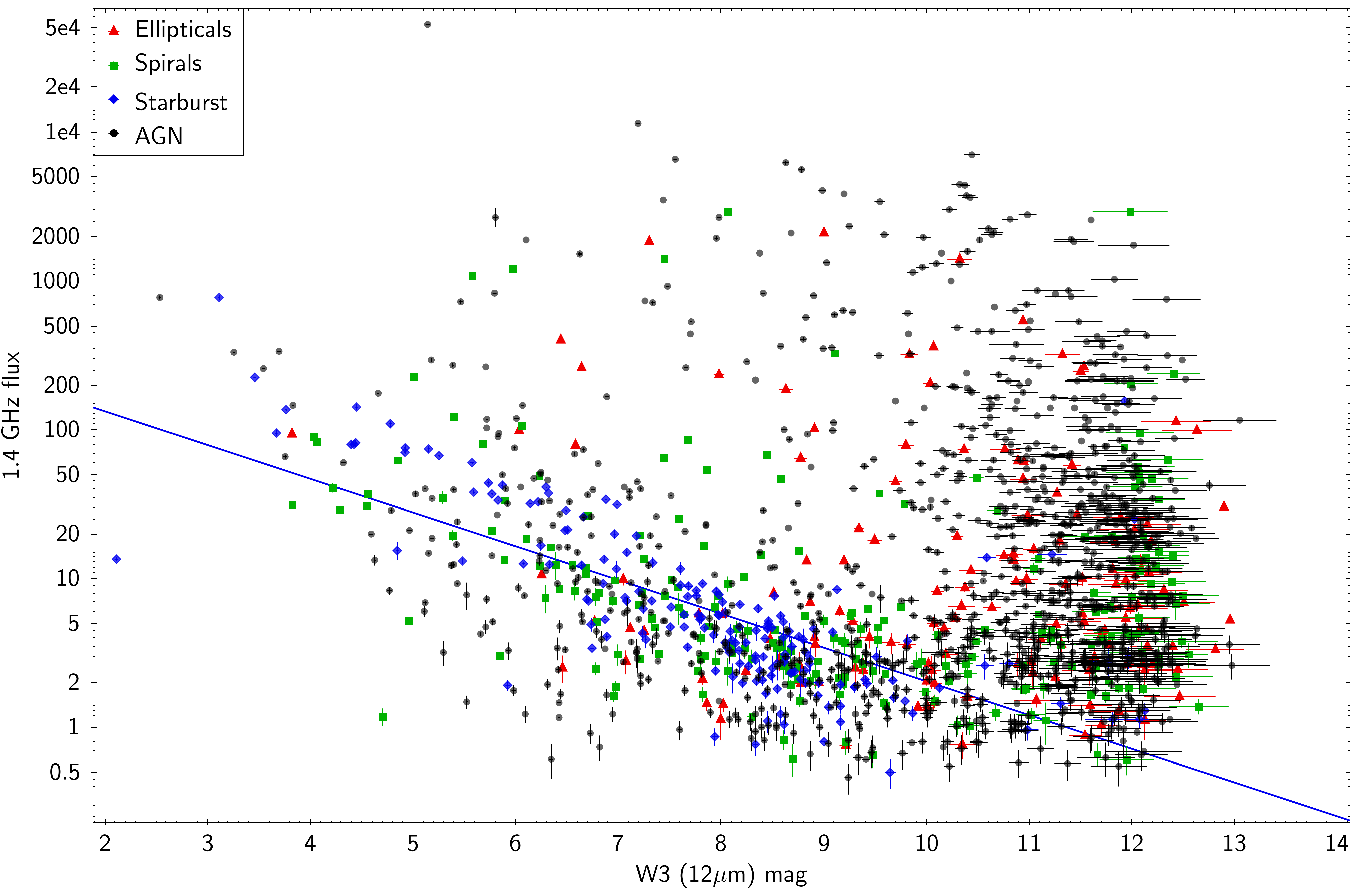}
\caption{1.4 GHz radio flux (mJy) versus W3 (12 $\mu$m) magnitude. The blue line represents the best linear correlation for the starburst sources ($r\sim 0.70$, where $r$ is the correlation coefficient). The colours and symbols follow the same scheme as those in Fig. \ref{C_C_all}. Please refer to Appendix \ref{ExtraFigures} for the W1 (3.4 $\mu$m) and W2 (4.6 $\mu$m) versions of this plot, and to Table \ref{nSources} for the source statistics.}\label{w3_combFlux}
\end{figure*}

\begin{figure*}
\centering
\includegraphics[width=0.95\textwidth]{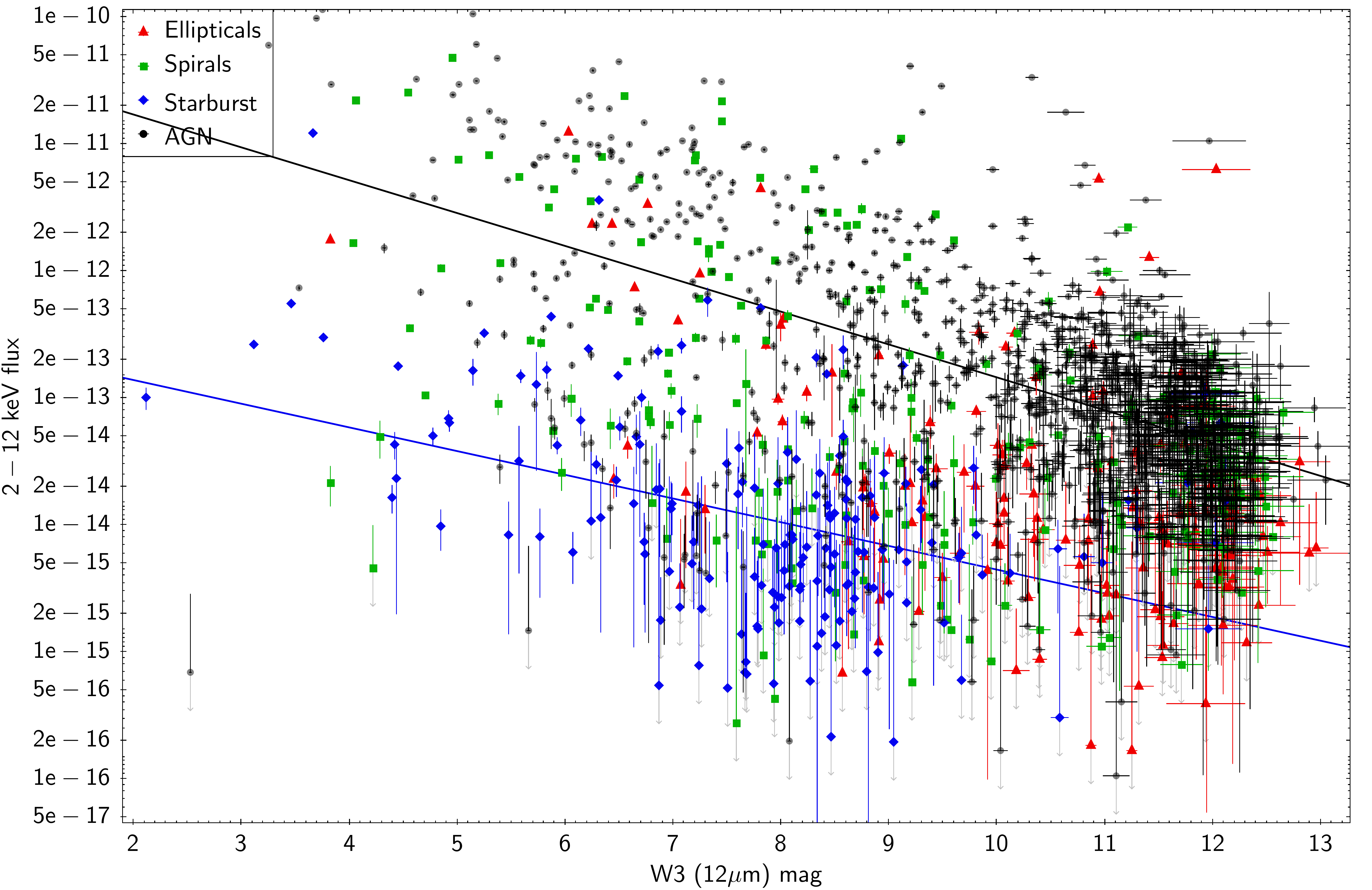}
\caption{Hard X-ray (2--12 keV) flux (erg cm$^{-2}$ s$^{-1}$) versus W3 (12$\mu$m) magnitude. The blue line represents the best linear correlation for the starburst sources ($r\sim 0.70$) and the black line the best attempt at a linear correlation for the AGN ($r\sim 0.55$). The colours and symbols follow the same scheme as those in Fig. \ref{C_C_all}. Upper limits (only for the X-rays) are represented with grey arrows. Please refer to Appendix \ref{ExtraFigures} for the equivalent plot for soft X-rays (0.2--2 keV), and to Table \ref{nSources} for the source statistics.}\label{w3_hardX}
\end{figure*}

To establish possible correlations between different types of activity in the MIXR sources, we plotted the radio and X-ray fluxes, and mid-IR magnitudes. As we introduced at the beginning of this section, in order to allow readers to establish an easier, more direct comparison, we have plotted the flux or magnitude values in the catalogues, with no transformations, other than the aperture corrections for WISE. Only the most relevant plots are displayed in this section, please refer to Appendix \ref{ExtraFigures} for details on the other flux correlations. 

To accurately triage sources according to their activity, we need to assess three properties: star formation, radiative AGN output, and kinetic (radio jet and lobes) AGN output. To do so we need to simultaneously consider where the sources fall on the three plots presented in this section, as well as the information from their mid-IR colours and the plots in section \ref{HR}.

Fig. \ref{w3_combFlux} shows the distribution of radio versus W3 flux for the populations we defined from Fig. \ref{C_C_all}. It is clear, at a first glance, that the starburst sources follow a correlation, which is likely to be a direct extension into the mid-IR of the well-known far-IR/radio correlation for star formation \citep[see e.g.][and sections \ref{LDiag} and \ref{RL_RQ_section} for more details]{Gruppioni2003}. A few of the spiral galaxies also follow this correlation, but a large fraction seem to prefer the locus inhabited by most of the AGN and the elliptical galaxies (which we know are likely to harbour radiatively inefficient AGN). Interestingly, many sources with the AGN classification also follow the star formation correlation: these are likely to be radio-quiet AGN. The correlation we derived from the starburst sources, plotted in Fig. 7, is slightly flatter than expected. This is probably caused by the presence of outliers, especially mis-classified AGN, which might be exerting a leverage on the fit, but we have not excluded these points, as doing so might introduce further bias in the subset. We have excluded outliers from the luminosity correlations in section \ref{LCorr}.

It is worth highlighting here that radio-quiet does not mean radio silent \citep[e.g.][]{Wong2016}: because the radio-loud/quiet classification is traditionally based on optical (or other bands) to radio flux ratios \citep[e.g.][]{Kellerman1989}, AGN with large radiative outputs, and small jets and lobes, are often classified as radio-quiet, as an increasingly large number of Seyfert galaxies shows \citep[e.g.][]{Hota2006,Gallimore2006,Croston2008,Mingo2011}. In this work, and in particular in section \ref{RL_RQ_section}, we refer to radio-quiet AGN as sources where the radio emission we detect from them is likely to originate mainly from stellar processes, accelerated particles in wind-driven shocks \citep[see][]{Nims2015,Zakamska2016} or, if arising from a jet and lobes, they are small and faint, and the AGN produces the bulk of its emission as radiative output in the other bands. Conversely, we refer to radio-loud AGN as those that have a substantial kinetic output in the form of jet and lobes, which we measure as radio emission well above the star formation correlation. The radiatively inefficient LERG/LINER sources also follow these criteria, so they are a subset of radio-loud AGN.

For the radio-loud AGN, as well as the potential LERG/LINER sources, there seems to be a wide range of possible radio fluxes for a given 12$\mu$m magnitude. This is partly due to the fact that most AGN have W3 magnitudes close to the detection limit, but it hints at what we observed in \citet{Mingo2014} for the 2Jy and 3CRR samples, when we found a large amount of scatter in the relation between radiative and kinetic output in radio-loud AGN, in apparent contradiction with the correlation proposed by \citet{Rawlings1991}. We will discuss this point in further detail in sections \ref{RL_RQ_section} and \ref{Power}.

Fig. \ref{w3_hardX} shows the hard X-ray versus W3 flux for our sources. There is a clear distinction between the starburst and AGN populations: they follow nearly parallel distributions, but the starburst galaxies have systematically lower X-ray fluxes. This is in agreement with what we know of the mid-IR/X-ray correlation for AGN and star-forming galaxies \citep[e.g.][]{Gandhi2009,Mateos2015}, and it illustrates why it is so difficult to distinguish the break between both populations in luminosity/luminosity plots. The presence of mis-classified sources in the AGN subset introduces scatter, and weakens the correlation we obtain, but the data show that it is clearly there. At this stage we have not wanted to re-classify sources based on their fluxes, we will do so after we obtain their luminosities, in section \ref{RL_RQ_section}.

The elliptical galaxies are systematically fainter in W3 than the starburst galaxies, and also have X-ray fluxes that are systematically lower than those of the AGN subset, reinforcing the conclusion that these sources harbour radiatively inefficient AGN. The spiral galaxies seem to be split between the AGN and the starburst loci, with a few sources falling in the gap between them. We repeated this same plot without the radio selection, and found that sources with spiral colours fill the entire gap between the AGN and starburst galaxies, making it impossible to distinguish between non-active spiral galaxies with different levels of star formation, and Seyferts with a range X-ray luminosities. Only with the radio selection is it possible to easily distinguish between Seyferts and non-active galaxies for sources in the spiral region of the WISE colour/colour plot.

There is a weak correlation between the X-ray and radio fluxes for the starburst sources, which appears mainly for the soft X-rays, but it is not very strong, especially if the sources with radio fluxes greater than 100 mJy are removed. The situation is also less clear for the other populations, even the AGN present a lot of scatter. Although we know that there is X-ray emission arising from the jet \citep{Hardcastle1999}, it appears mainly in the soft X-ray band. What is readily apparent in the X-ray/radio plots (Figs. \ref{combFlux_softX} and \ref{combFlux_hardX}) is the previously mentioned scatter between radiative and kinetic output that we also observed in the 2Jy and 3CRR sources, as well as the LERG/LINER nature of the elliptical sources.

Overall, these plots present a picture consistent with what we outlined in Table \ref{Activity} and section \ref{HR}, allowing us to diagnose the different types of activity present in each source type. These diagnostics need to be confirmed, however, using redshifts to derive the luminosities of the sources in each band. The redshifts will also help us determine the nature of any outliers in the diagnostic plots, as well as to assess the effects of evolution within the populations, which may not be negligible \citep[see e.g. the evolutionary tracks of][]{Assef2010}.

%

\section{SDSS redshifts}\label{z}

\begin{table*}\small
\caption{Number of sources in each subset. See also Table \ref{Activity} for the detailed WISE colour source classification. For all subsets we have applied the quality cuts for FIRST/NVSS and 3XMM, and the S/N cut for W1 and W2, but we treat the W3 S/N cut separately, as described in section \ref{WISE}. Throughout the text we refer to the full set of sources that passes the W3 S/N cut as W3 Sample (or W3 z sample for the subset that also have redshifts). For the source type subsets we use the prefix `z' for sources with redshifts, and the prefix `rejected' for sources that do not pass the W3 S/N cut, for example the galaxies with elliptical colours that have redshifts but do not pass the W3 S/N cut are referred to as `z Rejected Ellipticals' in the plots.}\label{nSources}
\centering
\begin{tabular}{lcccccc}\hline
Subset name&\multicolumn{6}{c}{Number of sources}\\\hline
&All sources&W3 S/N$\ge3$&W3 S/N$<3$&With $z$&With $z$ + W3 S/N$\ge3$&With $z$ + W3 S/N$<3$\\\hline
Full sample&2529&1575&954&1367&947&420\\
Ellipticals&203&145&58&137&94&43\\
Spirals&507&222&285&323&149&174\\
Starburst&268&174&94&168&114&54\\
AGN&1510&1008&502&721&577&144\\\hline
\end{tabular}
\end{table*}

To find redshifts for our sample, with as uniform a coverage as possible, the ideal choice is the Sloan Digital Sky Survey (SDSS), more so considering that FIRST was initially designed to cover the same sky area, so the overlap between MIXR and SDSS is very large ($\sim 70$ per cent in sky area). We decided to use the latest data release, DR 12 \citep{Gunn2006,Eisenstein2011AJ,Dawson2013}, to maximise the number of possible counterparts to the sources in our catalogue with photometric or spectroscopic redshifts. We did not use the ARCHES xmatch tool in this case, but rather a simple normalised distance histogram (between the \textsc{xmatch} averaged position for the WISE+3XMM+FIRST/NVSS MIXR sources, and the SDSS positions), making use of the astronomy software TOPCAT \citep{Taylor2005} to carry out the cross-match.

We initially selected SDSS sources within 10 arcsec of our merged catalogue positions. The distance distribution histogram (see Fig. \ref{SDSS_dist_histo} in the Appendix) showed a very clean selection at distances under 2--3 arcsec. To further test our selection we manually checked about 150 sources that had 4 or more SDSS matches, using the on-line SDSS finding charts, and found the nearest match to clearly be the best choice in nearly all cases (the other possible matches were stars or had much larger separations). We only found two potentially dubious cases, where the merged position coincided with an optical galaxy cluster, and the nearest and second-nearest match had similar separations, around 2--4 arcsec. We thus decided to only consider redshifts for matches with separations below 3 arcsec. We have included in the catalogue a column with the URL of the SDSS finding chart for each source (column SDSS\_URL), for the users to explore.

Roughly half of the sources with redshifts in SDSS had spectroscopic redshifts, so we used these values whenever possible. We used the sources that had both values to study the reliability of photometric redshifts, and found them to be fairly reliable in most cases. It is possible that the sources that only have photometric information, because they are fainter or more distant, also have less reliable redshifts. To mitigate this effect, and the fact that some photometric redshifts have fairly large error bars, we decided to take these uncertainties into account, when possible, when calculating the luminosities (see section \ref{LDiag}). 

Our manual check also confirmed that most optical spectroscopic classifications coincide with those we derived from mid-IR colours. The optical spectra revealed several broad-line AGN in objects classified as spirals in our sample, where there is also contribution from star formation. These objects tend to fall in the AGN locus in the X-ray/W3 diagnostic plot (Fig. \ref{w3_hardX}), but, notably, some of them fall in the star formation correlation for the Radio/IR plot (Fig. \ref{w3_combFlux}). If their luminosities are consistent with this assessment, these sources exhibit behaviour typical of radio-quiet AGN, as shown in the results of \citet{Padovani2011A,Bonzini2013}, where the bulk of the radio emission in (moderately luminous) radio-quiet AGN and star-forming galaxies is produced by star formation. 

The fraction of objects with SDSS counterparts is rather large, around $70$ per cent of the sample, although the fraction of objects with good redshift measurements (no upper limits, small separations) falls to $\sim50$ per cent, with the sources classified as AGN suffering the greatest loss. The limiting factor in our overall selection still seems to be the WISE W3 band, but the requirement of a mid-IR counterpart to the radio and X-ray selection clearly plays an important role on finding optical counterparts for our sources as well. The fraction of sources that pass the W3 cut and have redshifts in SDSS is $66$ per cent, but it is quite dependent on the source type. Table \ref{nSources} details the names, definition, and statistics of each subsample of sources.

\begin{figure}
\centering
\includegraphics[width=0.46\textwidth]{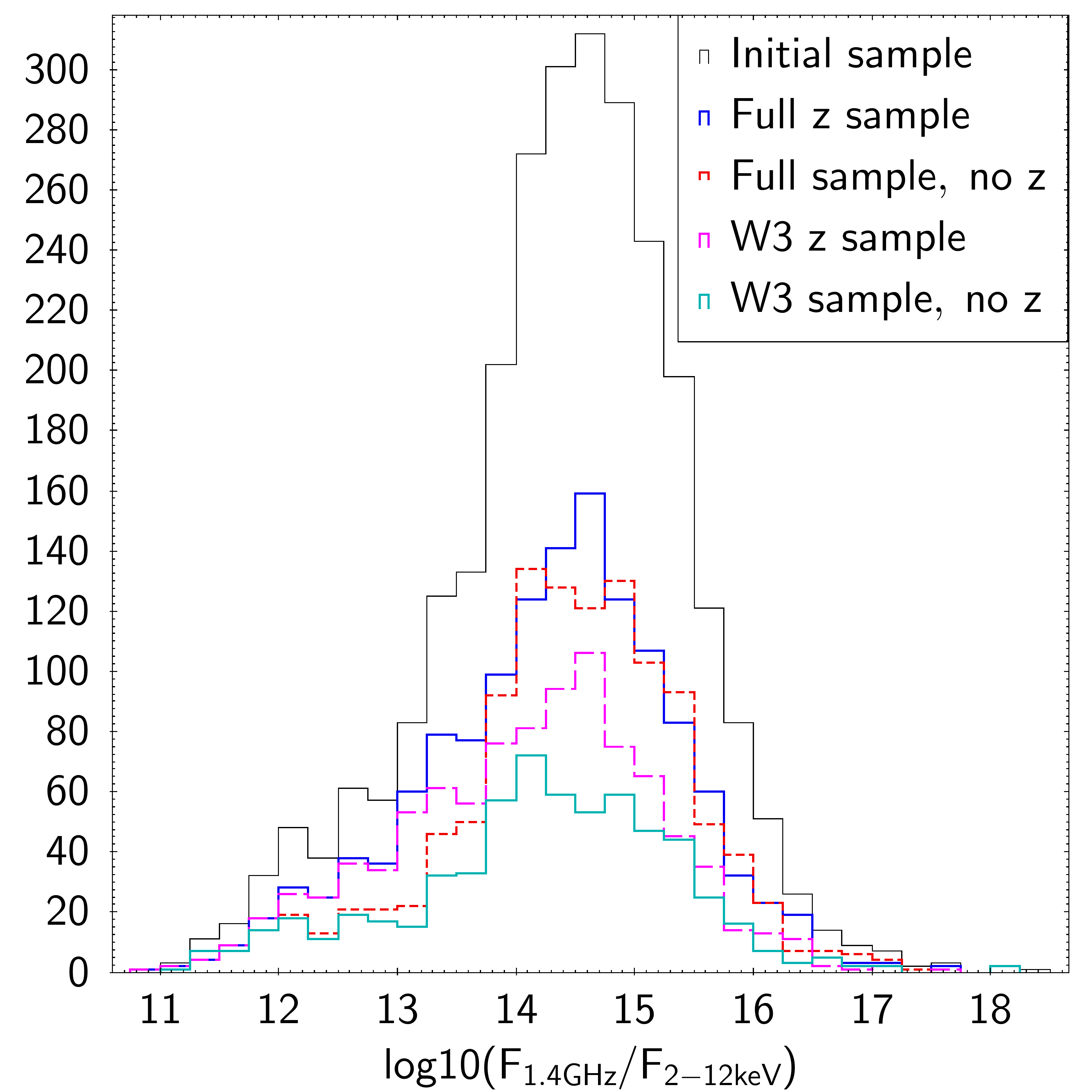}
\caption{Histogram of radio to X-ray fluxes to test possible selection biases introduced by SDSS and WISE. From top to bottom the curves represent all the sources in the initial sample (black, thin line) full z sample (blue line), full sample with no redshifts (red dashed line), W3 z sample (magenta line with longer dashes), and W3 sample with no redshifts (cyan line). See table \ref{nSources} for subsample definitions and statistics.}\label{SDSS_selection}
\end{figure}

We saw in section \ref{HR} (see Fig. \ref{Fr_Fx_W3} in particular) that the W3 selection introduces a slight skew in the distribution in terms of radio (or X-ray) loudness. Fig. \ref{SDSS_selection} is useful to also study the possibility of a selection bias introduced by the SDSS selection. The histograms represent the distribution of radio to hard X-ray flux for four different subsets of sources: all sources (initial sample, see Table \ref{nSources}), all sources with redshifts (full z sample), all sources with no redshifts (full sample, no z) all sources that pass the W3 S/N cut and have SDSS counterparts (W3 z sample), and all sources that pass the W3 cut and do not have SDSS redshifts (W3 sample, no z). At a glance the histograms look rather similar, but after carrying out a Kolmogorov--Smirnov test for the $F_{1.4 GHz}/F_{2-12keV}$ distributions for several subsets (Table \ref{KS_table}) we see that the W3 and redshift cuts do indeed change the shape of the original distribution. Interestingly, both cuts seem to skew the distribution in similar ways, as the `W3 z' and `W3 no z' distributions are the most similar in Table \ref{KS_table}. Overall, the W3 cut seems to have a larger effect on the distribution than the $z$ cut, but the combination of both seems to skew the distribution even further.

\begin{table}\small
\caption{Results of the Kolmogorov--Smirnov test for the various distributions. Please see Table \ref{nSources} for the statistics of each subset and Section \ref{Sample} for the subset definitions. The columns show, respectively, the subsets compared, the KS statistic, and the p-value (to test the null hypothesis probability). The `z' and `no z' denote whether a subset has or does not have SDSS redshifts.}\label{KS_table}
\centering
\begin{tabular}{lcc}\hline
Subsets tested&D&p\\\hline
Initial - W3&0.11&$1.76\times10^{-11}$\\
Initial - full z&0.06&$5.68\times10^{-3}$\\
Initial - full no z&0.05&$1.53\times10^{-2}$\\
Initial - W3 z&0.13&$1.86\times10^{-11}$\\
Initial - W3 no z&0.10&$7.78\times10^{-5}$\\
Full z - Full no z&0.11&$2.18\times10^{-7}$\\
W3 z - W3 no z&0.07&$5.14\times10^{-2}$\\
Full z - W3 z&0.08&$1.89\times10^{-3}$\\
Full no z - W3 no z&0.14&$2.64\times10^{-7}$\\\hline
\end{tabular}
\end{table}

This is to be expected, as we know that both cuts impose, in essence, a distance limit, but the W3 cut is more severe. We also know that AGN were radio-louder in the past \citep{Best2014,Williams2015}, and that most non-AGN galaxies in our sample must be at fairly low $z$, so the skew is consistent with what we expect. Looking at the histogram in Fig. \ref{SDSS_selection} the differences are subtle indeed, despite the numbers in Table \ref{KS_table}. There seems to be a slight bias in terms of X-ray loudness when only the SDSS selection is applied: the z sample histogram deviates more from the overall (full sample) distribution for sources with an intermediate radio/X-ray ratio. Looking at the sources that occupy this range of $F_{1.4 GHz}/F_{2-10 keV}$ in Fig. \ref{HR_all_2}, it seems that with the SDSS selection we must be eliminating some AGN and spiral galaxies, perhaps more distant or overall fainter in the optical than the others (see also section \ref{RL_RQ_section}). When comparing the `full z' and `full no z' histograms, they look very similar, indicating that the SDSS selection is mostly unbiased, except around $F_{1.4 GHz}/F_{2-12 keV}\sim10^{13}$, where more sources are preserved than discarded by the SDSS selection, meaning that there is a slight favouring of more X-ray bright sources with respect to the radio-bright sources on the other wing of the distribution.

For the distributions that apply the W3 S/N cut, `W3 z sample' and `W3 sample, no z', we see that the W3 cut is good at preserving the sources at intermediate values of $F_{1.4 GHz}/F_{2-12 keV}$, but it also appears to be more biased towards X-ray bright sources than the SDSS selection, eliminating a larger fraction of radio-bright sources.

\begin{figure}
\centering
\includegraphics[width=0.46\textwidth]{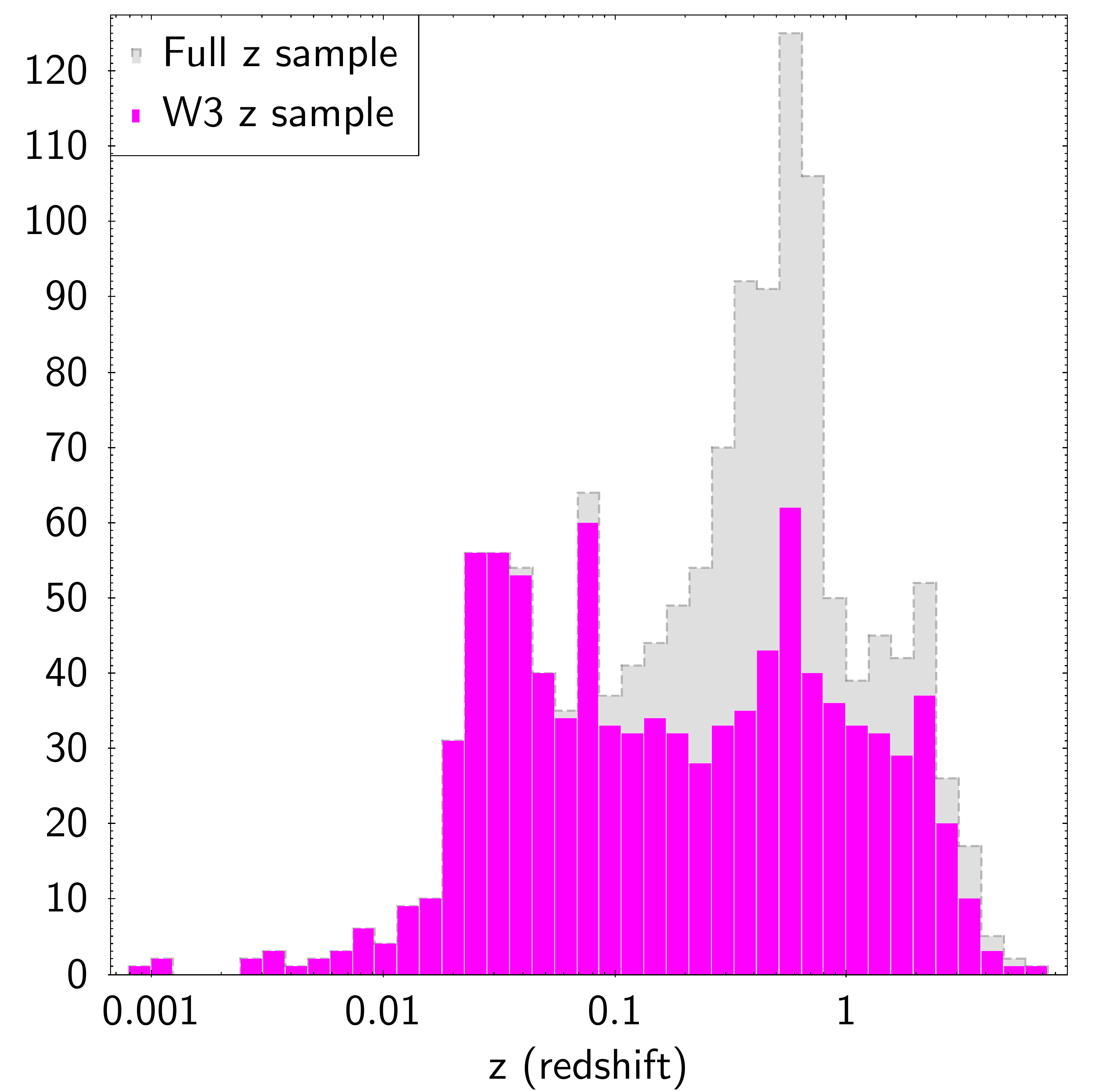}
\caption{Histogram of redshift distributions for the sources with SDSS redshifts. The dashed grey area represents the distribution of all the sources with SDSS redshifts (full z sample), while the full magenta bars show the distribution of sources with SDSS redshifts that also pass the W3 S/N cut (W3 z sample), as detailed in Table \ref{nSources}.}\label{z_histo_all}
\end{figure}

\begin{figure*}
  \begin{subfigure}[b]{0.5\linewidth}
    \centering
    \includegraphics[width=0.95\linewidth]{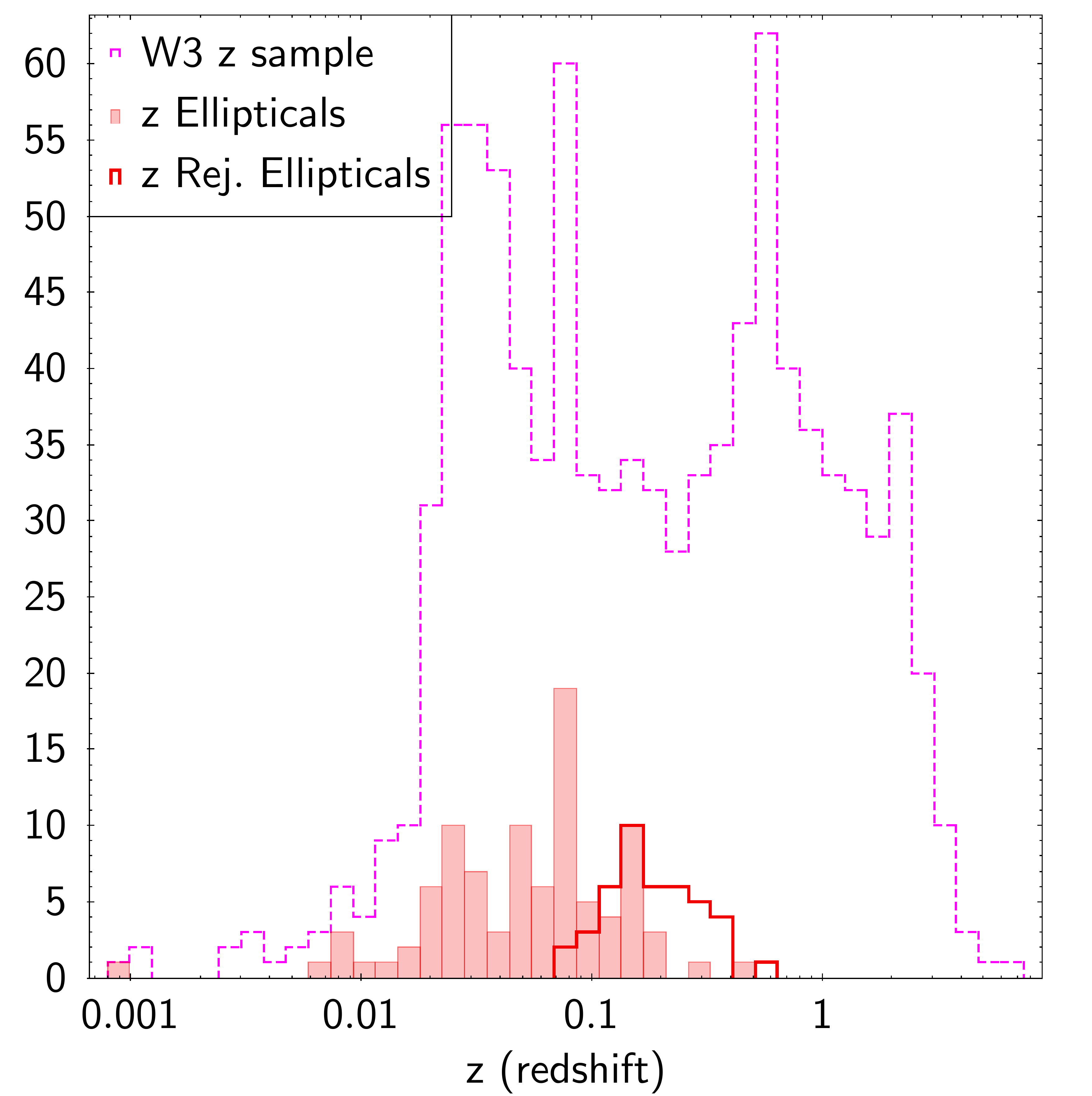} 
    \caption{Ellipticals} 
    \label{z_histo_ell} 
    \vspace{2ex}
  \end{subfigure}
  \begin{subfigure}[b]{0.5\linewidth}
    \centering
    \includegraphics[width=0.95\linewidth]{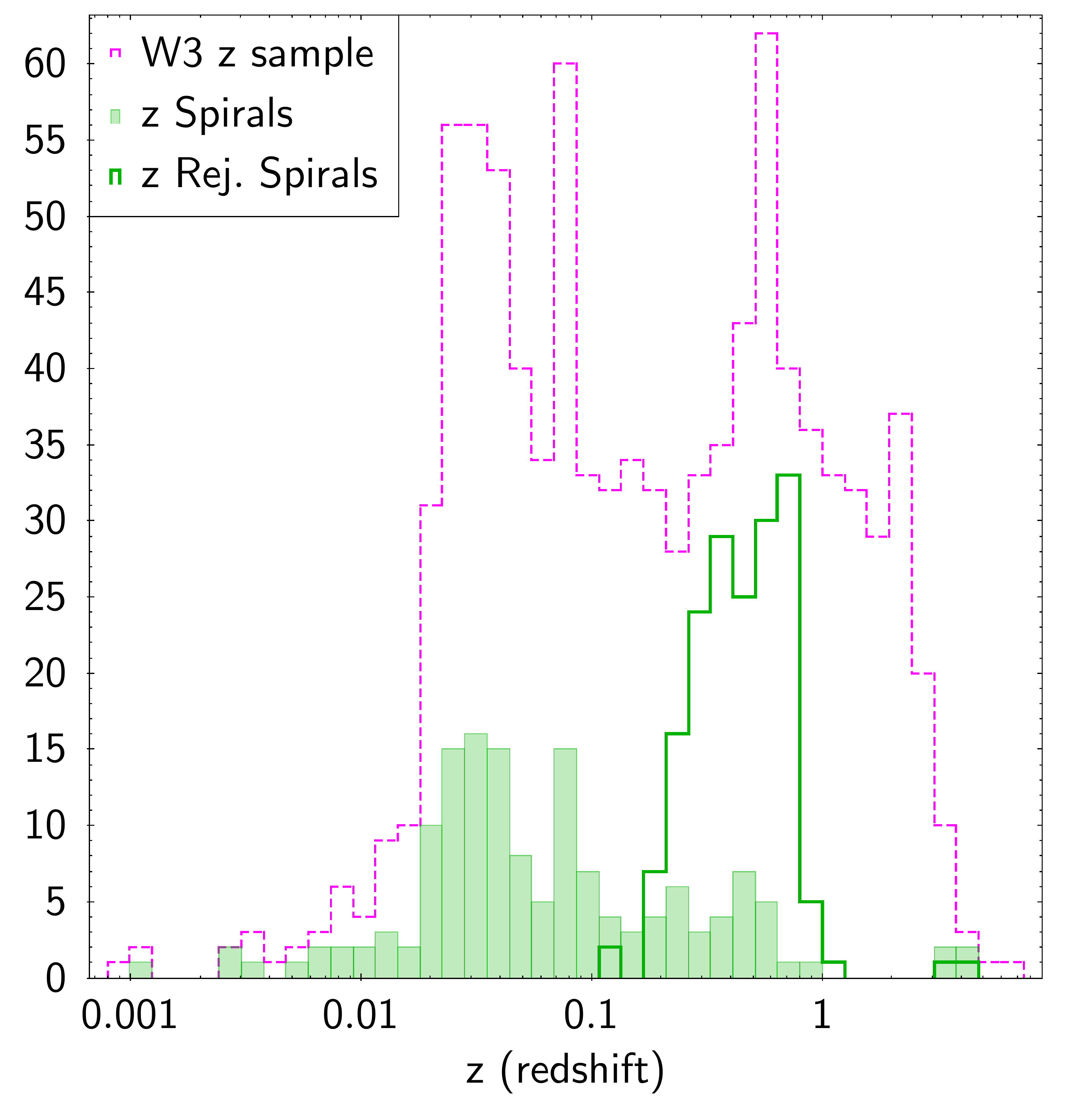} 
    \caption{Spirals} 
    \label{z_histo_spi} 
    \vspace{2ex}
  \end{subfigure}
  \begin{subfigure}[b]{0.5\linewidth}
    \centering
    \includegraphics[width=0.95\linewidth]{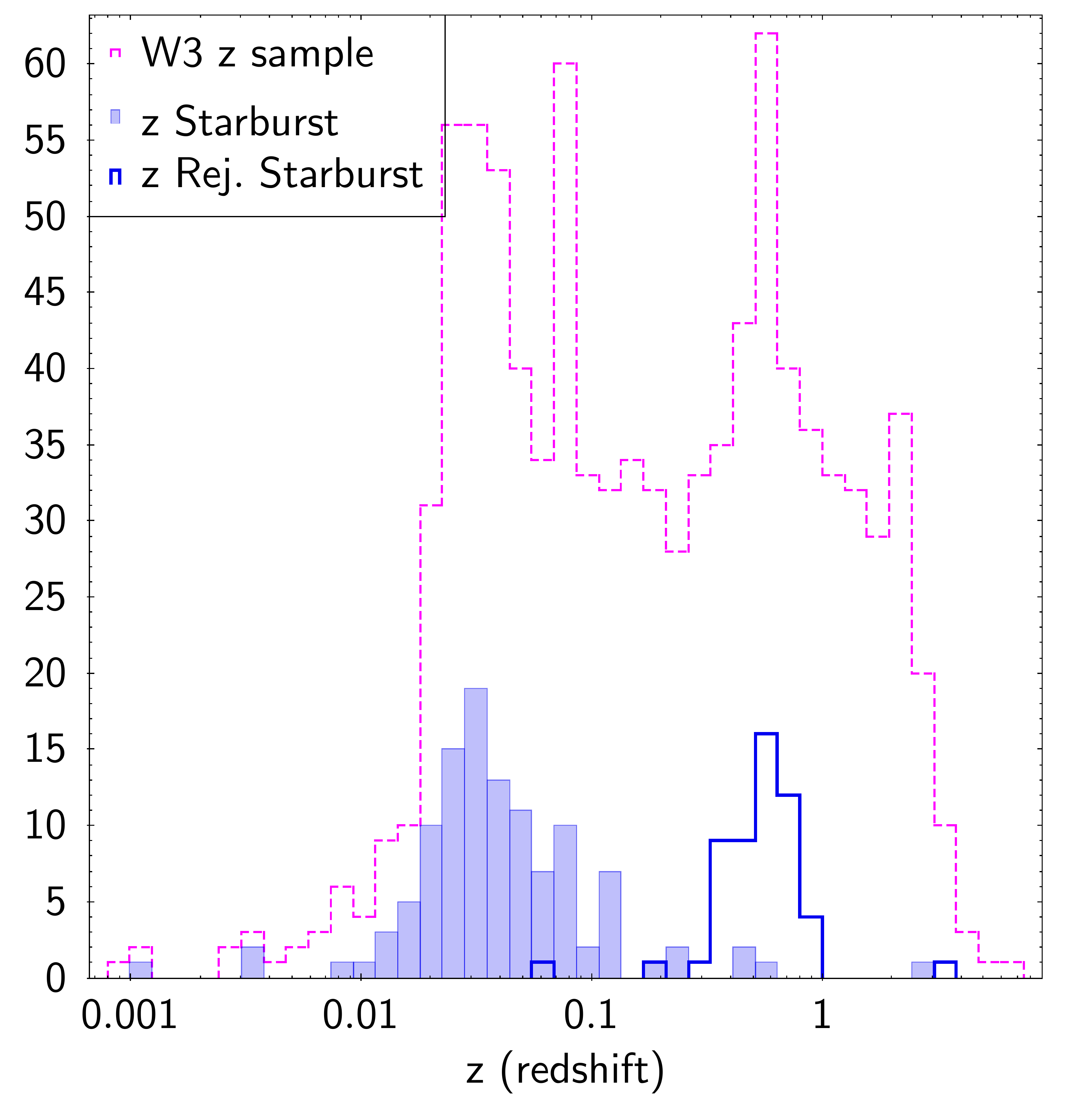} 
    \caption{Starburst} 
    \label{z_histo_sb} 
  \end{subfigure}
  \begin{subfigure}[b]{0.5\linewidth}
    \centering
    \includegraphics[width=0.95\linewidth]{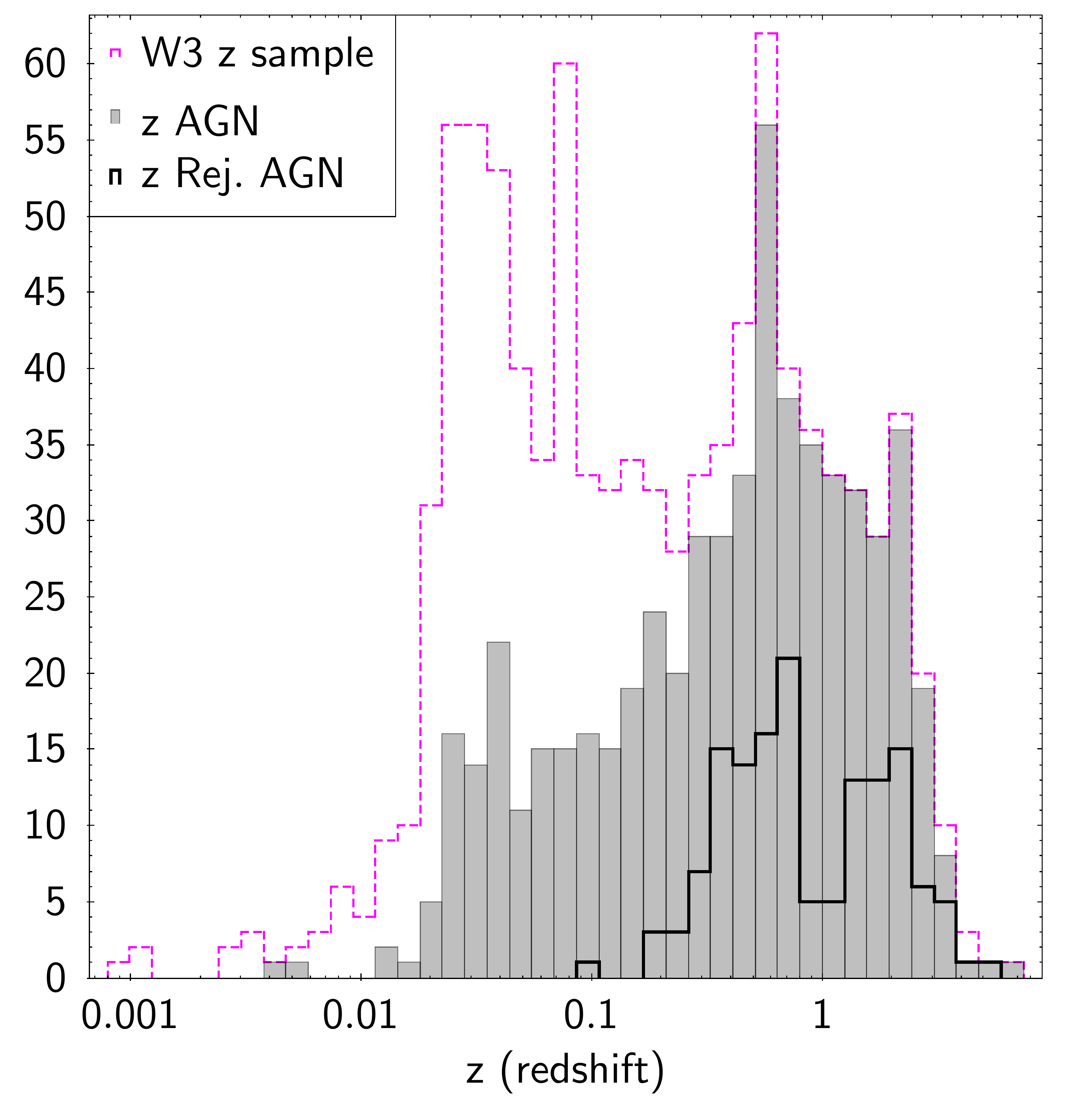} 
    \caption{AGN} 
    \label{z_histo_agn} 
  \end{subfigure}
  \caption{Histogram of redshift distributions for all the sources with SDSS redshifts (see Table \ref{nSources} for details). The dashed magenta outlines represent the distribution of all the sources with SDSS redshifts that also pass the W3 S/N cut (W3 z sample), and is plotted as a reference. For each subset, the full bars show the sources with SDSS redshifts that pass the W3 S/N cut, while the thick empty outlines show the sources that do not pass the W3 cut.}
  \label{z_histo_subsets} 
\end{figure*}

Fig. \ref{z_histo_all} shows the redshift distribution of sources for the full z sample (see Table \ref{nSources} for source statistics) and those on the W3 z sample. The W3 cut seems to preserve most sources at $z<0.1$, but there is a progressive and sharp increase in the number of sources lost at larger redshifts, confirming our earlier suspicions, with the greatest effect achieved at $z\sim 0.5$. After $z\sim$1--2 both distributions seem to converge again, indicating that the main limiting factor is not W3, but one (or more) of the other bands. The number of sources in both samples decreases very quickly after $z\sim 3$, which means that for the vast majority of our sources the colour cuts we have used for classification should be fairly reliable \citep[AGN, in particular, have bluer colours at larger $z$, see e.g.][]{DiPompeo2015}.

The redshift distributions for the different source subclasses are very different, as illustrated in Fig. \ref{z_histo_subsets} (where we have also plotted the W3 z sample for reference, as it serves to illustrate the relative contributions of each source type). For all the populations the W3 rejected sources can be found at higher $z$ than their W3 detected counterparts, in agreement with what we introduced in Section \ref{HR}. The ellipticals (Fig. \ref{z_histo_ell}) start disappearing from our sample at lower redshifts ($z\sim0.1$) than the other populations, as expected from radiatively inefficient, non-starforming sources, especially since we eliminated the clusters with our initial selection. The redshift distribution for the spirals (Fig. \ref{z_histo_spi}) reflects, again, the heterogeneous nature of this population, and confirms our suspicion that low-luminosity AGN with spiral mid-IR colours \citep[due to either star formation or evolutionary effects, as described by e.g.][]{Assef2010} are not detected by W3 even at moderate $z$. The starburst sources (Fig. \ref{z_histo_sb}) show a markedly bimodal distribution in terms of the W3 filter. It is difficult to speculate how much of this effect arises from genuinely different underlying populations, but it is likely that the W3 rejected sources with $z>0.1$ contain AGN. The AGN sources (Fig. \ref{z_histo_agn}) are still the largest population in our sample, despite the trim suffered by the cross-correlation with SDSS (see Table \ref{nSources}), as for this population, the W3 and redshift cuts seem to mostly overlap, probably because optically bright AGN are also expected to have a substantial contribution from the torus to the W3 band. As expected, the $z$ distribution for the AGN peaks at the highest value of all the populations, and is also the broadest.

\begin{figure}
\centering
\includegraphics[width=0.46\textwidth]{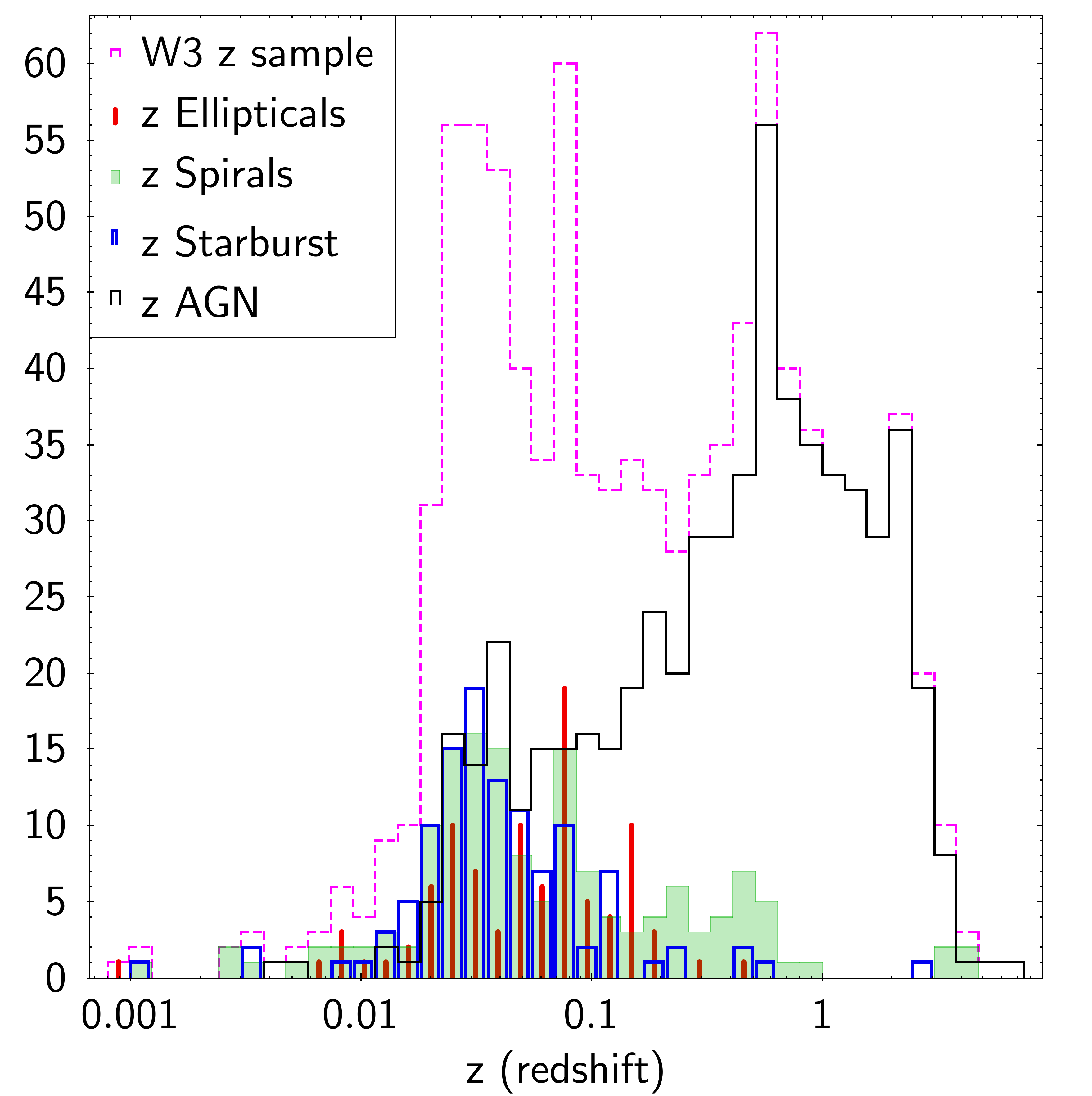}
\caption{Summary of the relative contributions (Figs. \ref{z_histo_ell} to \ref{z_histo_agn}) to the W3 z sample (dotted magenta outline). Elliptical galaxies are represented with vertical red lines, spirals with full green bars, starburst with empty blue bars, and AGN with a continuous black distribution, following the colour scheme of Fig. \ref{C_C_all}.}\label{z_histo_pass}
\end{figure}

Fig. \ref{z_histo_pass} summarises the results of Figs. \ref{z_histo_ell} to \ref{z_histo_agn}, by showing the relative contributions of the different source populations to the overall W3 z sample, and explaining some of the bimodality that appears in the latter. Regular galaxies and LERGs are likely to contribute $\sim 50$ per cent to the first peak of the W3 z sample distribution, with radiatively efficient AGN gradually taking over and making up most of the second peak of the distribution. The W3 cut clearly eliminates some of the sources that would fill the gap between both peaks of the distribution, but it is likely that the other selections, especially the radio, also contribute to create the bimodal shape.

%

\section{Luminosity diagnostics}\label{LDiag}

We have used different techniques to correct the flux densities to rest-frame for each wavelength, in order to obtain the respective luminosities. With these luminosity plots we can determine how true the sources in each population are to the labels we assigned to them based in Fig. \ref{C_C_all}.

For the radio we assumed a spectral index ($\alpha$, where $S_{\nu}\propto\nu^{-\alpha}$) of 0.8, which is consistent with what is found in most star-forming galaxies \citep[e.g.][]{Magnelli2015,Magliocchetti2014}, and a large fraction of AGN. Using a single value of $\alpha$ for AGN is not ideal, as this population can exhibit quite substantial variation in their spectral indices, but without data at other frequencies it is a necessary compromise, and, as we will see in section \ref{RL_RQ_section}, moderate changes in $\alpha$ ($\pm 0.2$) have very little impact on our results. We had initially planned to include low-frequency data from VLSSr \citep[the VLA Low-frequency Sky Survey Redux, ][]{Lane2012} in our sample, to more accurately estimate the spectral index and jet kinetic energy for each source class, but doing so would have drastically reduced the number of sources (we only found $\sim$150 VLSSr-MIXR matches within a 30 arcsec radius, for our full sample). We consider that this assumption does not introduce a larger degree of uncertainty than any of the others we have used throughout this work.

For the WISE values we first calculated the flux densities using the zero point magnitude values given in the on-line documentation and the work of \citet{WISE2010}, adopting the additional colour correction for starburst sources. We then used SED (spectral energy distribution) fitting software to correct the flux densities to rest-frame values. We used the SED code developed by Ruiz et al. (2016, in prep.), which uses additional torus templates and stellar emission, as well as the simple templates for elliptical, spiral, starburst and AGN galaxies from the SWIRE library \citep{Polletta2007}. We included all the SDSS (aperture and reddening corrected) magnitude measurements in the fit, to better constrain the contribution from the host. Once we obtained the redshift-corrected fluxes from the SED curves and filter profiles, we calculated the luminosities and extrapolated the flux and redshift errors to estimate their uncertainties. 

The X-ray corrections were somewhat more problematic. The fluxes in the 3XMM DR5 source catalogue are calculated with the same method established for 2XMM \citep{Watson2009}, which assumes a series of corrections based on a power law fit with $\Gamma = 1.7$ and a fixed foreground $N_H$ column \citep[see also][]{Mateos2009}. Working with this assumption would introduce a bias, as a fraction of our sources are likely to deviate quite substantially from this model, especially at low energies \citep[see e.g. the work by ][on the XMM-Newton spectral fit database]{Corral2015}. The alternative would be to use the detections version of the catalogue, which lists the count rates and instruments used for each observation of each source, and use different models for each population. This solution, however, would still have required a model assumption, as it would not be possible to obtain reliable spectra for the faintest sources to carry out proper spectral fitting. It would also have required assumptions on the instrument observation modes and responses to use in each case, a work that was already done to obtain the 3XMM fluxes. As such, we decided to work with the catalogue fluxes, using a model very similar to that assumed by \citet{Watson2009}, but slightly more flexible, and to limit our luminosities to the (rest-frame) 2--10 keV range, where divergences from our assumed model should be minor.

To calculate the 2--10 keV luminosities we used the X-ray spectral analysis tool XSPEC, with an X-ray model consisting of a foreground absorption $N_H$ column (\textit{tbabs}) set to the Galactic value, an intrinsic absorption column (\textit{ztbabs}), and a powerlaw with $\Gamma = 1.7$. We used the method of \citet{Willingale2013} to calculate the Galactic extinction column. This method is innovative in that it takes into account both the atomic (HI) and the molecular (H$_{2}$) Hydrogen absorption columns; the first is calculated from the 21 cm Leiden/Argentine/Bonn maps of \citet{Kalberla2005}, while the second is obtained using the dust maps of \citet{Schlegel1998} and constraints from Gamma-ray burst afterglows detected by Swift. We used the abundance values from \citet{Wilms2000} and cross-sections from \citet{Verner1996}. We fixed the foreground and intrinsic $N_H$, and the powerlaw slope, for each source and set the powerlaw normalisation to 1.0. , calculated the 2--12 keV flux, and used its ratio to the catalogue 2--12 keV flux (SC\_EP\_FLUX\_4 + SC\_EP\_FLUX\_5 = F$_{2-4.5 keV}$ + F$_{4.5-12 keV}$) to rescale the 2--10 keV luminosity.

To fully appreciate the range of uncertainty present in our X-ray luminosity estimations, rather than just propagating the redshift and flux errors, we calculated lower and upper boundaries for each source including a variation in the intrinsic $N_H$ between zero and $10^{22}$ cm$^{-2}$, with the nominal value at $10^{21}$ cm$^{-2}$. Thus, for the upper luminosity boundary we used the highest intrinsic $N_H$ value ($10^{22}$ cm$^{-2}$), the highest possible redshift ($z + z_{err}$), and the largest flux value (F$_{2-12 keV}$ + F$_{2-12 keV, \ err}$), while for the lowest boundary we used the lowest intrinsic $N_H$ (zero), the lowest redshift ($z - z_{err}$), and the lowest flux value (F$_{2-12 keV}$ - F$_{2-12 keV, \ err}$). Our chosen range of $N_H$ does not encompass heavily obscured and Compton-thick sources. Although we know that WISE is very good at detecting obscured AGN \citep[e.g][]{Stern2012,Assef2013,Mateos2013}, we only expect a fraction of them to be detected by the other catalogues we used for our sample, in particular 3XMM. To be safe, however, we excluded most remaining potential cases from our sample with our WISE colour cuts (see section \ref{C_C}), as it would mean imposing a very restrictive criterion, which only affects a fraction of AGN \citep[e.g.][]{Wilkes2013}, over the entire sample. As we will see in section \ref{RL_RQ_section}, a few potentially Compton-thick sources may remain, but the fraction is very low and should not affect our conclusions.

Unlike for the radio and mid-IR, we do have some upper limits for a few X-ray fluxes, which we have represented with down-pointing arrows on our plots. This is due to the fact that in 3XMM only a detection in one of the bands (and a minimum overall detection likelihood) is required, upper limits can be derived for the other bands, while for the radio there is a single band, and our S/N filters in WISE ensure that we have detections in all the bands involved. For the X-ray upper limits we have nonetheless calculated positive errors as well, so as to fully reflect the other sources of uncertainty (redshift and $N_H$). The larger flux errors and the introduction of a further uncertainty (from the $N_H$ values) also contribute to the X-ray data having, in most cases, larger error bars than the mid-IR and radio data.

\subsection{Luminosity correlations}\label{LCorr}

\begin{figure*}
\centering
\includegraphics[width=0.95\textwidth]{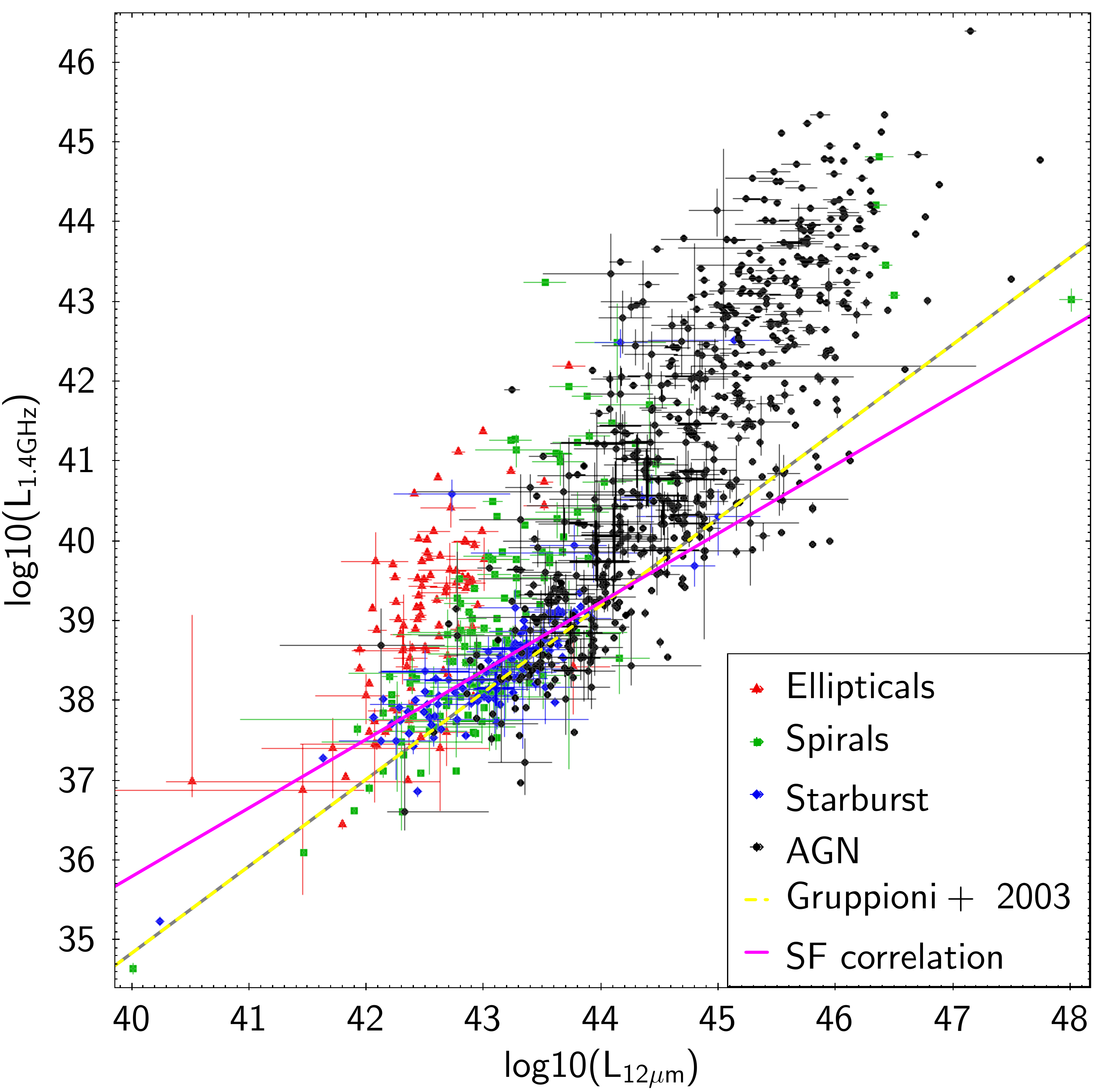}
\caption{1.4 GHz versus 12$\mu$m luminosity for the W3 z sample subsets listed in Table \ref{nSources}. Colours and symbols as in Fig. \ref{C_C_all}. The yellow and grey dotted line shows the radio/mid-IR star formation correlation of \citet{Gruppioni2003}; The magenta line shows our best correlation fit for the starburst sources (eq. \ref{SFcorr1}).}\label{LR_LI}
\end{figure*}

\begin{figure*}
\centering
\includegraphics[width=0.95\textwidth]{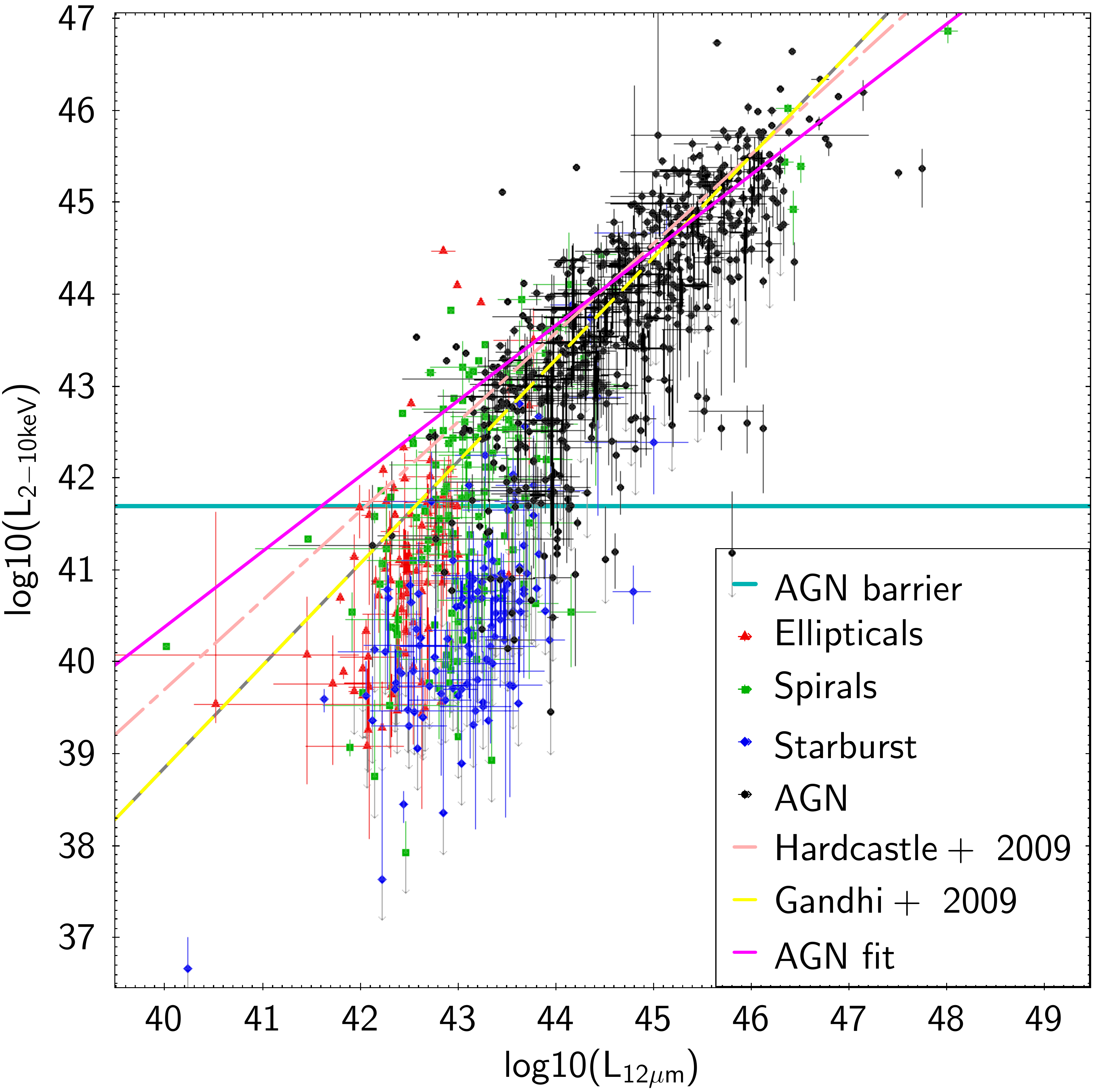}
\caption{2--10 keV versus 12$\mu$m luminosity for the W3 z sample subsets listed in Table \ref{nSources}. Colours and symbols as in Fig. \ref{C_C_all}. The horizontal cyan line indicates the X-ray luminosity above which sources are classified as AGN (which we have set at $5\times10^{41}$ erg s$^{-1}$, see the text). Upper limits (only for the X-rays) are indicated with grey arrows. The dashed pink line shows shows the X-ray/mid-IR AGN correlation of \citet{Hardcastle2009}. The yellow and grey dotted line shows the X-ray/mid-IR AGN correlation of \citet{Gandhi2009}. The magenta line shows our best correlation fit for the AGN sources (eq. \ref{AGNCorr1}).}\label{LX_LI}
\end{figure*}

\begin{figure*}
\centering
\includegraphics[width=0.95\textwidth]{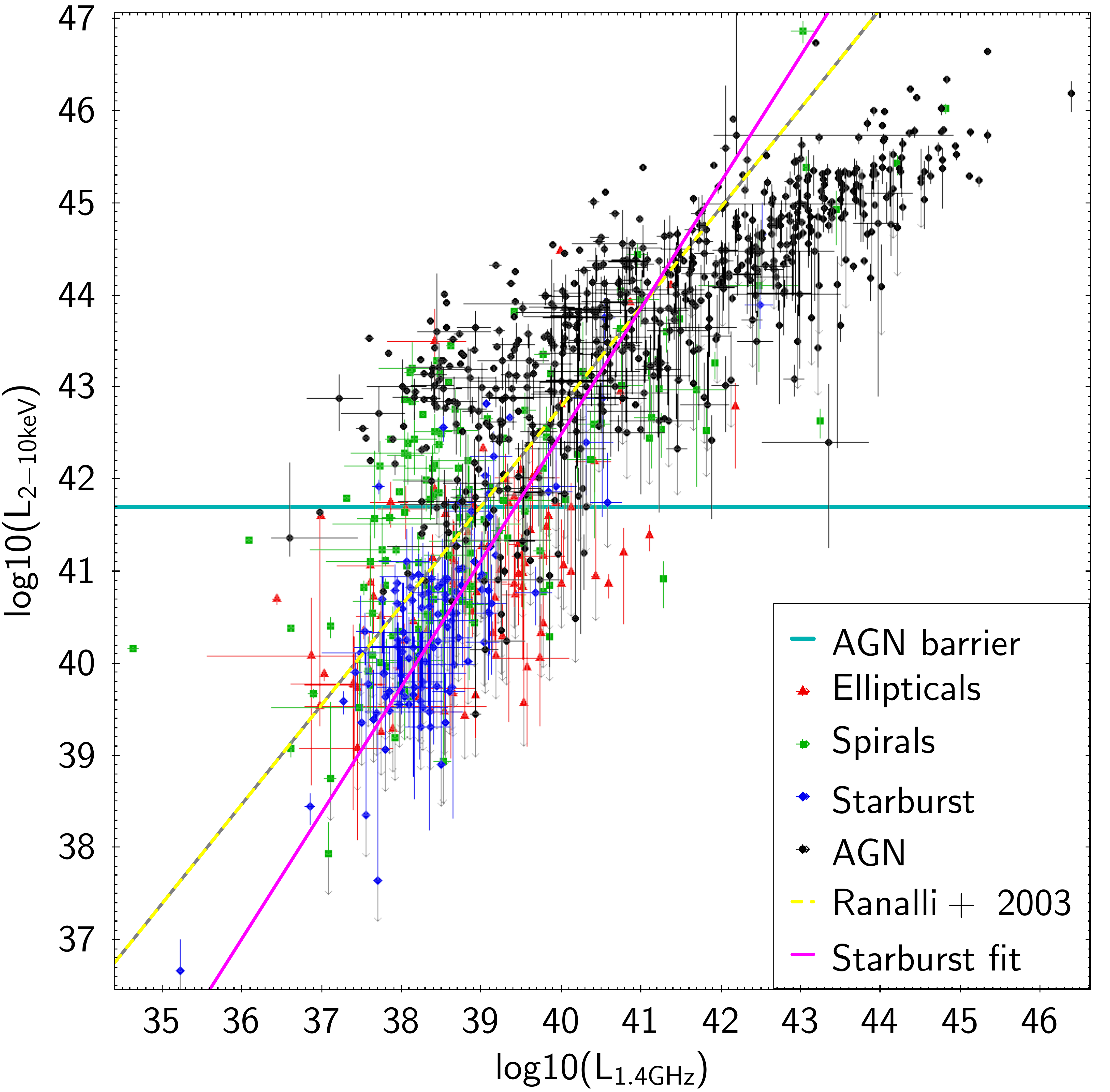}
\caption{2--10 keV versus 1.4 GHz luminosity for the W3 z sample subsets listed in Table \ref{nSources}. Colours and symbols as in Fig. \ref{C_C_all}. The horizontal cyan line indicates the X-ray luminosity above which sources are classified as AGN (which we have set at $5\times10^{41}$ erg s$^{-1}$, see the text). Upper limits (only for the X-rays) are indicated with grey arrows. The yellow and grey dotted line shows the X-ray/radio star formation correlation of \citet{Ranalli2003}. The magenta line shows our best correlation fit for the starburst sources (eq. \ref{XSFCorr}).}\label{LX_LR}
\end{figure*}

There is a danger, when assessing luminosity/luminosity plots, to forget that both quantities have a common dependence with redshift. This is evident when comparing Figs. \ref{LR_LI} to \ref{LX_LR} with their flux counterparts in section \ref{FCorr}. For this reason, we have carried out a partial correlation analysis, to test the strength of the luminosity correlations for various subsets. Partial correlation analysis measures the degree of association of two variables (in our case, the two luminosities), when the effect of a third variable (in our case, redshift) is removed. The method we have used for this work is based on Kendall's rank correlation coefficient, which is non-parametric, meaning that it does not rely on any assumptions for the two variables tested. The derivation is described in detail by \citet{Akritas1996}. An advantage of this method is that it works with censored data (upper limits), allowing us to keep our X-ray upper limits. We found the results to agree quite well with those of \citet{Hardcastle2006,Hardcastle2009}, and \citet{Mingo2014}. The most relevant results are listed in Table \ref{parcorr}.

\begin{table}\small
\caption{Results of the partial correlation analysis for the most relevant source subsets. The number of sources in each subset is given in column 3. Column 4 lists the values for Kendall's $\tau$; column 5 shows the square root of the variance; in column 6, $\tau/\sigma$ gives an idea of the strength of the correlation between the luminosities in column 1 in the presence of redshift. We consider the correlation significant if $\tau/\sigma>3$.}\label{parcorr}
\centering
\begin{tabular}{llcccc}\hline
L tested&Subset&n&$\tau$&$\sigma$&$\tau/\sigma$\\\hline
\multirow{2}{*}{L$_{1.4 GHz}$-L$_{12 \mu m}$}&Starburst&114&0.58&$6.36\times10^{-2}$&9.15\\
&AGN&577&0.27&$2.38\times10^{-2}$&11.35\\\hline
\multirow{2}{*}{L$_{2-10 keV}$-L$_{12 \mu m}$}&Starburst&114&0.21&$4.81\times10^{-2}$&4.36\\
&AGN&577&0.30&$2.26\times10^{-2}$&13.50\\\hline
L$_{2-10 keV}$-L$_{1.4 GHz}$&AGN&577&0.31&$2.33\times10^{-2}$&13.50\\\hline
\end{tabular}
\end{table}

Fig. \ref{LR_LI} shows the 1.4 GHz luminosity versus the 12 $\mu$m luminosity for all the sources in the W3 z sample, using the same classifications we derived from Fig. \ref{C_C_all}. As we had already anticipated from the flux distributions in Fig. \ref{w3_combFlux}, and as we can see in Table \ref{parcorr}, the starburst sources, as well as some of the spirals, follow a strong correlation. We find $\tau / \sigma =9.13$, a highly significant correlation. Some of the AGN seem to follow an extension of this correlation (radio-quiet AGN), while others have larger radio luminosities (radio-loud AGN), that also seem to span a wide range of values. The elliptical galaxies and some of the spirals also seem to have a radio excess with respect to the starburst sources, indicating that they host radio-loud AGN given that, as we remarked in section \ref{FCorr}, the correlation derived from the starburst sources represents the maximum degree of star formation we can detect with the flux limits of FIRST and NVSS. 

We calculated the radio/mid-IR star formation correlation \citep[an extension of the radio/FIR correlation originally described by][]{VDKruit1973,Condon1982,DeJong1985} \citep[see also e.g. the NVSS/IRAS results of][]{Yun2001} for the starburst sources in Fig. \ref{LR_LI}. For this and all subsequent linear fits we used the Bayesian MCMC (Markov chain Monte Carlo) code developed by \citet{Hardcastle2009}, which can work with upper limits, when present. We excluded the three most obvious outliers (which have high redshifts and AGN-like X-ray luminosities). The resulting correlation will also be used in section \ref{RL_RQ_section}, as a baseline to establish the break between star-forming sources and radio AGN. The correlation we found is:
\begin{equation}\label{SFcorr1}
\log(L_{1.4 GHz})=(0.86\pm0.04) \log(L_{12 \mu m}) + (1.4\pm1.5)
\end{equation}
The MCMC fit also provides a measure of the intrinsic scatter, $\Delta_{LR-LI}=0.54\pm0.05$ (in linear units), such that e.g. a $3\sigma$ distance from the line fit would be the equivalent of multiplying the linear equivalent of eq. \ref{SFcorr1} by $(1+\Delta_{LR-LI})^{3}$ (see also Section \ref{RL_RQ_section} for more details). 
We have also plotted in Fig. \ref{LR_LI} the correlation originally obtained by \citet{Gruppioni2003}, which would translate in our units as:
\begin{equation}\label{GruppioniCorr}
\log(L_{1.4 GHz})=(1.09\pm0.05) \log(L_{12 \mu m}) -(8.76\pm0.54)
\end{equation}
Our results are not entirely consistent with those of \citet{Gruppioni2003}. We find a flatter slope, but this could be due to the different selection criteria and redshift ranges covered by both samples, as well as the limited range of luminosities spanned. In terms of redshift-corrected fluxes, the slope in eq. \ref{SFcorr1} corresponds to a value of the IR/radio flux ratio $q_{12}\sim0.78$, which seems compatible with the results obtained (at 24 $\mu$m) from Spitzer data \citep[e.g.][]{Appleton2004,Garrett2015}. The FIR/radio star formation correlation extrapolates linearly quite well into the mid-IR since, even though both the IR and the radio can underestimate star formation at low galaxy luminosities, they do so in a way that the correlation is preserved \citep{Bell2003}. However, recent results show that there may be a dust temperature dependence \citep{Smith2014}, so the results need to be carefully checked for each sample.

It is interesting to note in Fig. \ref{LR_LI} that although not many sources with QSO-like 12 $\mu$m luminosities ($\ge10^{45}$ erg s$^{-1}$) seem to follow the extrapolation of the star-formation correlation, as most luminous sources also seem to be fairly radio-loud, there are indeed a few that do so. This is probably one of the factors driving the correlation we see in Table \ref{parcorr}. Even without a detailed analysis of their star formation rate it is difficult to see how such high radio luminosities could be achieved purely through star formation, and indeed if these sources, which predominantly inhabit the higher end of our redshift distribution, would be detected at all in FIRST/NVSS based solely on their star formation. It is very possible that their emission is also arising from jets and lobes, albeit less powerful ones than those of their more radio-luminous counterparts, or that shocks driven by powerful radiative winds are producing relativistic particles that, in turn, produce synchrotron radio emission, as suggested by \citet{Zakamska2016,Nims2015}, or a combination of both factors. In any case, it is clear that it might not be wise to use the mid-IR/radio star formation correlation to draw conclusions on the star formation rate of very luminous AGN.

Our conclusions are reinforced by what we observe in Fig. \ref{LX_LI}, which shows the 12 $\mu$m luminosity versus the 2--10 keV luminosity for our sources. As we saw in our previous work, all the sources over the AGN barrier seem to follow a fairly tight correlation (Table \ref{parcorr}), which holds even when we consider the common dependence with redshift, with a few outliers that may suffer from beaming (if they have an X-ray excess) or heavy obscuration (if they have an IR excess). This makes sense, as both the mid-IR and the X-rays are expected to be very good proxies for AGN activity. We have plotted a horizontal line at $L_{2-10 keV}=5\times10^{41}$ erg s$^{-1}$ as a reference, to indicate the point above which the X-ray emission we observe is most likely to originate in AGN activity (only very high star formation rates, $>100$ M$_{\odot}$ yr$^{-1}$, can produce X-ray luminosities around this break without any AGN contribution). The exact luminosity at which this happens is a matter of debate, as many (radio-quiet) AGN studies place it at $10^{42}$ erg s$^{-1}$, but we have seen in our previous work that the break between HERGs (high excitation or radiatively efficient radio galaxies) and LERGs seems to occur closer to $10^{41}$ erg s$^{-1}$, so we have plotted the line at an intermediate value. We discuss some of the implications of this choice in section \ref{RL_RQ_section}.

We see that, despite some of them following the tail of the starburst sources on the previous plot, the vast majority of the AGN have X-ray luminosities that leave no doubt as to the nature of the X-ray emission. The few outliers we see are likely luminous infrared galaxies that have W1-W2 values slightly larger than 0.5, as the diagram from \citet{Lake2012} shows that there is some overlap between the populations in the WISE colour/colour diagram. A surprisingly large fraction of spiral galaxies also seem to harbour X-ray luminous AGN, more than was immediately apparent from the fluxes in Fig. \ref{w3_hardX}. Looking again at Fig. \ref{LR_LI}, the fraction of spiral galaxies that also have substantial radio emission seems to be smaller than that of spirals with bright X-ray AGN, thus the number of fairly radio-loud Seyferts we find is not large, but it proves again that these sources do indeed exist. In fact, given that the spiral sources are hit the hardest by the W3 S/N cut, it seems that we really need deeper mid-IR (and X-ray) data to properly study this population.

We carried out a MCMC linear fit to the AGN sources, using the method of \citet{Hardcastle2009}, as described above, to compare our results with those obtained by \citet{Hardcastle2009} and \citet{Gandhi2009} for radio-loud AGN and Seyfert galaxies. We excluded three outliers with X-ray luminosities $4\sigma$--$5\sigma$ above the correlation, which artificially steepened it (it is possible that the redshifts or fluxes for these sources are not entirely correct, or that they are relativistically beamed). Although it is clear at this point that several of the sources in our `spiral' category harbour Seyfert nuclei, we decided to work exclusively with the sources labelled as `AGN', for consistency with our earlier selections. The correlation of \citet{Hardcastle2009} is:
\begin{equation}\label{HcastleCorr}
\log(L_{12 \mu m})=(0.97^{+0.23}_{-0.12})\log(L_{2-10 keV})+(0.91^{+5.35}_{-10.13})
\end{equation}
which has large constraints due to the lower number of sources in the 3CRR sample. Translated to our units, the correlation of \citet{Gandhi2009} is:
\begin{equation}\label{GandhiCorr}
\log(L_{12 \mu m})=(1.11\pm0.07)\log(L_{2-10 keV})-(5.54\pm0.05)
\end{equation}
The best linear fit that we obtained for all the AGN was:
\begin{equation}\label{AGNCorr1}
\log(L_{12 \mu m})=(0.82\pm0.01)\log(L_{2-10 keV})+(7.6\pm0.5)
\end{equation}
which is flatter than those of \citet{Hardcastle2009} and those of \citet{Gandhi2009}, as can clearly be seen in Fig. \ref{LX_LI}. This difference may be due, at least in part, to our X-ray luminosity cut, as well as the different selections arising from the instruments employed in each sample, the fact that we are including quasars, but excluding `spiral' AGN in our correlation, and the presence in our sample of some sources with high mid-IR luminosities, which could be contributing to the flat slope. The scatter for our correlation is (in linear units) $\Delta_{LX-LI}=1.25\pm0.08$, larger than that found by \citet{Hardcastle2009} ($0.32\pm0.05$), which is consistent with the fact that with MIXR we have sampled a broader range of luminosities and populations. 

Back in Fig. \ref{LX_LI}, the starburst sources seem to follow a distribution parallel to that of the AGN, as we saw on the flux plot of Fig. \ref{w3_hardX}, although there is some scatter, and several of the starburst sources have only upper limits for their X-ray fluxes; this is reflected by the relatively weak correlation in Table \ref{parcorr}. The elliptical galaxies seem to fall off both the AGN and the star formation correlations, which is consistent with what we have previously observed in LERGs \citep{Hardcastle2009,Mingo2014}.

On Fig. \ref{LX_LR} we can see the 2--10 keV luminosity versus the 1.4 GHz luminosity for our sources. We have again drawn the horizontal `AGN barrier' at $L_{2-10 keV}=5\times10^{41}$ erg s$^{-1}$, as a reference. There seems to be a broad correlation for the AGN and the spiral sources with high X-ray luminosities (Seyferts), which was not evident on the flux plot of Fig. \ref{combFlux_hardX}, and which seems to hold in the presence of redshift (Table \ref{parcorr}). We will see in section \ref{RL_RQ_section} that this correlation is much weaker when we also consider the AGN with W1-W2$<0.5$ on the WISE colour/colour plot. The quiescent spirals, ellipticals and starburst galaxies fall off the correlation for AGN, as expected. Again, the ellipticals display higher radio luminosities than the other two groups, arising from the presence of radiatively inefficient AGN.

Although the X-ray/radio star formation correlation is not very strong for our sample ($r\sim0.5$), we carried out a MCMC linear fit to the starburst sources, again excluding three very X-ray bright outliers, to compare the results with those of \citet{Ranalli2003}, who found (in our units):
\begin{equation}
\log(L_{2-10 keV})=(1.08\pm0.09) \log(L_{1.4 GHz}) -(0.4\pm2.7)
\end{equation}\label{RanalliCorr}
Our best fit shows:
\begin{equation}
\log(L_{2-10 keV})=(1.37\pm0.06) \log(L_{1.4 GHz}) -(12.3\pm2.5)
\end{equation}\label{XSFCorr}
Our result is steeper than that of \citet{Ranalli2003}, but compatible if we consider the large underlying uncertainties and the small range of luminosities covered, as well as the fact that we probed lower X-ray luminosities with our sample \citep[see also the more recent results of][]{Ranalli2012}. The scatter for this correlation is (in linear units) $\Delta_{LX-LR}=1.2\pm0.3$, also rather large.

Figs. \ref{LR_LI} to \ref{LX_LR} reinforce the conclusions we reached in section \ref{FCorr} about the nature of the emission in the different source populations, suggesting that our early diagnostics were correct. The spiral galaxies, being a mixed population, require extra care for classification. Data from the next generation of radio and X-ray instruments will prove invaluable to better study this area of the WISE colour/colour plot, but deeper surveys will also reveal additional complexity: our MIXR sample, complex as it already is, is still dealing with relatively shallow flux limits in all the bands, particularly in the radio, and picking up fairly bright sources. Further data need to be obtained, and new techniques need to be developed to better study host-dominated AGN, particularly those with radio jets and lobes. This is a fairly neglected population in AGN studies, that may provide key clues to further our understanding of the life cycles of AGN feedback and its impact on the star formation activity of AGN hosts, as we have seen in sources such as Circinus \citep{Mingo2012} and NGC 6764 \citep{Croston2008}.

%

\section{Revisiting the diagnostics: radio-loudness, accretion mode, and star formation}\label{RL_RQ_section}

\begin{figure*}
\centering
\includegraphics[width=0.95\textwidth]{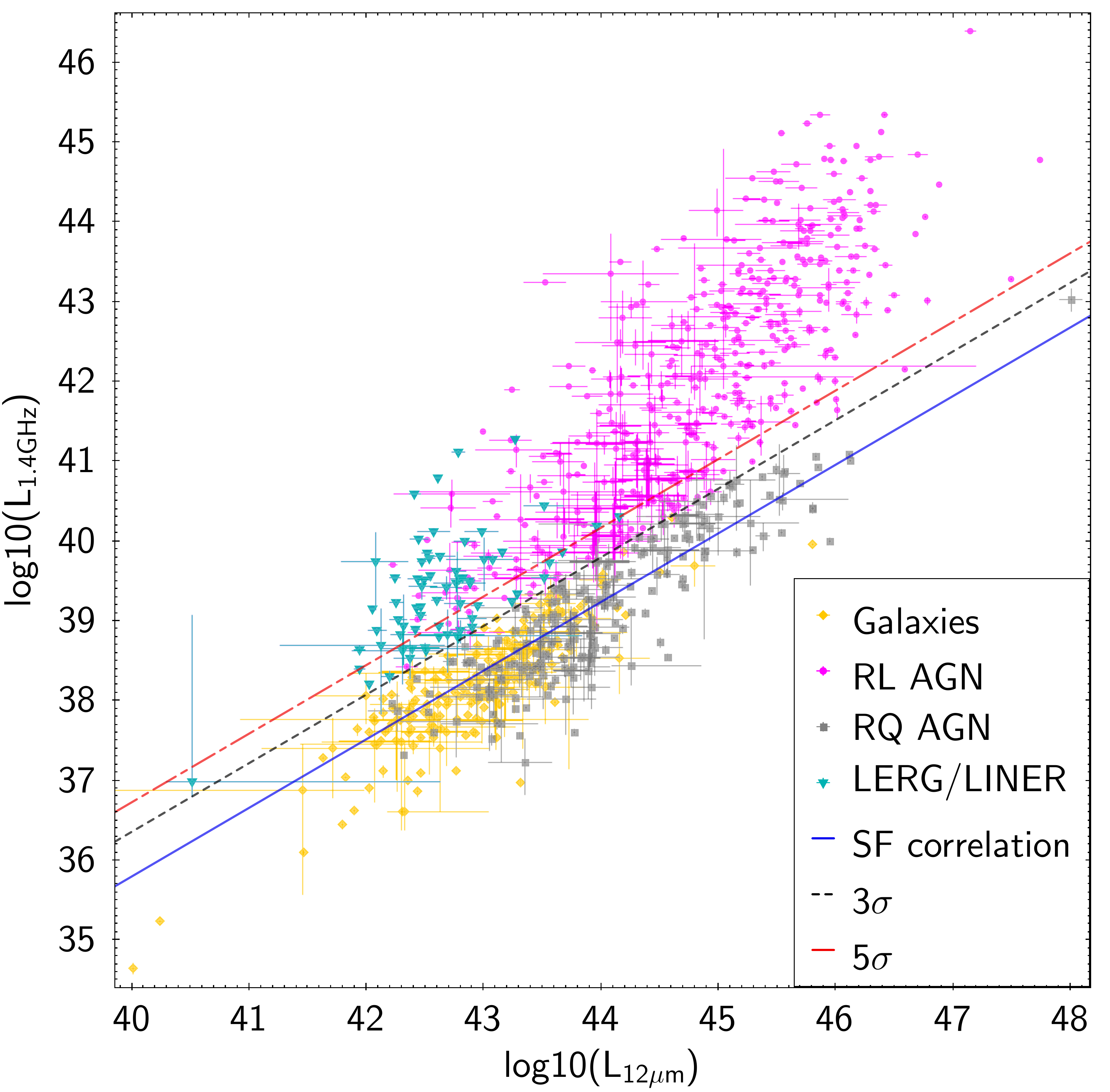}
\caption{1.4 GHz versus 12 $\mu$m luminosity log/log plot for the MIXR sources, illustrating the radio/IR star formation correlation, and the method we used to define the radio-loud and radio-quiet AGN, LERG/LINER and galaxy samples. Only the sources that pass the WISE W3 cut are plotted. The equation for the $3\sigma$ line is: $\log(L_{1.4 GHz})=0.86 \log(L_{12 \mu m}) + 2.0$. We have also plotted a line at 5$\sigma$ above the star formation correlation, for reference (equation: $\log(L_{1.4 GHz})=0.86 \log(L_{12 \mu m}) + 2.3$).}\label{RL_RQ}
\end{figure*}

\begin{table*}\small
\caption{Number of sources in each subset after the new classifications defined from Fig. \ref{RL_RQ}. See also Table \ref{nSources} for the statistics with the old source classifications.}\label{nSources_RL_RQ}
\centering
\begin{tabular}{llcc}\hline
Subset name&Description&\multicolumn{2}{c}{Number of sources}\\\hline
&&W3 S/N$\geq 3$&W3 S/N$< 3$\\\hline
RL AGN&$\log(L_{1.4 GHz})/\log(L_{12 \mu m})\geq(0.86+2.0)$&505&413\\
RQ AGN&$\log(L_{1.4 GHz})/\log(L_{12 \mu m})<(0.86+2.0)$; $L_{2-10 keV}\geq 5\times 10^{41}$ erg s$^{-1}$&218&0\\
LERG/LINER&subset of RL AGN with $L_{2-10 keV} < 5\times 10^{41}$ erg s$^{-1}$&69&34\\
Galaxies&$\log(L_{1.4 GHz})/\log(L_{12 \mu m})<(0.86+2.0)$; $L_{2-10 keV} < 5\times 10^{41}$ erg s$^{-1}$&211&2\\\hline
\end{tabular}
\end{table*}

In the previous sections we have shown that our early diagnostics are very efficient to pre-classify sources based on their various fluxes, mid-IR colours, and hardness ratios. However, and as introduced in section \ref{FCorr}, within each of the groups we defined from Fig. \ref{C_C_all} we see a range of properties that point to a mix of underlying populations (see also Table \ref{Activity}). Now that we know how to identify the different types of emission in each of them, it might be more efficient to redefine the populations based on their physical properties, rather than their mid-IR colours.

Based on their activity we can distinctly identify non-active star-forming galaxies, radio-quiet AGN, and radio-loud AGN. As a reminder of what we introduced in section \ref{FCorr}, we refer to radio-quiet AGN as sources where the radio emission we detect is likely to be originated mainly from stellar processes, accelerated particles in wind-driven shocks \citep[see][]{Nims2015,Zakamska2016} or, if arising from a jet and lobes, they are small and faint, and the AGN produces the bulk of its emission as radiative output in the other bands. Conversely, we refer to radio-loud AGN as those that have substantial kinetic output in the form of jet and lobes, which we measure as radio emission well above the star formation correlation. The radiatively inefficient LERG/LINER sources also follow these last criteria, so they are a subset of radio-loud AGN.

We used the radio/mid-IR star formation correlation we derived in section \ref{LCorr} (eq. \ref{SFcorr1}) as the basis of our new classification. As illustrated in Fig. \ref{RL_RQ}, any sources with radio emission in excess of $3\sigma$ over the correlation most likely harbour luminous radio jets and/or lobes, and thus we can classify them as `RL AGN' (this category includes both LERGs/LINERs and the radio-loud fraction of the radiatively efficient AGN, or HERGs). $\sigma$ is defined from the MCMC linear fit intrinsic scatter, $\Delta_{LR-LI}$, as detailed in eq. \ref{SFcorr1}, and the equation for the $3\sigma$ line is given in the caption of Fig. \ref{RL_RQ}. In Figs. \ref{LX_LI} and \ref{LX_LR} we used a barrier to define a reference X-ray luminosity above which a source must host an X-ray AGN (L$_{2-10 keV}=5\times10^{41}$ erg s$^{-1}$); applying that barrier to the data in Fig. \ref{RL_RQ}, we can distinguish between `galaxies' and `RQ AGN', and also define the subset of RL AGN that are clearly LERG/LINER sources. Our `radio-loud barrier' is fairly conservative, and likely to classify as non-active galaxies several sources with weak but non-negligible jets and lobes, but such an approach might be necessary for the brighter, QSO-like sources, where some radio emission could arise in radiatively-driven shocks, as mentioned above. Even though the break between LERG and HERG for our sources is determined by the X-ray luminosity, we can also see it clearly in the mid-IR in Fig. \ref{RL_RQ}, and it happens at the same range of mid-IR luminosities as those observed by \citet{Gurkan2014}. The statistics for the new source subsets are listed in Table \ref{nSources_RL_RQ}.

The percentage of non-active galaxies in our sample is $\sim17$ per cent ($\sim27$ per cent if we only include the sources that pass the W3 cut, which eliminates mostly RL AGN, see Table \ref{nSources_RL_RQ}). This fraction is smaller than those found with FIRST and optical data by \citet{Magliocchetti2002,Ivezic2002}, but that is not entirely surprising, considering that we are using flux cuts in more wavelengths, and NVSS data as well as FIRST. Comparing Figs. \ref{RL_RQ} and \ref{LR_LI}, we can also see that the LERG/LINER in our sample inhabit mostly sources with elliptical colours (although there are a few outliers).

\begin{table*}\small
\caption{Results of the partial correlation analysis for the most relevant source subsets. The number of sources in each subset is given in column 3. Column 4 lists the values for Kendall's $\tau$; column 5 shows the square root of the variance; in column 6, $\tau/\sigma$ gives an idea of the strength of the correlation between the luminosities in column 1 in the presence of redshift. We consider the correlation significant if $\tau/\sigma>3$. Please see the source classifications and statistics in Table \ref{nSources_RL_RQ}. HERG are RL AGN that are not LINERs/LERGs.}\label{parcorr2}
\centering
\begin{tabular}{llcccc}\hline
L tested&Subset&n&$\tau$&$\sigma$&$\tau/\sigma$\\\hline
\multirow{4}{*}{L$_{1.4 GHz}$-L$_{12 \mu m}$}&HERG&436&0.32&$2.94\times10^{-2}$&10.93\\
&RQ AGN&218&0.49&$4.95\times10^{-2}$&9.90\\
&HERG+RQ AGN&654&0.24&$2.19\times10^{-2}$&11.04\\
&LERG/LINER&69&0.29&$7.87\times10^{-2}$&3.69\\\hline
\multirow{4}{*}{L$_{2-10 keV}$-L$_{12 \mu m}$}&HERG&436&0.39&$2.78\times10^{-2}$&13.85\\
&RQ AGN&218&0.23&$4.18\times10^{-2}$&5.60\\
&HERG+RQ AGN&654&0.35&$2.18\times10^{-2}$&15.78\\
&Galaxies&211&0.11&$3.63\times10^{-2}$&2.94\\\hline
\multirow{4}{*}{L$_{2-10 keV}$-L$_{1.4 GHz}$}&HERG&436&0.38&$2.72\times10^{-2}$&13.95\\
&RQ AGN&218&0.13&$3.88\times10^{-2}$&3.40\\
&HERG+RQ AGN&654&0.29&$2.12\times10^{-2}$&13.78\\
&Galaxies&211&0.12&$3.42\times10^{-2}$&3.42\\\hline
\end{tabular}
\end{table*}

When we extended our partial correlation analysis to the new classifications we also found some interesting results (Table \ref{parcorr2}). Our results illustrate how important large statistics are when studying populations with the amount of scatter we observe. The RQ AGN have more scatter and lower statistics in the relation between L$_{2-10 keV}$ and L$_{12 \mu m}$ that their RL counterparts, but when added to the radiatively efficient RL AGN (HERGs), they strengthen the overall correlation, which is consistent with our previous results on the 2Jy and 3CRR samples \citep{Hardcastle2006,Hardcastle2009,Mingo2014}, and it highlights the fact that both quantities are very good proxies for radiatively efficient accretion. In the relation between L$_{1.4 GHz}$ and L$_{12 \mu m}$, the RQ AGN follow a slightly different correlation from that of the HERGS (as evident in fig. \ref{RL_RQ}), and thus they don't add much to the overall AGN correlation. The correlation between L$_{2-10 keV}$ and L$_{1.4 GHz}$ shows a similar situation, due to the much larger scatter and a lower reange of L$_{1.4 GHz}$ covered by the RQ AGN with respect to their RL counterparts. These results reinforce our conclusions about the scatter in the relation between radiative and kinetic power in AGN, which we discuss in detail in section \ref{Power}. The L$_{2-10 keV}$--L$_{1.4 GHz}$ correlation for the galaxies is fairly weak, as expected from our earlier results. The low number of sources and the presence of many upper limits in the X-rays are diluting the underlying star formation correlation.

\begin{figure}
\centering
\includegraphics[width=0.46\textwidth]{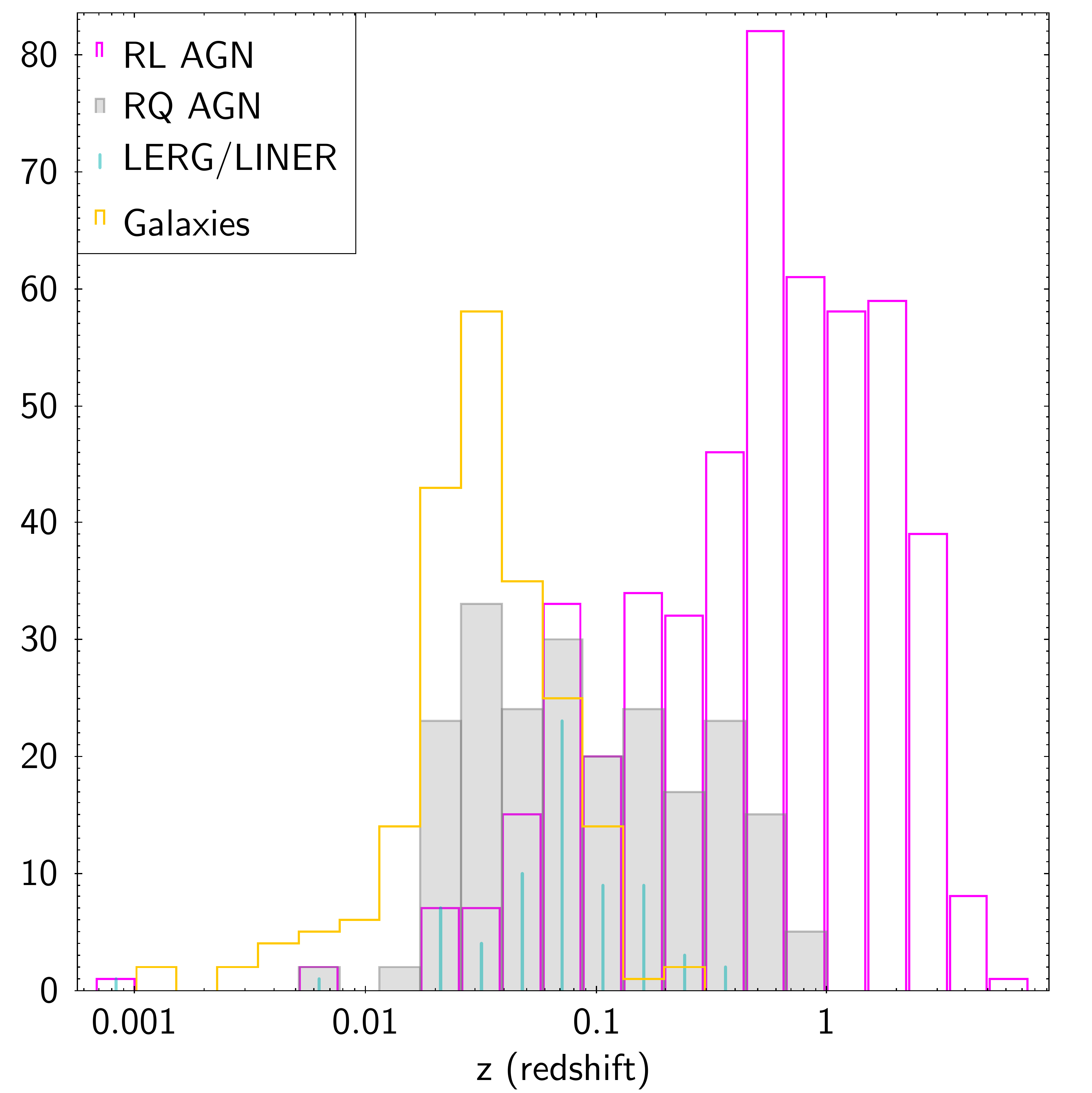}
\caption{Redshift distribution histogram for the samples defined in Fig. \ref{RL_RQ}. Only sources that pass the W3 cut are considered.}\label{z_histo_RL_RQ}
\end{figure}

\begin{figure}
\centering
\includegraphics[width=0.46\textwidth]{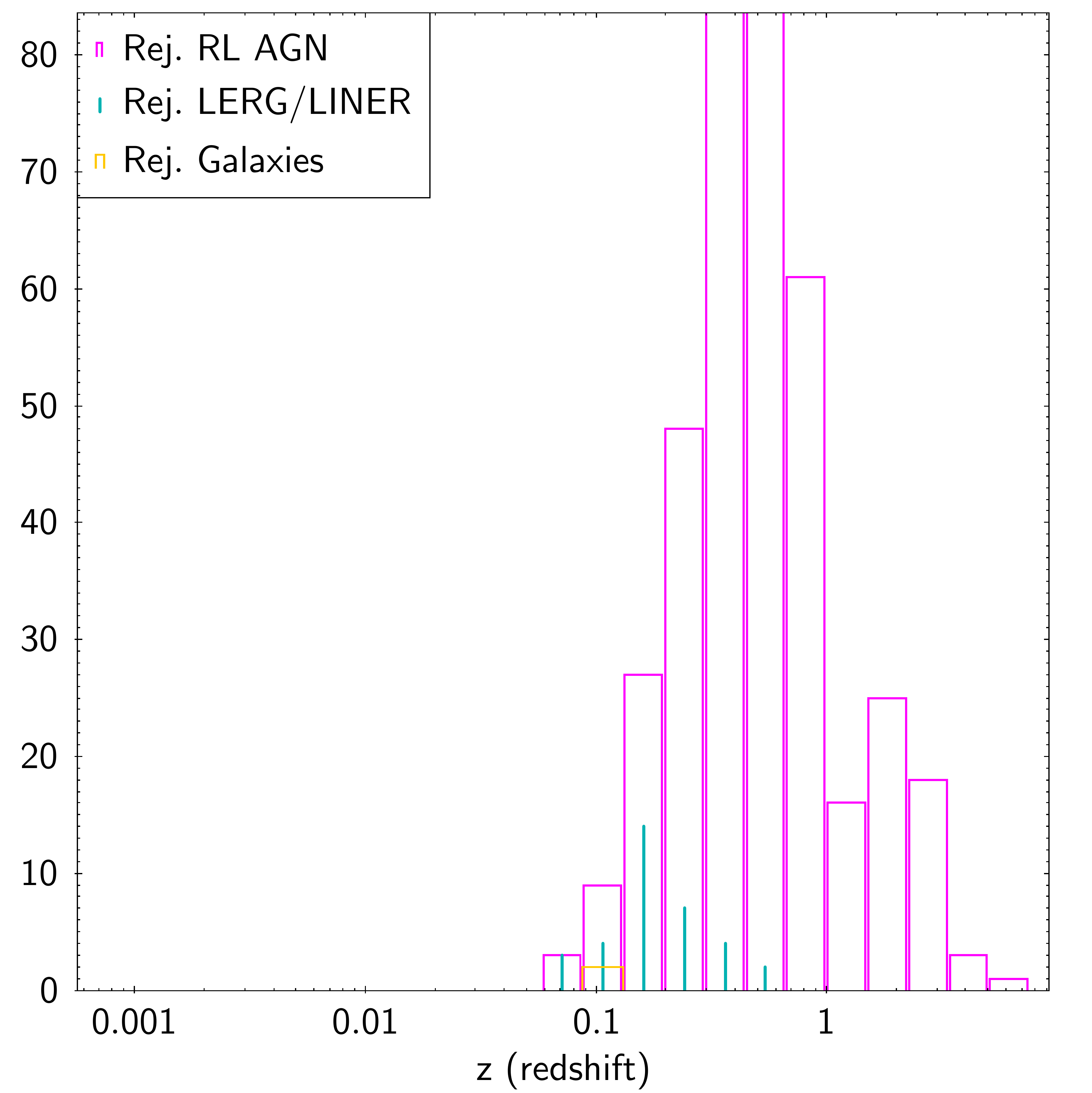}
\caption{Redshift distribution histogram for the sources that do not pass the WISE W3 S/N cut, using the sample selection criteria defined in Fig. \ref{RL_RQ}.}\label{z_histo_RL_RQ_rej}
\end{figure}

Figs. \ref{z_histo_RL_RQ} and \ref{z_histo_RL_RQ_rej} show the redshift distributions with the new classifications, for sources that pass and do not pass the W3 S/N cut, respectively. What immediately draws attention in Fig. \ref{z_histo_RL_RQ} are the very different redshift distributions for RL and RQ AGN, with the former spanning a broader range and peaking at higher $z$ than the latter. This is a selection effect caused by our radio selection, and, as we will see on section \ref{Eddington}, it has some repercussions for the luminosities and Eddington rates we observe for both populations. It is also interesting that the number of RL AGN drops quite quickly below $z\sim0.1$, while that of RQ AGN does not, and that almost all the RL AGN we detect at these low redshifts are LERG/LINER sources. This is probably caused both by selection and evolutionary effects, as the number and power of RL AGN drop quite quickly at lower $z$, and LERGs come to dominate the RL AGN population at low redshifts \citep{Best2014,Williams2015}. Unsurprisingly, the star-forming, non-active galaxies have a redshift distribution that peaks at lower $z$ values than any of the others, and all but disappear after $z\sim0.1$. 

It is also interesting that there are no RQ AGN and two galaxies that do not pass the W3 cut; the bulk of sources lost in this manner are RL AGN  at $z\sim$0.1--2, as we will see in the next sections, most likely Seyfert-like sources with fairly luminous radio structures.

\begin{figure}
\centering
\includegraphics[width=0.46\textwidth]{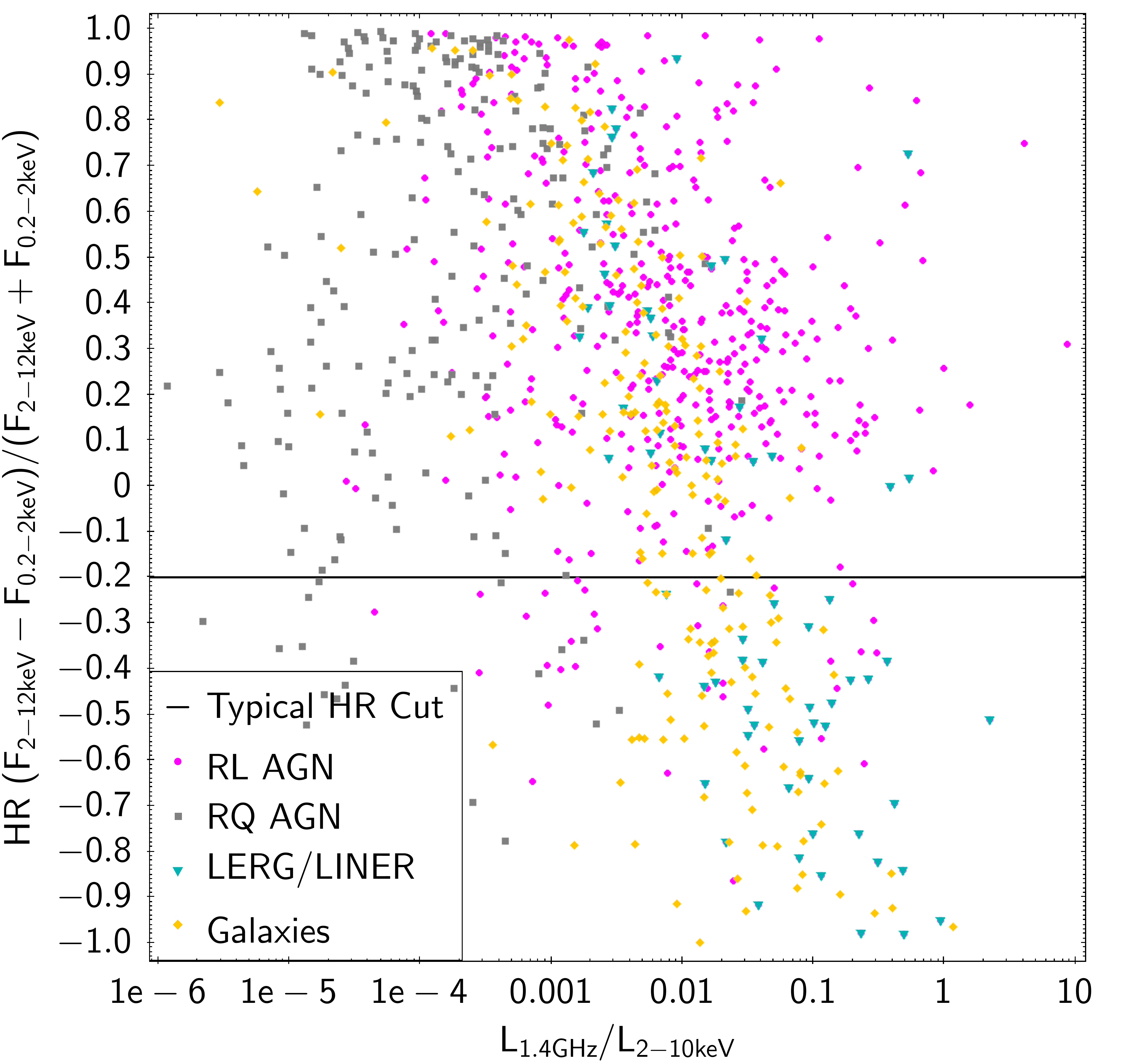}
\caption{Hardness ratios versus the ratio of $L_{1.4 GHz}/L_{2-10 keV}$ for the new samples defined in Fig. \ref{RL_RQ} (only those that pass the W3 cut).}\label{HR_RL_RQ}
\end{figure}

Fig. \ref{HR_RL_RQ} is essentially the equivalent of Fig. \ref{HR_all_2} with the new classifications and using luminosities instead ot fluxes. This plot shows even more clearly than those in section \ref{HR} that a mere hardness ratio cut is not enough to eliminate contamination from non-active galaxies. Overall, the RQ AGN have slightly higher hardness ratios than the RL ones, and the LERGs/LINERs are, as a population, softer than the others, which is consistent with what we know about the radio/soft X-ray correlation for jet emission \citep{Hardcastle1999}, and with the radiatively inefficient nature of the LERG. The fraction of galaxies with high hardness ratios (HR$>0$) is still large after our re-classification, $\sim57$ per cent. It is possible that some of these sources are harbouring low-luminosity AGN. Lowering the AGN barrier to L$_{2-10 keV}=10^{41}$ erg s$^{-1}$ would result in $\sim23$ per cent of the galaxies (and also $\sim 48$ per cent of the LERG/LINER) in Fig. \ref{HR_RL_RQ} being re-classified as AGN. Most of these would-be-AGN galaxies ($\sim 85$ per cent, which correspond to $\sim 19$ per cent of all galaxies) have HR$>0$. While this would not completely solve the conundrum of the `hard' galaxies, as the overall fraction of galaxies with HR$>0$ is much larger, it certainly sheds some light on the fraction of possible low-L AGN that might be present in these sources, even considering the uncertainties. By contrast, raising the threshold to L$_{2-10 keV}=10^{42}$ erg s$^{-1}$ would re-classify $\sim 4$ per cent of the RL AGN as LERG/LINER and $\sim 13$ per cent of the RQ AGN as galaxies.

\begin{figure}
\centering
\includegraphics[width=0.46\textwidth]{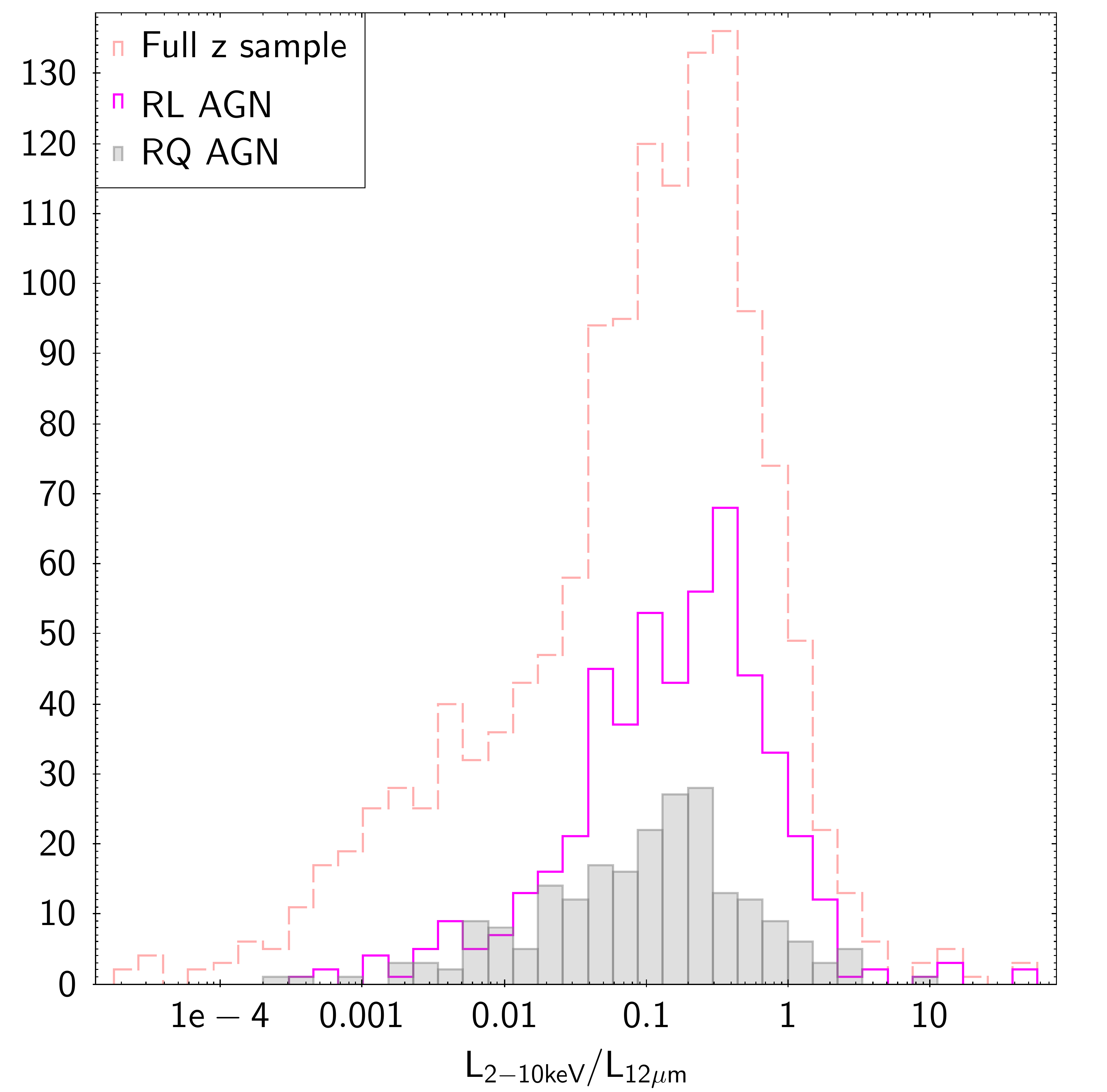}
\caption{$L_{2-10 keV}$/$L_{12 \mu m}$ as a proxy for AGN intrinsic obscuration, for the RL and RQ AGN. The distribution for the full z sample (all sources with redshifts, regardless of their W3 S/N, see Table \ref{nSources}) is also plotted as a dashed line, for reference. Please see also Appendix \ref{ExtraFigures} for the equivalent plot using the 4.6 $\mu$m luminosities.}\label{CT_RL_RQ_1}
\end{figure}

Now that we have obtained a reasonably clean separation between the AGN and the star-forming galaxies, we can also test our assumption that we have a very small fraction of obscured AGN. As shown by e.g. \citet{Alexander2008,Georgantopoulos2011}, the $L_{2-10 keV}$/$L_{6  \mu m}$ ratio is a good proxy to detect Compton-thick sources. We have not extrapolated the corresponding 6 $\mu$m luminosities, as all our other calculations are based on WISE, rather than Spitzer, but we can use our 12 $\mu$m luminosities to compare our results with those of \citet{Rovilos2014}, as shown in Fig. \ref{CT_RL_RQ_1}. We see that the fraction of sources with $L_{2-10 keV}$/$L_{12 \mu m}\leq 0.01$ (potentially Compton-thick AGN) is fairly small, more so if we consider the uncertainties at lower X-ray luminosities (see the error bars in Fig. \ref{LX_LI}), meaning that our chosen range of $N_H$ was appropriate, as such a small fraction of (potentially) Compton-thick AGN should not bias our results. The distributions for RL and RQ AGN are fairly similar, but the former tend to have marginally larger values of $L_{2-10 keV}$/$L_{12 \mu m}$. This could be caused by the fact that the RL sources have additional (soft) X-ray emission arising from the jet, some of which could be contaminating the 2--10 keV band, particularly in beamed objects. We have not plotted the LERGs/LINERs for clarity, but they would occupy the left-most end of the RL AGN distribution. Although radiatively inefficient sources produce soft X-ray emission related to the jet, but no substantial emission in the mid-IR because they have no tori, our LERG/LINER sources still show mid-IR emission from the old stellar population in the host galaxy (arising from the R-J tail of stars, not heated dust in the ISM), as we saw for the 2Jy and 3CRR sources \citep{Mingo2014}, and as studied in detail by e.g. \citet{Mason2012}.

\begin{figure*}
\centering
\includegraphics[width=0.92\textwidth]{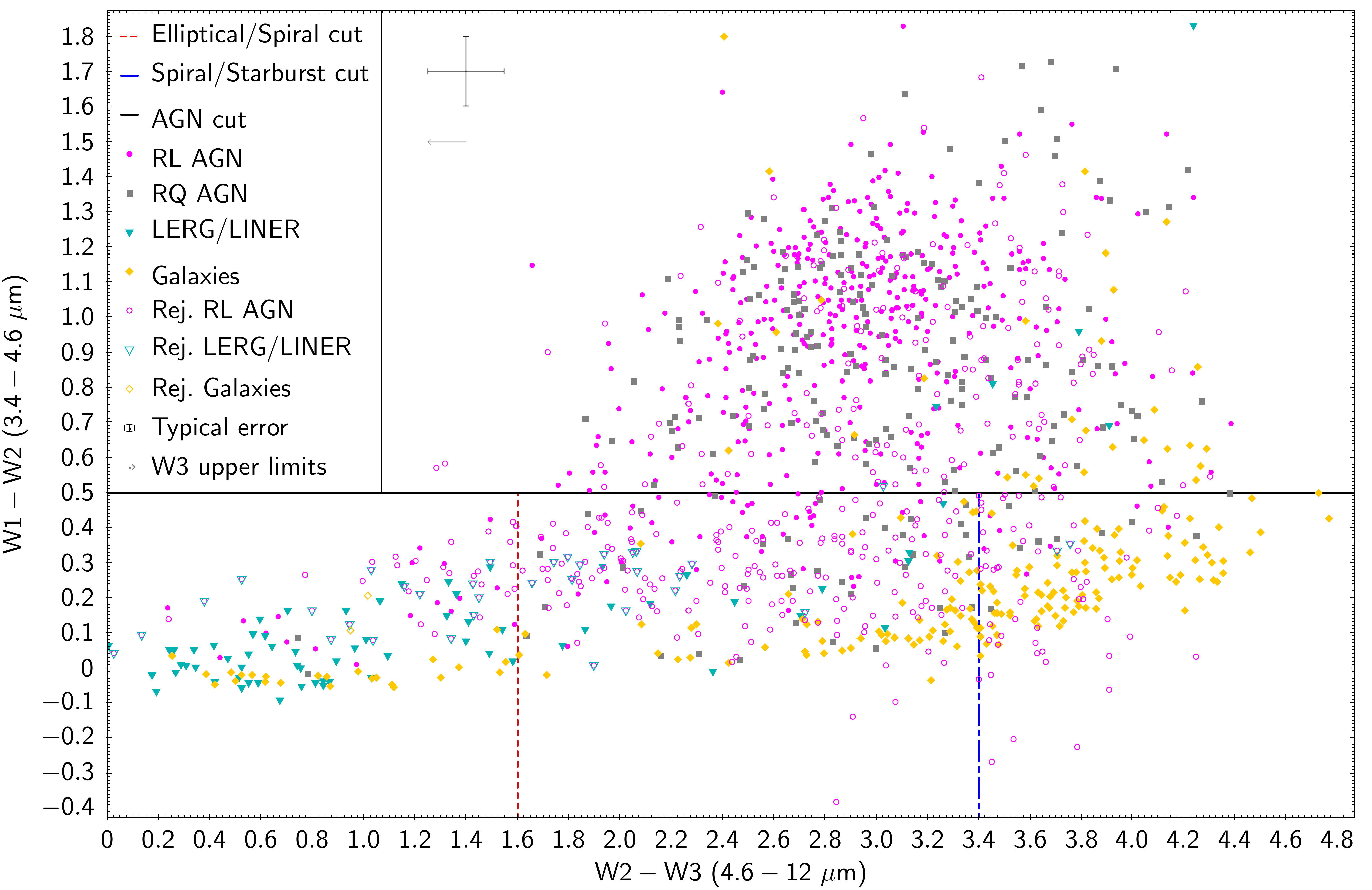}
\caption{WISE colour/colour diagram, illustrating the mid-IR colours of the new subsets defined from Fig. \ref{RL_RQ}, both for the sources that do and do not pass the W3 S/N cut. The cross and arrow next to the legend indicate the typical size of the errors and the direction of the W3 upper limits (for the sources that do not pass the W3 cuts). We have plotted lines to indicate the cuts that we used to separate the various populations in Fig. \ref{C_C_all}, as a reference.}\label{C_C_all_RL_RQ}
\end{figure*}

Finally, Fig. \ref{C_C_all_RL_RQ} illustrates the distribution of our newly defined subsets (both with and without the W3 cut) on the WISE colour/colour diagram. This Figure illustrates the extent to which mid-IR AGN selections that use WISE are biased against low-luminosity AGN. In Fig. 3 of the work of \citet{Gurkan2014}, a large fraction of NLRGs (narrow-line radio galaxies, roughly speaking, the radio-loud equivalent to Seyferts 1.5--2) fall below the $W1-W2 \ge 0.5$ cut, with roughly half of them falling outside the more conservative wedge of \citet{Mateos2012}. We see a very similar situation for the RL AGN in Fig. \ref{C_C_all_RL_RQ}. Our plot also illustrates how the low sensitivity of W3 makes the situation even worse for the lower luminosity RL AGN, as most of them do not pass the S/N cut. The situation does not look as dramatic for the radio-quiet sources on the plot, but that is probably due to the smaller number and redshift range of RQ AGN we are able to detect in our sample; it is very likely that a similar fraction of low-luminosity, intermediate Seyferts are also excluded from WISE-selected AGN samples. Conversely, the wedge of \citet{Mateos2012} (see their Fig. 2) would be very efficient, with our sample, at eliminating contamination from red, non-active galaxies, whereas a simple $W1-W2 \ge 0.5$ cut would not be sufficient. 

It might be interesting to test with an independent sample if the W3 cut has as dramatic an impact on low-luminosity RQ AGN as it does on the (non-LERG/LINER) RL AGN in our sample. These low-luminosity RQ sources are scarce in the MIXR sample because we require a radio detection, and if they are truly radio-quiet, they can only be detected based on their star formation, which limits their host type and redshift. The RQ sources at the same luminosity and redshift ranges as our W3 rejected RL AGN should also have similar W3 fluxes and S/N values, otherwise it would mean that the accretion structures are different for both populations. Extrapolating the fraction of W3 eliminated RL AGN to their RQ counterparts, even without considering the various AGN selection wedges and cuts, it becomes clear that the fraction of AGN excluded in WISE-selected samples is not trivial, by any means.

The bottom line is that, while mid-IR AGN selections, and in particular thorough methods like those of \citet{Mateos2012,Secrest2015}, are very good at selecting clean samples of bright AGN, one must keep in mind that they are biased against lower luminosity sources, particularly at higher redshifts \citep[see also e.g.][]{Rovilos2014}, and thus their conclusions cannot be extrapolated to the entire AGN population. This bias is made much worse by the fact that many of these low luminosity sources are too faint for the W3 and W4 bands. Auxiliary methods, like our radio selection, can be used to partly rectify this bias, but deeper observations are also needed.

%

\section{Eddington rates}\label{Eddington}

In this section we aim to test the Eddington rates for our redefined populations (excluding the non-active galaxies), to assess whether there are any systematic differences with the results of \citet{Mingo2014,Gurkan2014,Best2012}. To do so, we need to first calculate the bolometric luminosities and black hole masses for our sample.

We calculated the bolometric luminosity for our sources from the X-ray 2--10 keV rest-frame corrected fluxes, using the correlations of \citet[][eq. 21 ]{Marconi2004}:
\begin{equation}\label{Lbol}
\log(L/L_{2-10 keV})=1.54+0.24 \mathcal{L} +0.012 \mathcal{L}^{2} - 0.0015 \mathcal{L}^{3}
\end{equation} where $\mathcal{L}=(\log(L)-12)$, and $L$ is the bolometric luminosity in units of $L_{\odot}$. The bolometric luminosities used for the data points from \citet{Mingo2014} and \citet{Punsly2011} were derived from [OIII] measurements, which, unfortunately, we do not have for the entirety of our current sample. However, we do not expect the results to be systematically different, especially considering the uncertainties involved. We calculated the $L_{bol,X}/L_{bol,[OIII]}$ ratio for the 2Jy, 3CRR and Punsly sources, and found it to be 0.98, with a standard deviation of 0.11.

Given that only about half of our sources have spectroscopic information, and because we want to use a consistent method for the entire sample, we cannot use optical line widths to calculate the black hole masses. We thus derived B band magnitudes using the equations of \citet{Jester2005} and the redshift-corrected fluxes for the g and r bands of SDSS, and calculated the black hole masses using the relations from \citet{Graham2007}. This method for deriving the black hole mass, originally derived from the work of \citet{Kormendy1995,Magorrian1998} is not accurate when the source is a QSO, as the emission from the QSO completely dominates over that of the host. We eliminated the potential QSOs by applying a (conservative) luminosity cut at $L_{2-10 keV}=5\times10^{44}$ erg s$^{-1}$, based on where we found the break between broad-line radio galaxies and QSOs in our previous work \citep{Mingo2014,Hardcastle2009}, and excluding sources above this luminosity. After this cut we were left with 347 RL AGN and 215 RQ AGN (and 372 W3 rejected RL AGN). When we compare these numbers with those in Table \ref{nSources_RL_RQ}, we clearly see that the overwhelming majority of the QSO-like sources in our sample classify as radio-loud (see also section \ref{Power}).

Although some recent results \citep{Kormendy2013b,Kormendy2013} cast doubt on both the $M_{BH}$/L relation and the M-$\sigma$ relation of \citet{Ferrarese2000,Gebhardt2000}, they are still the best indirect methods we have to estimate the black hole masses of AGN.

\begin{figure}
\centering
\includegraphics[width=0.46\textwidth]{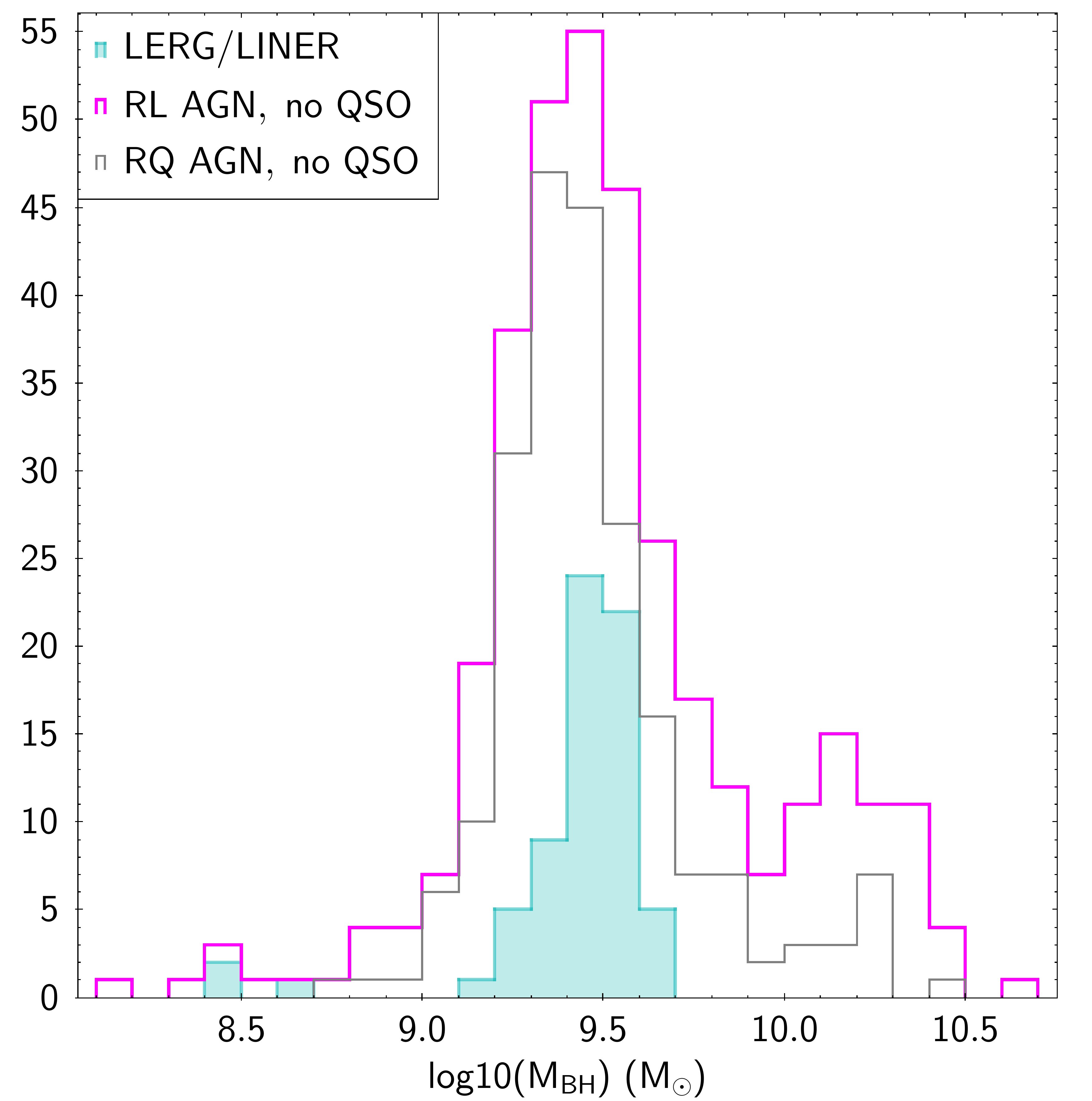}
\caption{Black hole masses for the source subsets defined from Fig. \ref{RL_RQ}, excluding QSOs. Only the sources that pass the W3 S/N cut are considered.}\label{BH_masses_RL_RQ_noQSO}
\end{figure}

The distribution of inferred black hole masses is rather similar for the RQ and RL AGN in our sample, as Fig. \ref{BH_masses_RL_RQ_noQSO} illustrates. Although the former seem to peak at slightly lower masses and show a narrower distribution than the latter, the difference is very subtle, and only evident when we consider that the LERG/LINER sources, which are a subset of the RL AGN, take up a large fraction of the lower-mass end of the distribution. This is quite consistent with the fact that RL AGN tend to inhabit hosts with larger $M_{BH}$, but the black hole mass distributions of RL and RQ sources do not look very different in our sample. 

\begin{figure}
\centering
\includegraphics[width=0.46\textwidth]{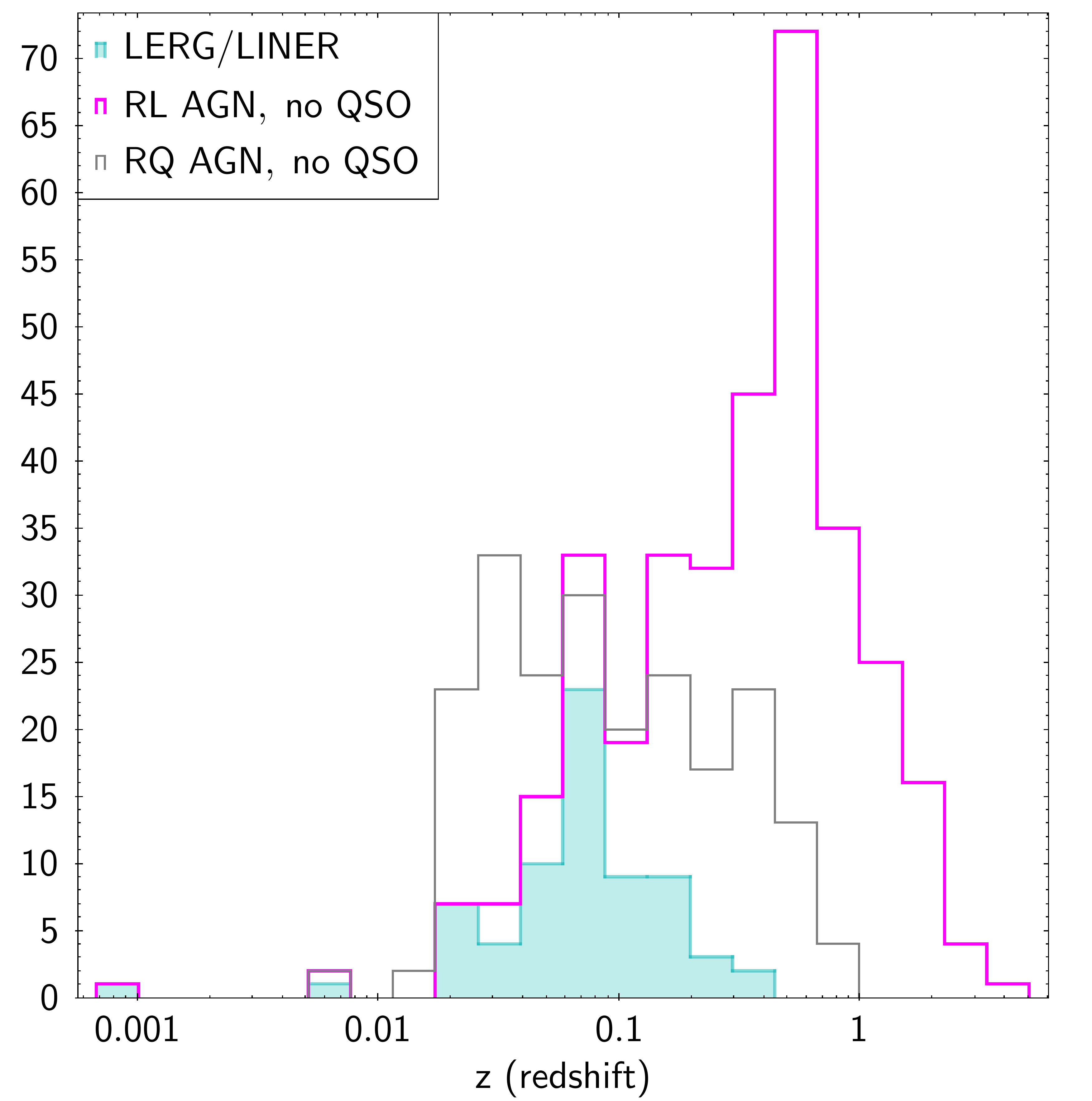}
\caption{Redshift distributions for the source subsets defined from Fig. \ref{RL_RQ}, excluding QSOs. Only the sources that pass the W3 S/N cut are considered.}\label{z_histo_RL_RQ_noQSO}
\end{figure}

Fig. \ref{z_histo_RL_RQ_noQSO}, showing the distribution of sources in $z$, helps us to understand why the difference between the black hole mass distributions for RL and RQ AGN is not more evident. As we saw in Fig \ref{z_histo_RL_RQ}, because of a combination of evolution effects and our radio selection, we are selecting RQ AGN at lower redshifts than the RL AGN. Eliminating the QSOs from our sample mitigates this bias, as luminous QSOs tend to appear at larger $z$, but it does not eliminate it completely. Moreover, if we consider that the LERGs now make up a larger fraction of the RL AGN, the difference between the radiatively efficient RL and RQ source distributions is even more marked.

That difference in redshift distributions, and thus black hole masses and luminosities, has a clear impact on the results we obtain for the Eddington rates, as shown in Fig. \ref{Ledd_RL_RQ_noQSO} (See Appendix \ref{ExtraFigures} for equivalents to Figs. \ref{Ledd_RL_RQ_noQSO} and \ref{Ledd_RL_RQ_noQSO_rej} with that include the jet power). While the LERG/LINER sources have Eddington rates comparable to those we obtained for the 2Jy and 3CRR sources \citep{Mingo2014}, and compatible with the results of e.g. \citet{Best2012}, the HERG (RQ AGN + non-LERG RL AGN), in bulk, seem to have higher Eddington rates than the radio-quiet sources.

We highlight that this is not likely to be an underlying physical difference, but a selection effect. With the catalogues we use to build our sample, we are selecting bright radio-loud sources at high redshift (thus with still growing black holes) and faint radio-quiet sources at low redshift (where the black hole masses are larger). This should serve as a warning when comparing samples of radio-loud and radio-quiet sources: a comparison based purely on one criterion, be it luminosity, redshift, or black hole mass, is unlikely to truly compare similar sources. Several factors need to be taken into account to minimise bias.

\begin{figure}
\centering
\includegraphics[width=0.46\textwidth]{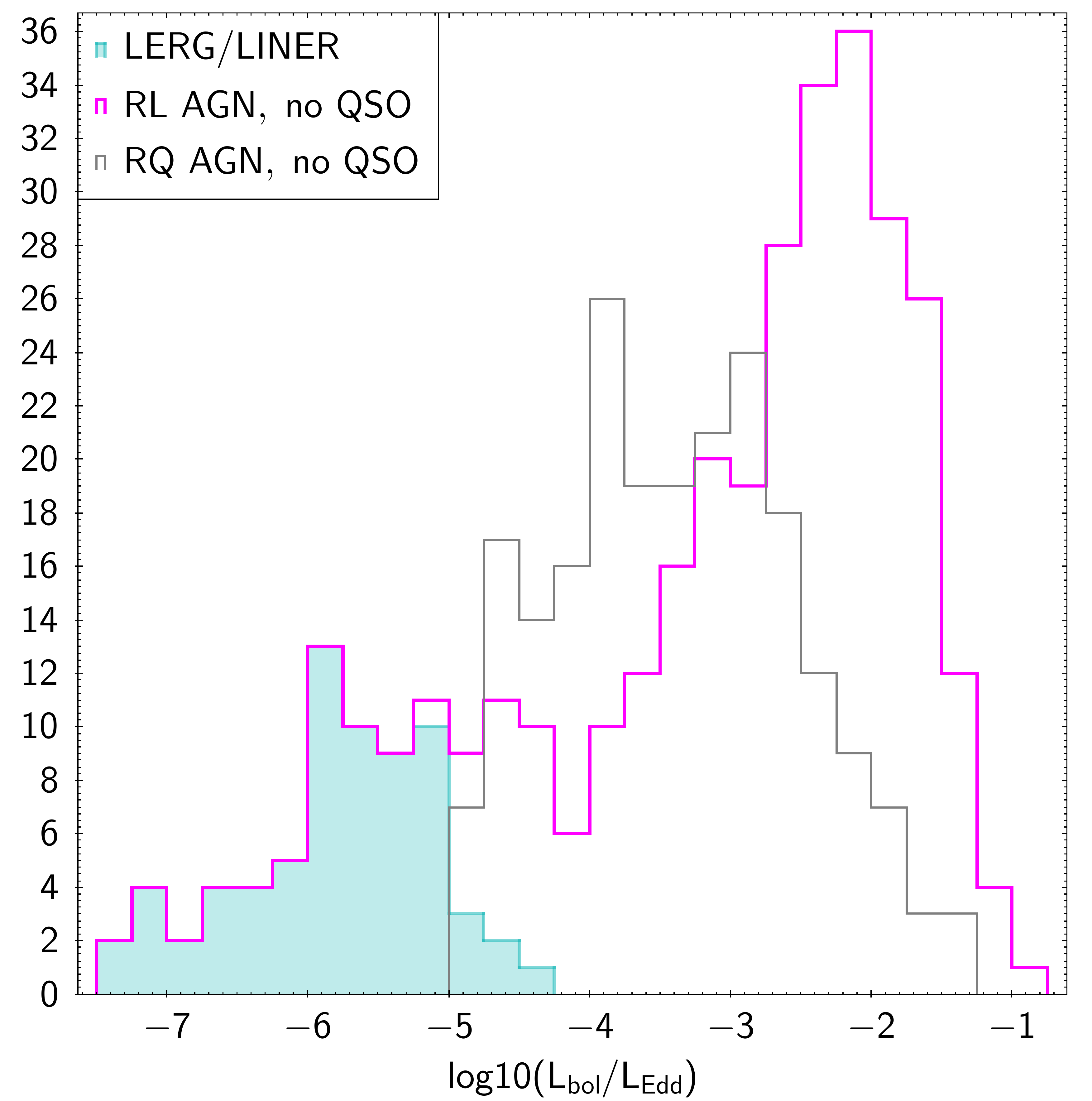}
\caption{Eddington rates for the source subsets defined from Fig. \ref{RL_RQ}, excluding QSOs. Only the sources that pass the W3 S/N cut are considered (see also Appendix \ref{ExtraFigures} for an equivalent Figure including the jet power in the Eddington luminosity).}\label{Ledd_RL_RQ_noQSO}
\end{figure}

Fig. \ref{Ledd_RL_RQ_noQSO_rej} shows that, by eliminating most of the Seyfert-like RL AGN with the W3 cut, we are also essentially eliminating those RL sources that are most similar to the RQ AGN in terms of their Eddington rate (though not entirely in terms of $z$, see again Fig. \ref{z_histo_RL_RQ_rej}). 

\begin{figure}
\centering
\includegraphics[width=0.46\textwidth]{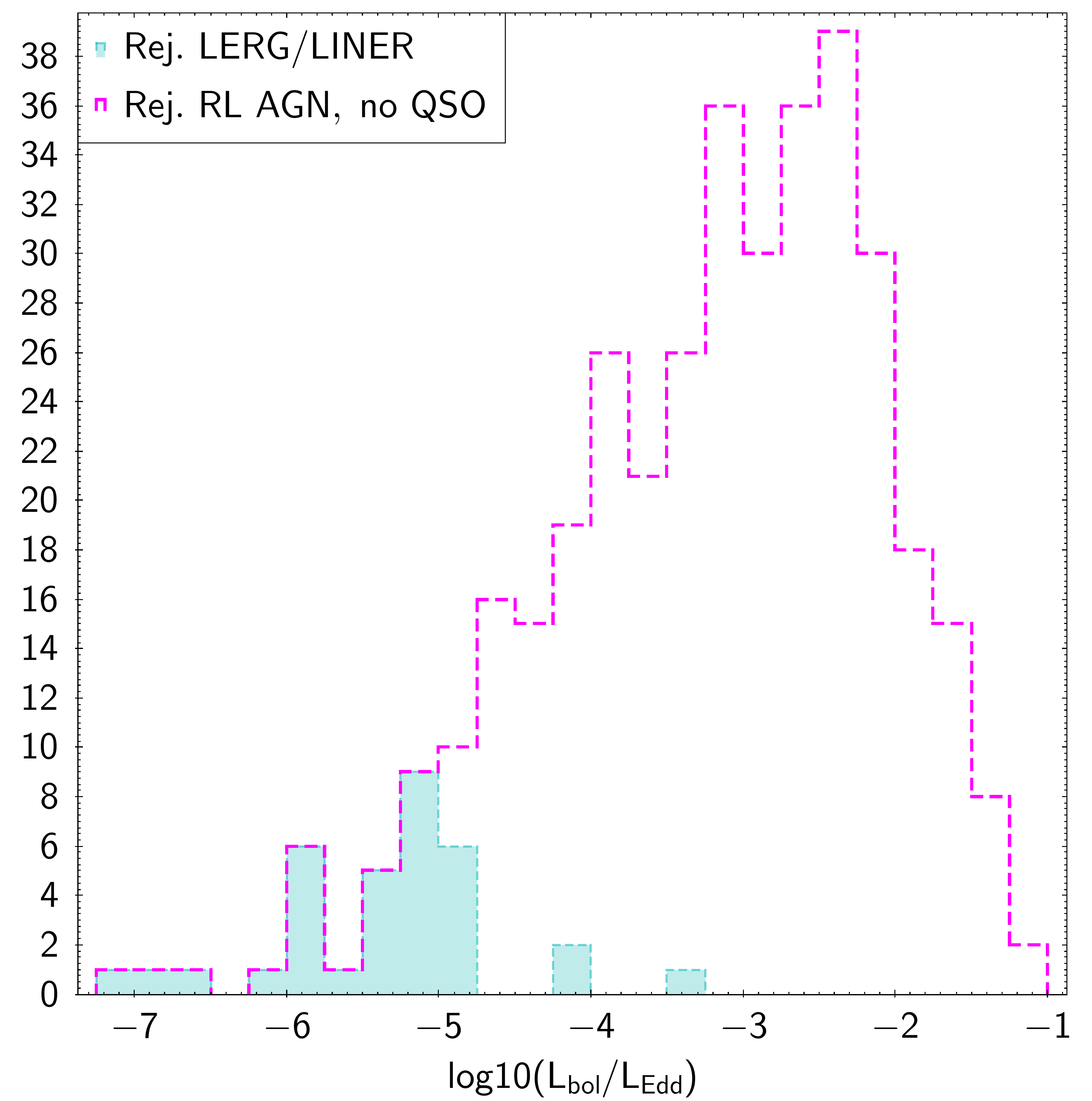}
\caption{Eddington rates for the LERG/LINER and radio-quiet AGN that do not pass the W3 S/N cut. QSOs are excluded (see also Appendix \ref{ExtraFigures} for an equivalent Figure including the jet power in the Eddington luminosity).}\label{Ledd_RL_RQ_noQSO_rej}
\end{figure}

%

\section{Jet vs radiative output}\label{Power}

In our previous work \citep[e.g.][]{Hardcastle2009,Mingo2014} we have observed that the `radio loudness' of a source is not easily determined based just on its radiative power. More than a sharp dichotomy, the radio-loud/quiet transition seems to be gradual, hinting at underlying mechanisms that regulate how accretion power is transformed into radiative output (luminosity) and kinetic output (jets or winds). However, this effect is very difficult to observe with small samples that use monochromatic flux limits or that study just a subset of AGN. MIXR, due to its mixed population, is ideally suited to study the relationship between jet and radiative output from LERGs, through Seyferts, to QSOs, and across a wide range of radio powers. 

The question of how the jet and the radiative output are regulated is a very complex one. While a parallel between X-ray binaries and AGN is frequently drawn \citep[e.g.][]{Connolly2016}, there are some important considerations to take into account. LERG can be likened to the low/hard, radiatively inefficient state in low-mass X-ray binaries \citep{Fender2014,Yuan2014}, while, on a first approach, the parallel of the radiatively efficient transition state between the low/hard and high/soft states in low-mass X-ray binaries (LMXRB) would be the HERG (non-LERG, radiatively efficient RL AGN). In LMXRB only sources in the high/soft state can produce winds, and they never have jets. However, we know that there there are many (radiatively efficient) Seyferts and QSOs with jets, sometimes coexisting with winds \citep[e.g.][]{Nesvadba2008,Mullaney2013,Collet2016,Harrison2015}, and that they show examples of both steady, slow jets and fast, relativistic ones. To complicate matters further, we also now know that the radio luminosity we measure depends on the environment around the host \citep{Hardcastle2013,Hardcastle2014,English2016}. And it is also important to keep in mind that the timescales involved in AGN activity are very long; as such, low frequency radio observations, used to study the jet efficiency, are often reflecting the activity level of the AGN on scales of Myr, while X-ray and even mid-IR observations provide measurements of the core activity on much shorter timescales (essentially instantaneous, in the case of the X-rays).

The fact that we can observe the same type of jets in LERG and HERG means that the accretion disc \citep[in the classical][sense, at least]{Shakura1973} cannot be the element responsible for jet generation. In the model of \citet{Blandford1977}, a jet is expected when both the spin of the black hole and the magnetisation of the surrounding material are high. It is possible that in AGN, because of the larger volumes of gas and more inhomogeneous feeding rates compared to XRBs, it is feasible to accrete enough magnetic flux to launch a jet even in the radio-quiet regime. The role of the spin has been recently brought to light \citep[see e.g.][]{McNamara2011,Done2014,Done2016}, especially for sources with very powerful jets \citep{Tchekhovskoy2011}. It is also possible, however, that the magnetic flux accretion alone is the main driving mechanism, and that the episodic accretion of hot or cold gas is what truly drives the difference between inefficient and efficient sources \citep{Hardcastle2007b,Sikora2013,Ineson2013, Ineson2015}, or it could be a combination of both, at least in some sources \citep[see e.g.][]{Nemmen2014,Nemmen2015}. 

What we observe is a large scatter in the plots of e.g. \citet{Punsly2011,Mingo2014}, where for a given bolometric luminosity (radiative output) there is a wide range of possible jet powers, and vice versa. This contradicts the conclusions of \citet{Rawlings1991}, who established a tight correlation between both quantities. It is possible that their correlation arises as a selection effect, as most of the work carried out on the topic involves flux-limited samples that only select a particular subset of the radio-loud AGN population, in particular, in the case of \citet{Rawlings1991}, the most luminous radio-loud sources in the Universe. The recent Fermi results of \citet{Chen2015}, for example, seem to agree with \citet{Rawlings1991}, but if we plotted them together with our current or previous results, the scatter in the plot would be too large to support a strong correlation. 

\begin{figure*}
\centering
\includegraphics[width=0.92\textwidth]{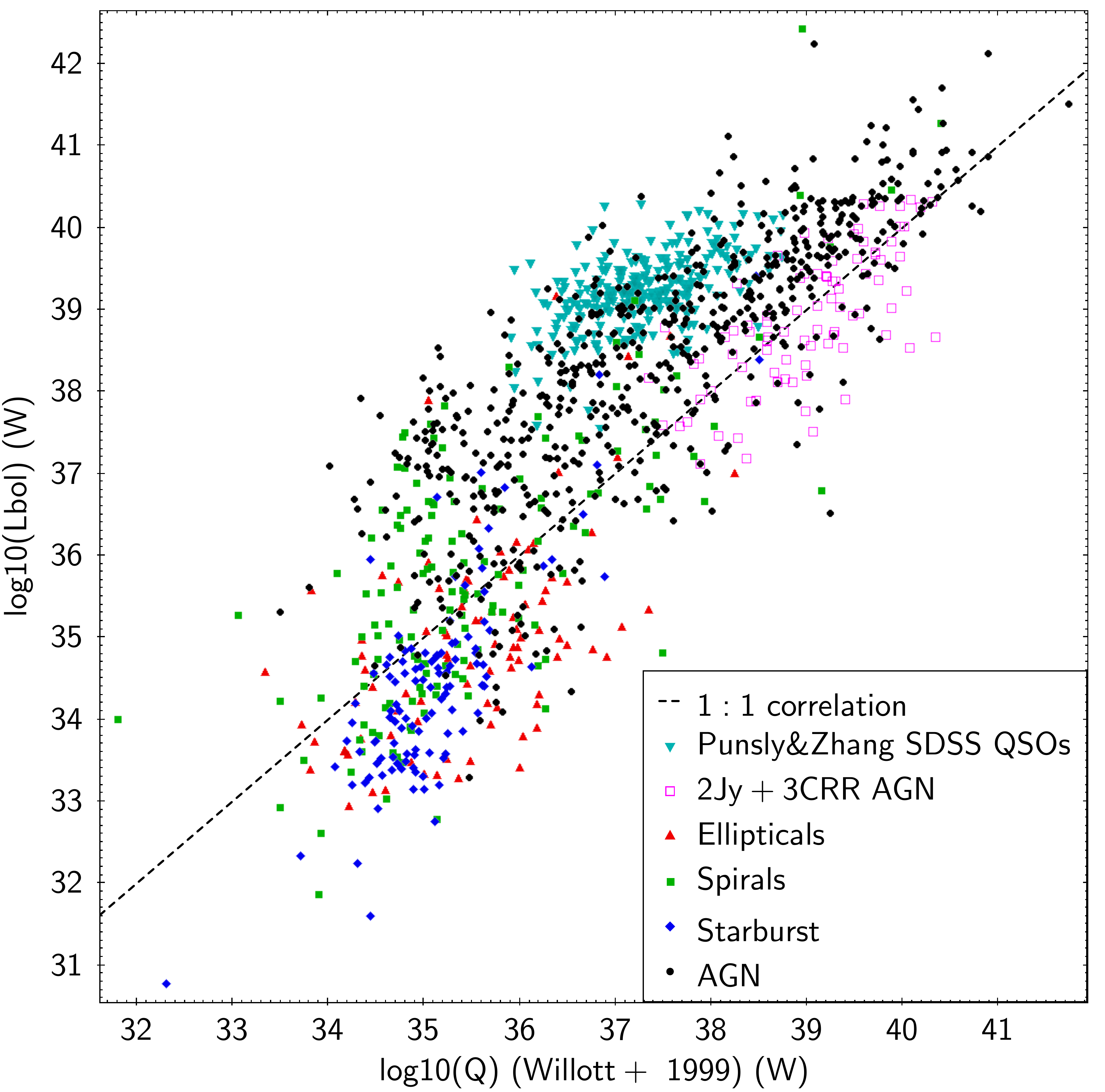}
\caption{Radiative ($L_{bol}$) versus jet ($Q$) output for our sources (colours and symbols as in Fig. \ref{C_C_all}). The magenta squares represent the 2Jy and 3CRR sources from \citet{Mingo2014}, and the cyan inverted triangles represent the SDSS QSOs from \citet{Punsly2011}. Error bars are omitted for clarity, their sizes are comparable to the X-ray luminosity error bars in Figs. \ref{LX_LI} and \ref{LX_LR}.}\label{Punsly}
\end{figure*}

\begin{figure*}
\centering
\includegraphics[width=0.92\textwidth]{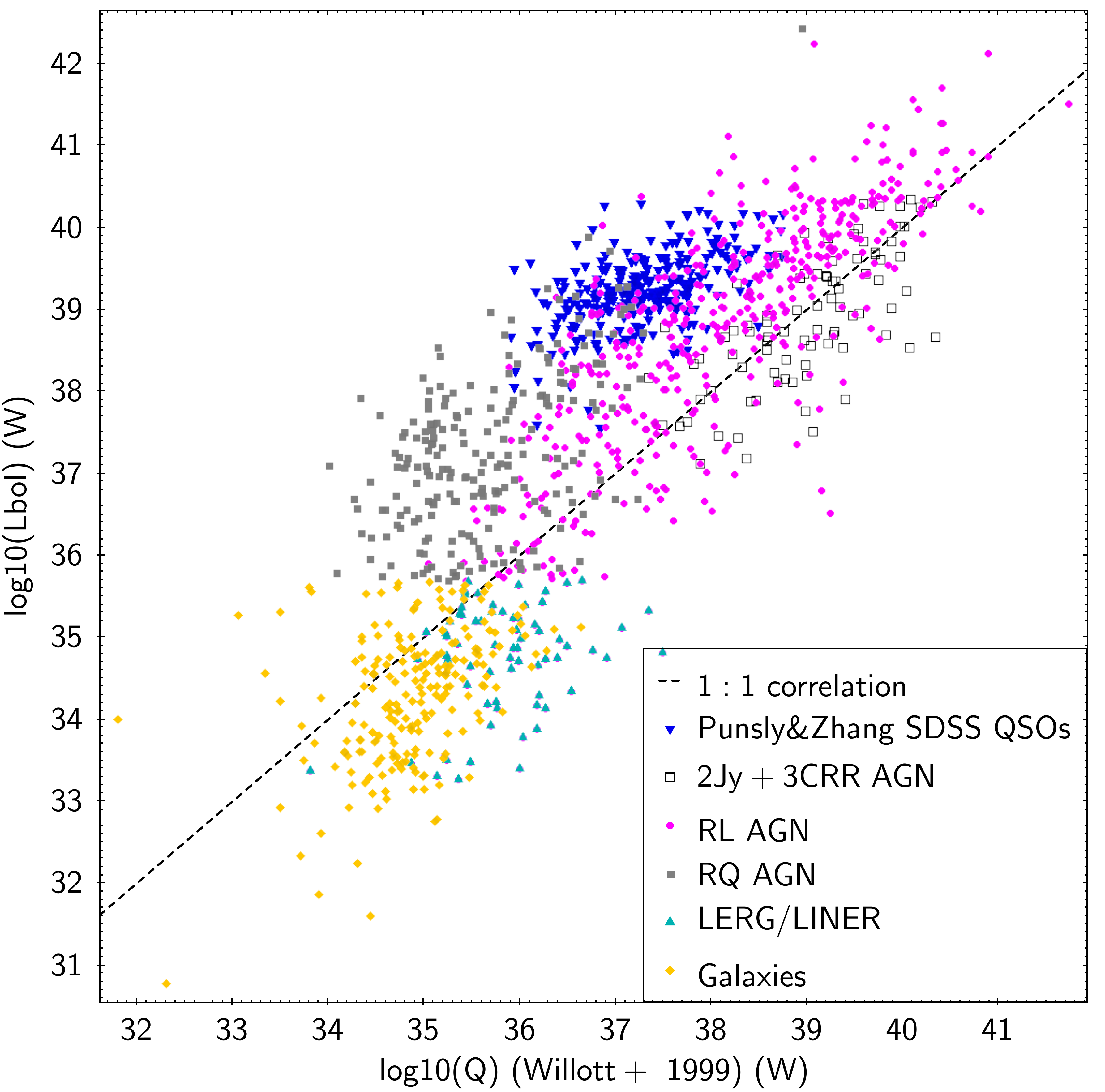}
\caption{Radiative versus jet output for the MIXR sources that pass the W3 cut, as in Fig. \ref{Punsly}, but using the new classifications from Fig. \ref{RL_RQ}. Again, we have plotted the 2Jy and 3CRR sources from \citet{Mingo2014}, and the SDSS QSOs from \citet{Punsly2011}, for reference. Error bars are omitted for clarity, their sizes are comparable to the X-ray luminosity error bars in Figs. \ref{LX_LI} and \ref{LX_LR}.}\label{Punsly2}
\end{figure*}

The MIXR sample is ideal to test our previous conclusions about the correlation of \citet{Rawlings1991}, as it contains a large number of sources, with a large range of radio powers and bolometric luminosities. In Figs. \ref{Punsly} and \ref{Punsly2} we have plotted the previous data points from \citet{Mingo2014}, which include the 2Jy and 3CRR sources as well as the SDSS quasars from \citet{Punsly2011}, and all the sources from our current sample, using both the WISE source classifications and those we derived in section \ref{RL_RQ_section}. Please see again Tables \ref{nSources} and \ref{nSources_RL_RQ} for the statistics of each population. Although it is difficult to see in Figs. \ref{Punsly} and \ref{Punsly2}, due to the large number of points, there is an overlap between the MIXR, 2Jy+3CRR, and \citet{Punsly2011} sources. The MIXR sources also bridge the gap between the other two samples, which have more restrictive selection criteria: powerful radio sources for the 2Jy+3CRR sample, SDSS-classified QSOs with no X-ray selection in the case of the sample from \citet{Punsly2011}.

The calculation of the bolometric luminosity for our sources is described in section \ref{Eddington} (eq. \ref{Lbol}). For the jet output we used again the method by \citet[][eq. 12]{Willott1999} that we applied in \citet{Mingo2014}:
\begin{equation}\label{Willott}
Q=3 \times 10^{38}f^{3/2}L_{151}^{6/7} W
\end{equation} where $L_{151}$ is the luminosity at 151 MHz, in units of $10^{28}$ W Hz$^{-1}$ sr$^{-1}$, and f=15 \citep[see the discussion by][for the origin and possible values of f]{Hardcastle2007b}. We derived the 151 MHz fluxes by extrapolating the rest-frame corrected 1.4 GHz fluxes, using a spectral index of 0.8 for all the sources (see the discussion in section \ref{LDiag}). We calculated, and present here, $Q$ and $L_{bol}$ in Watts, rather than erg s$^{-1}$, for easier comparison with previous work.

\begin{figure}
\centering
\includegraphics[width=0.46\textwidth]{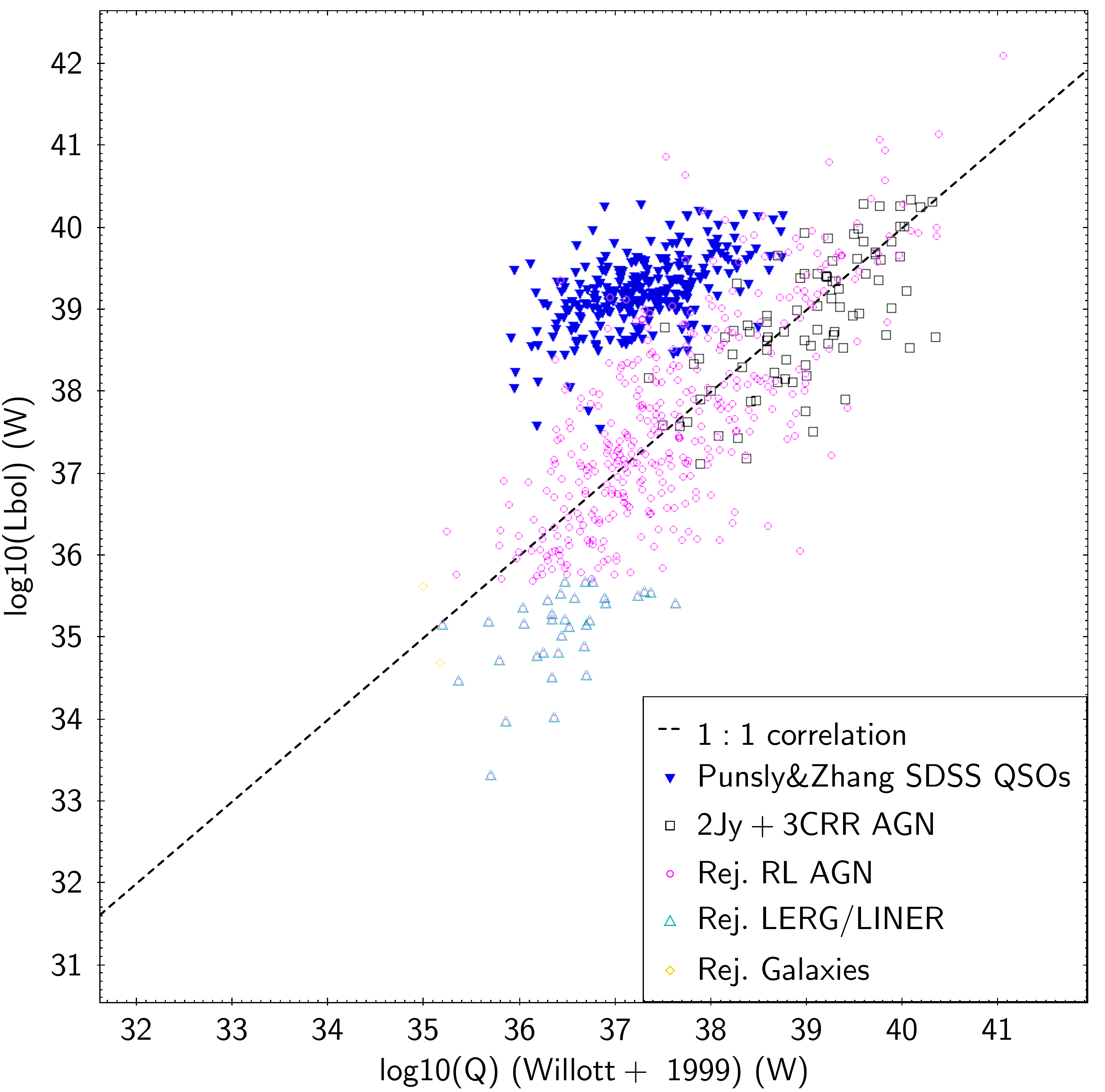}
\caption{Same as Fig. \ref{Punsly2}, for sources that do not pass the W3 cut. No RQ AGN and only 2 Galaxies are rejected by the W3 cut, see Table \ref{nSources_RL_RQ}. Error bars are omitted for clarity, their sizes are comparable to the X-ray luminosity error bars in Figs. \ref{LX_LI} and \ref{LX_LR}.}\label{Punsly2_rej}
\end{figure}

Although star-forming galaxies should not be plotted in Fig. \ref{Punsly}, as their radio emission does not originate from AGN activity, we have included the elliptical, spiral, and starburst sources on the plot, to show where mis-classified sources might lie (as we have seen, using the WISE colour/colour classification some AGN are classified as non-active galaxies, and vice versa). The difference between the loci for the bulk of the starburst and elliptical populations also illustrates how the latter have excess radio emission that cannot be accounted for by even the most powerful star formation, confirming again that our elliptical galaxies host LINER/LERG AGN. Interestingly, a few spiral galaxies also seem to have excess radio emission. Using the new classifications in Fig. \ref{Punsly2}, we see a much clearer separation between the LERG and the non-active galaxies. The overlap between RL and RQ AGN in this plot reflects the degree of uncertainty present in both quantities plotted, but also the fact that the transition between both classes is gradual, rather than abrupt. Fig. \ref{Punsly2} also shows that most of the sources with QSO-like luminosities ($\sim10^{45}$ erg s$^{-1}$, or $\sim10^{38}$ W, and above) are classified as RL AGN.

Figure \ref{Punsly2_rej} is interesting, because it illustrates exactly what types of radio source are eliminated by the W3 cut: RL AGN with Seyfert-like radiative luminosities, but jet outputs that rival those of the SDSS QSOs. These sources are very interesting, as they, presumably, do not have the high accretion rate of luminous QSOs available to produce jets, and yet they can output a similar amount of radio power. It is also now clear that the RQ AGN with similar bolometric luminosities pass the W3 cut because they are found at lower redshifts (see Figs. \ref{z_histo_RL_RQ} and \ref{z_histo_RL_RQ_rej}). 

In light of the large scatter in Figs. \ref{Punsly} and \ref{Punsly2}, and knowing that both quantities have a common dependence with redshift, we carried out a partial correlation test on  $L_{bol}$ and $Q$, including all the sources from \citet{Punsly2011} and \citet{Mingo2014}. We found a strong correlation in the presence of redshift ($\tau/\sigma = 16.94$), although it is partially driven by the brightest sources (removing the QSOs lowers the ratio to $\tau/\sigma = 12.29$), as we observed previously for the 2Jy and 3CRR sources. This is reassuring, as, intuitively, we do expect these quantities to be tied through their common dependence on accretion, but it is important to note that we are still working with flux-limited samples: our plot is not complete, especially on the left side (where the low radio luminosity sources we are not detecting with FIRST/SDSS would likely fall). However strong the results of the partial correlation, it is undeniable that the amount of scatter we observe (4--5 orders of magnitude in each direction) presents a very different picture from that of the tight correlation of \citet{Rawlings1991}. A critical question in radio-loud AGN research remains: what is mediating between the radiative and jet output in AGN, and how can we measure it?

Some care must be taken when considering measurements of the kinetic output of an AGN, even beyond any scatter we may have introduced by using a single spectral index to extrapolate our fluxes (e.g. increasing/decreasing the spectral index to 1.0/0.6 would increase/decrease $Q$ by a factor of $\sim$1.5--2, the effect increasing with $z$). The recent results of \citet{Godfrey2016} show that the observed correlation between radio power and $Q$ on which the work of \citet{Cavagnolo2010} is based may not be as strong as previously believed, due to effects such as the different behaviour of low (FRI) and high (FRII) power radio galaxies \citep[see e.g.][]{Croston2005,Godfrey2013}, the dependence of radio luminosity with environmental density \citep[see e.g.][]{Hardcastle2013,Hardcastle2014,English2016}, and the common distance dependence of $Q$ and the radio luminosity \citep{Godfrey2016}, on which the correlation is based. However, the $Q$ parameter of \citet{Willott1999} is based on an analytical model, and even a different dependence of $Q$ with $L_{radio}$ for FRI and FRII is unlikely to decrease the scatter in Fig. \ref{Punsly} to the extent needed to be consistent with the correlation of \citet{Rawlings1991}, especially if high and low power sources are considered separately.

It is possible that regulating mechanisms for the jet, such as the spin, mentioned above, are introducing scatter between $Q$ and $L_{bol}$ along the X axis of Figs. \ref{Punsly} to \ref{Punsly2_rej}, although the interplay between black hole spin, disc magnetic fields, accretion rate, and jet properties is likely to be fairly complex in and of itself \citep[e.g.][]{Hawley2015}. Most importantly, we must also consider the aforementioned uncertainty in the relationship between $Q$ and the measured radio luminosity (which depends on factors such as the age and the environment of the source). Along the Y axis, it is possible that long-term variability is an important factor; as mentioned earlier, $L_{bol}$ is essentially an instantaneous measurement of the AGN power, while $Q$ (or any $L_{radio}$ at low frequencies) is a time-averaged measurement on timescales of up to a few Myr. 

Fig. \ref{Punsly2} is also useful to understand the role of the individual selections in sampling different regions on the $Q/L_{bol}$ diagram: the QSOs of \citet{Punsly2011} were optically selected; the 3CRR and 2Jy sources of \citet{Mingo2014} were radio selected, and our MIXR sample was X-ray selected (additional constraints driven by the other catalogues we used, as we have seen). The fact that the three samples cover different ranges of $Q$ values (with some overlap) indicates that there are genuine differences in the underlying $Q$ (the SDSS QSOs and the 2Jy+3CRR AGN span a relatively similar range of luminosities), so differences in $Q$ must play a role in the scatter we observe on the $Q/L_{bol}$ diagram. However, the large range of $Q$ values covered by the individual populations, in particular the MIXR RQ AGN, seems to indicate that changes in $L_{bol}$ may also play a role, whose importance we need to assess.

If variability on $\sim$Myr timescales is indeed a factor introducing scatter in Fig. \ref{Punsly2}, it might have very powerful implications for our understanding of AGN feedback and its impact on the star formation history of AGN hosts. Recent works like those of \citet{Hickox2014,Stanley2015} highlight the difficulty of studying the interplay of AGN activity and star formation when the AGN is varying, although they might be correlated on longer timescales \citep[e.g.][]{DelVecchio2014} due to their mutual dependence on reservoirs of cold gas \citep[see also][]{Wild2010}. However, if Fig. \ref{Punsly2} is to be believed, the radiative output of AGN can vary by far more than the two orders of magnitude generally considered in these works, and, more importantly, so can their jet output, which is more likely to influence the host on large scales. 

Although there is increasing evidence for `radiative mode' (or `radio-quiet mode') winds and powerful outflows, and a lively ongoing debate on their impact on AGN hosts \citep[see e.g. the review by][and references therein]{King2015}, it is not clear yet whether these winds can significantly affect star formation beyond the bulge, even in the most powerful sources, and how ubiquitous they really are \citep[see e.g.][]{VillarMartin2016}. We know, however, that small radio sources are very efficient at transporting enormous amounts of energy to the interstellar medium (ISM) through jet-ISM driven shocks \citep[see e.g. the energy calculations for NGC 3801 by][]{Croston2007}, with a much larger potential to disrupt star formation on galaxy scales. But that effect on star formation, either as triggering or quenching, takes several Myr to become observable, by which point the radio lobes have long faded out. Even considering star formation on longer timescales ($\sim100$ Myr), such an injection of energy has the potential to alter the overall energy budget, especially if there are periodic recurrences. Recent results show as well that at $z$ greater than $\sim1$, and unlike in the local Universe, the hosts of moderately-powerful radio-loud AGN are very actively star-forming \citep{Magliocchetti2016}, but are these sources truly the ancestors of local radio-loud AGN? If both the bolometric luminosity and the jet output of an AGN can vary by 3--5 orders of magnitude in the space of a few Myr, and we cannot extrapolate their life cycles from those we know about from LMXRB, and if the star formation rates measured are influenced by consecutive radio outflows \citep[see e.g.][on the recurrence of radio outflows]{Saikia2010} that are no longer detectable, and completely unrelated to the current radiative and jet properties of the AGN, how can we analyse the interplay between AGN activity and star formation?

We clearly need to better understand the life cycles of the radio-loud phase of AGN, and to start taking into account the `radio mode' feedback for small, host-scale sources, as well as the larger cluster-sized ones, in simulations of galaxy dynamics and evolution. The first step could be an assessment of how many sources with radiatively driven winds also have radio emission, as recent evidence seems to point out to a frequent coexistence of both \citep[see e.g.][]{Nesvadba2008,Collet2016,Harrison2015}, but we still need to assess whether in the most luminous sources the radio emission is actually produced by jets and lobes or just by particles accelerated in wind-driven shocks \citep[e.g.][]{Nims2015,Zakamska2016}, and what fraction of the observed outflows is actually produced by star formation, rather than the AGN \citep{Sarzi2016}. We also need to better understand the conditions for a coexistence of small radio outflows and star formation in gas-rich hosts \citep[e.g.][]{Frank2016}, as recent studies highlight how challenging it is both to trace black hole growth in low-luminosity AGN with starforming hosts \citep{Jones2016} and to measure star formation in brighter AGN \citep{Symeonidis2016}, even without considering their radio properties. Although we detect a number of radio-loud sources in star-forming hosts (spiral galaxy colours), we are severely hindered by the lack of sensitivity of W3 and do not have the statistics to study this population individually. A dedicated study of Seyfert-type sources with radio emission would be necessary to assess what the true fraction of radio-loud star-forming galaxies is, and how they differ from the more radio-loud gas-depleted systems, to assess the dynamical impact of the jets and lobes on the host and its star formation properties \citep[see also][]{Kaviraj2015,Kaviraj2015b}.

\begin{figure}
\centering
\includegraphics[width=0.46\textwidth]{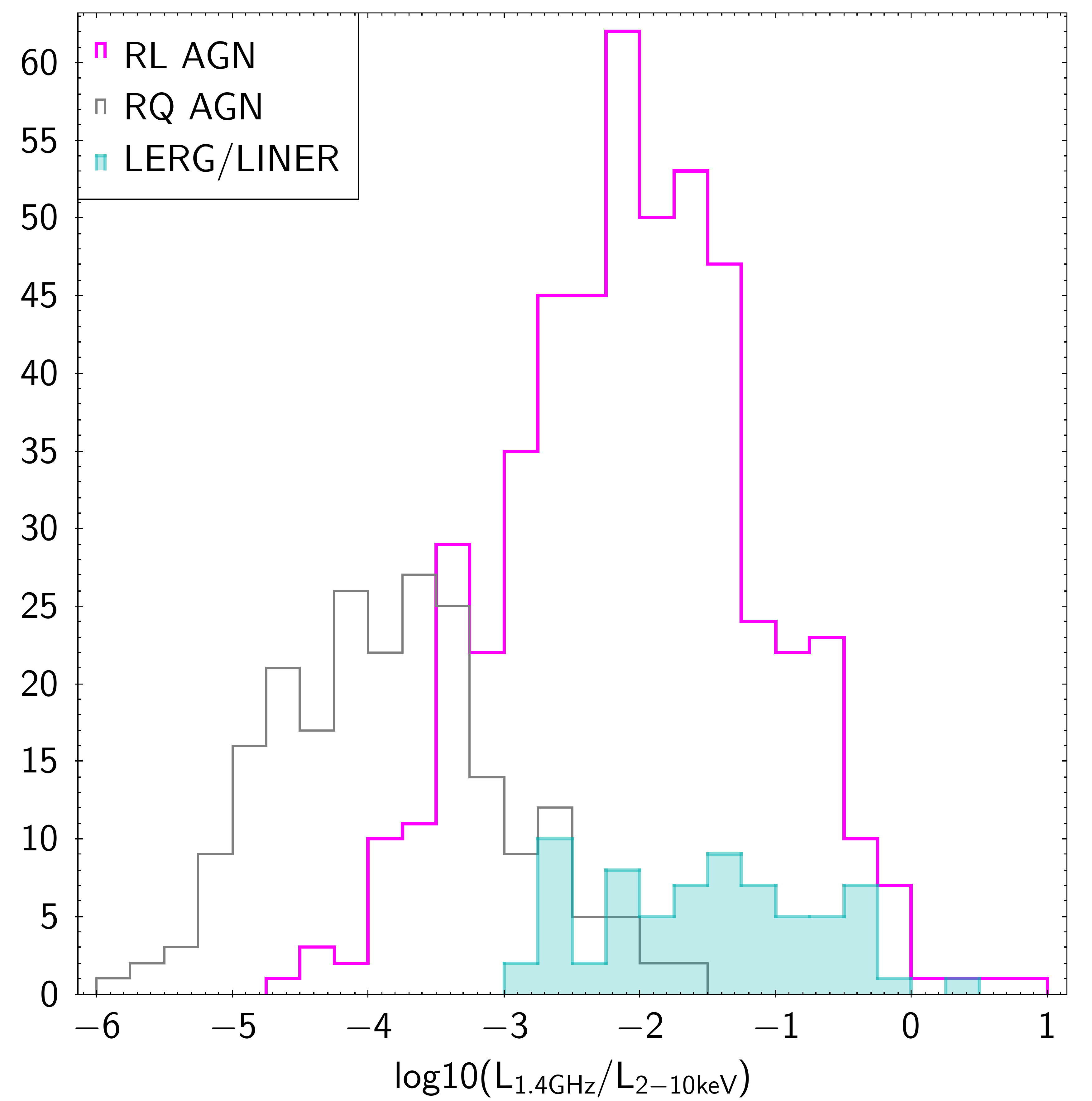}
\caption{Histogram of the L$_{1.4 GHz}$/L$_{2-10 keV}$ ratio for the MIXR RL AGN, RQ AGN, and LERG, showing the overlap between the distributions.}\label{RL_RQ_dichotomy}
\end{figure}

It also seems clear that we need to revisit the radio-loud/radio-quiet dichotomy. While the radio-loud and radio-quiet definitions can work well for cases where one regime (kinetic vs radiative) dominates, the definitions based on flux ratios can be very misleading, and there is increasing evidence that transition between both regimes is not as abrupt as we thought in the past (see Fig. \ref{RL_RQ_dichotomy}). 

To further study this effect, we carried out a few quick tests on the 80 sources between the $3\sigma$ and $5\sigma$ lines in Fig. \ref{RL_RQ}, to assess where they stood in terms of the RL/RQ division. We checked the properties of these sources in terms of hardness ratio, redshift, black hole mass, Eddington rate, $L_{X}/L_{radio}$, and $L_{bol}/Q$. Overall, the 24 sources with $L_{X}<5\times10^{41}$ erg s$^{-1}$ seemed to show characteristics more similar to those of our LERG/LINER sources than those of our galaxies (with a few outliers), in agreement with our choice of the $3\sigma$ line as a division between both populations. For the AGN ($L_{X}\geq 5\times10^{41}$ erg s$^{-1}$) the division is less clear: overall, the 56 sources exhibited characteristics clearly in the transition between the RL and the RQ AGN populations (with a few outliers as well), for all the parameters we tested. Although the number of sources was, unavoidably, too small to carry out proper statistical tests, this `intermediate' behaviour is consistent with the overlap between both populations that we observe in Fig. \ref{RL_RQ_dichotomy}.

It is quite clear that the RL and RQ regimes are very complicatedly interwoven, and we need to better understand their relationship. With the wealth of radio data that will be made available with ongoing and future surveys, it might finally be possible to revisit the RL/RQ classifications and define better criteria to assess the interplay of the radiative and kinetic output in AGN, and its effect on AGN hosts.

%

\section{Summary and Conclusions}\label{Conclusions}


We have used the ARCHES xmatch statistical tool to create a large cross-correlated sample of AGN and star-forming galaxies, using the largest, most uniform catalogues available in the Mid-IR (WISE), X-rays (3XMM) and Radio (FIRST+NVSS) bands. The MIXR sample we thus obtain provides efficient and broad-reaching diagnostic tools to classify sources based on their type of activity (radio-loud and radio-quiet AGN, and star formation), even in the absence of redshifts. The techniques we have developed for MIXR can be used to triage sources for any extragalactic sample with measurements that can be translated to these bands, paving the way for classification techniques that will allow us to fully exploit the vast amounts of data that the next generation of instruments will make available.

We pre-classify our sources based on their mid-IR colours, using the WISE colour/colour plot and the results of \citet{Lake2012}, as elliptical, spiral, starburst and AGN sources. While these initial classifications provide a general idea of the type of underlying activity we can expect in our sources, there is a great deal of overlap between populations (see Table \ref{Activity}). We use first flux and magnitude plots, and then luminosity plots, to triage our sources based on their emission in each band, clearly separating star-forming, non-active galaxies \citep[for which we recover the radio/IR correlation of][]{DeJong1985,Appleton2004,Garrett2015} from radio-loud AGN, both of the radiatively efficient and inefficient \citep[LERG/LINER, see][]{Narayan1995} varieties, and from radio-quiet AGN, where the bulk of the radio emission we detect is produced by star formation, or particle acceleration in shocks \citep{Zakamska2016,Nims2015}, but which could also host minor jets and lobes. 

Our results show that WISE-colour selected AGN samples are heavily biased against Seyfert-type, moderate- to low-luminosity AGN. This selection bias occurs in two ways: WISE colour cuts such as those of \citet{Assef2010,Mateos2012,Stern2012} are very efficient at selecting clean samples of luminous AGN, but necessarily omit those sources where the host contributes a substantial fraction of the total emission, as only with additional proxies, such as radio emission, it is possible to distinguish between non-active galaxies and AGN at low luminosities; the WISE W3 (and W4) band is also not sensitive enough to detect faint AGN at redshifts beyond $\sim0.1$, which is particularly detrimental to radio-loud Seyfert-like sources, as these tend to appear at higher redshifts than radio-quiet sources of similar bolometric luminosity. In fact, our sample size is cut by $\sim40$ per cent simply by imposing requirement for a signal/noise of 3 in W3, with radio-loud AGN suffering the bulk of the cut (we lose another $40$ per cent of the sample when requiring SDSS redshifts for the second part of our diagnostics). 

We find that RL and RQ AGN of similar bolometric luminosities and Eddington rates are found at different redshifts, with the RL sources being found at slightly larger $z$, and our sources become `radio-louder' with increasing redshift, up to our detection limit. Our sample is biased against RQ AGN, as we require a radio detection, limiting the redshift more quickly for RQ sources than for RL ones. As a consequence, when considering both populations as a whole, our RL AGN are more luminous and have larger Eddington rates. This is clearly very likely to be a selection effect, and it illustrates one of the easiest causes of bias that can be incurred when comparing RL and RQ AGN: it is not enough to match both samples exclusively on luminosity, redshift, or Eddington rate; all these variables (plus their environments) must be taken into account to ensure that we are comparing like with like.

Perhaps the most crucial result of this work is the confirmation of the scatter we observed in the 2Jy and 3CRR sources \citep{Mingo2014} between their radiative (bolometric luminosity) and kinetic (jet) output, in contradiction with the tightness of the long-standing correlation of \citet{Rawlings1991}. These two quantities must have a common underlying mechanism, as they are both tied to accretion, but either jet regulating mechanisms \citep[e.g.][]{Done2014,Tchekhovskoy2011,Hawley2015}, dispersion in the jet power/L$_{radio}$ relationship, long-term AGN variability \citep[e.g.][]{Hickox2014,Stanley2015}, or a combination of all three, are introducing the 4--5 order of magnitude scatter we observe in our plots.

Given what we know about the potential impact of small-scale radio sources on the energetics of their hosts \citep[e.g.][]{Croston2007}, and the recently found coexistence of radiative winds and radio outflows \citep{Nesvadba2008,Harrison2015}, which has no parallel in X-ray binaries, we may need to reassess what we know about the interplay between AGN activity and star formation. If both the bolometric luminosity and the jet output of an AGN can vary by 3--5 orders of magnitude in the space of a few Myr, and we cannot extrapolate their life cycles from those we know about from LMXRB, how can we analyse the interplay between AGN activity and star formation? Although star formation occurs on longer timescales, jet-driven shocks can carry enough energy, far enough into the ISM, to potentially change the course of an on-going episode of star formation. However, we may not be able to detect whether the star formation rates we measure are influenced by consecutive radio outflows that have long faded out, and are completely unrelated to the current radiative and jet properties of the AGN because of the short timescale and wide range of AGN variability.

Radio-loud Seyferts may hold the key both to understanding the details of the jet-ISM interaction, and the mechanisms regulating the jet. Some of these sources can produce jet outputs similar to those of luminous QSOs, but at values of $L_{bol}$ and $L_{Edd}$ that are orders of magnitude lower. Unfortunately, these are exactly the sources that W3 is not sensitive enough to reliably detect, as they have similar mid-IR luminosities to those of radio-quiet Seyferts at $z\sim$0.1--0.3, but are far more distant ($z\sim$0.8--1). 

We clearly need to better understand the life cycles of the radio-loud phase of AGN, both from a theoretical and from an observational point of view. A sensible first step might be to assess the fraction and properties of sources with radio emission in samples that do not use radio selections, supplemented with dedicated studies of moderate- to low-luminosity AGN, to establish larger samples that we can systematically study from a broad perspective that includes the hosts. We could also focus on samples such as those of e.g. \citet{Lonsdale2015}, for which the radio emission is compact, and the AGN and star formation are acting on similar spatial and time scales. We have also seen that it might be time to revisit and redefine the radio-loud/quiet classifications, as we have shown that the distribution of sources in terms of L$_{1.4 GHz}$/L$_{2-10 keV}$ displays a gradual transition between both regimes, rather than a dichotomy, showing that in many AGN there is a coexistence of two complicatedly interwoven regimes (kinetic and radiative), both with the potential to influence the host galaxy in different ways. 

%

\subsection*{Acknowledgements}

We thank the anonymous referee for insightful and constructive comments that greatly improved the paper. We also thank the TOPCAT developer Mark Taylor for programming and releasing a software patch so that we could improve our Figures as the referee requested, and Dr B. Punsly, for allowing us to plot the data of \citet{Punsly2011} again in Figs. \ref{Punsly} to \ref{Punsly2_rej}. This work has made use of data/facilities and financial support from the ARCHES project (7th Framework of the European Union n$^{\circ}$ 313146). SM, FJC and AR acknowledge financial support by the Spanish Ministry of Economy and Competitiveness through grant AYA2012-31447, which is partly funded by the FEDER programme. FJC also acknowledges financial support through grant AYA2015-64346-C2-1-P (MINECO/FEDER). This work is based on observations obtained with \textit{XMM-Newton}, an ESA science mission with instruments and contributions directly funded by ESA Member States and NASA. It also makes use of data products from the Wide-field Infrared Survey Explorer, which is a joint project of the University of California, Los Angeles, and the Jet Propulsion Laboratory/California Institute of Technology, funded by the National Aeronautics and Space Administration. We acknowledge the use of the FIRST and NVSS catalogues, provided by the NRAO. Optical magnitudes and redshifts were obtained from the Sloan Digital Sky Survey Data Release 12. Funding for the Sloan Digital Sky Survey IV has been provided by the Alfred P. Sloan Foundation and the Participating Institutions. SDSS-IV acknowledges support and resources from the Center for High-Performance Computing at the University of Utah. The SDSS web site is \url{www.sdss.org}.

\bibliographystyle{mnras}
\bibliography{BMingo_MIXR}

\appendix

\section{Extra Figures}\label{ExtraFigures}

We have included here all the plots that are not fundamental to the main core of the paper, but would still be useful to the readers, as they provide additional information.

\begin{figure*}
\centering
\includegraphics[width=0.85\textwidth]{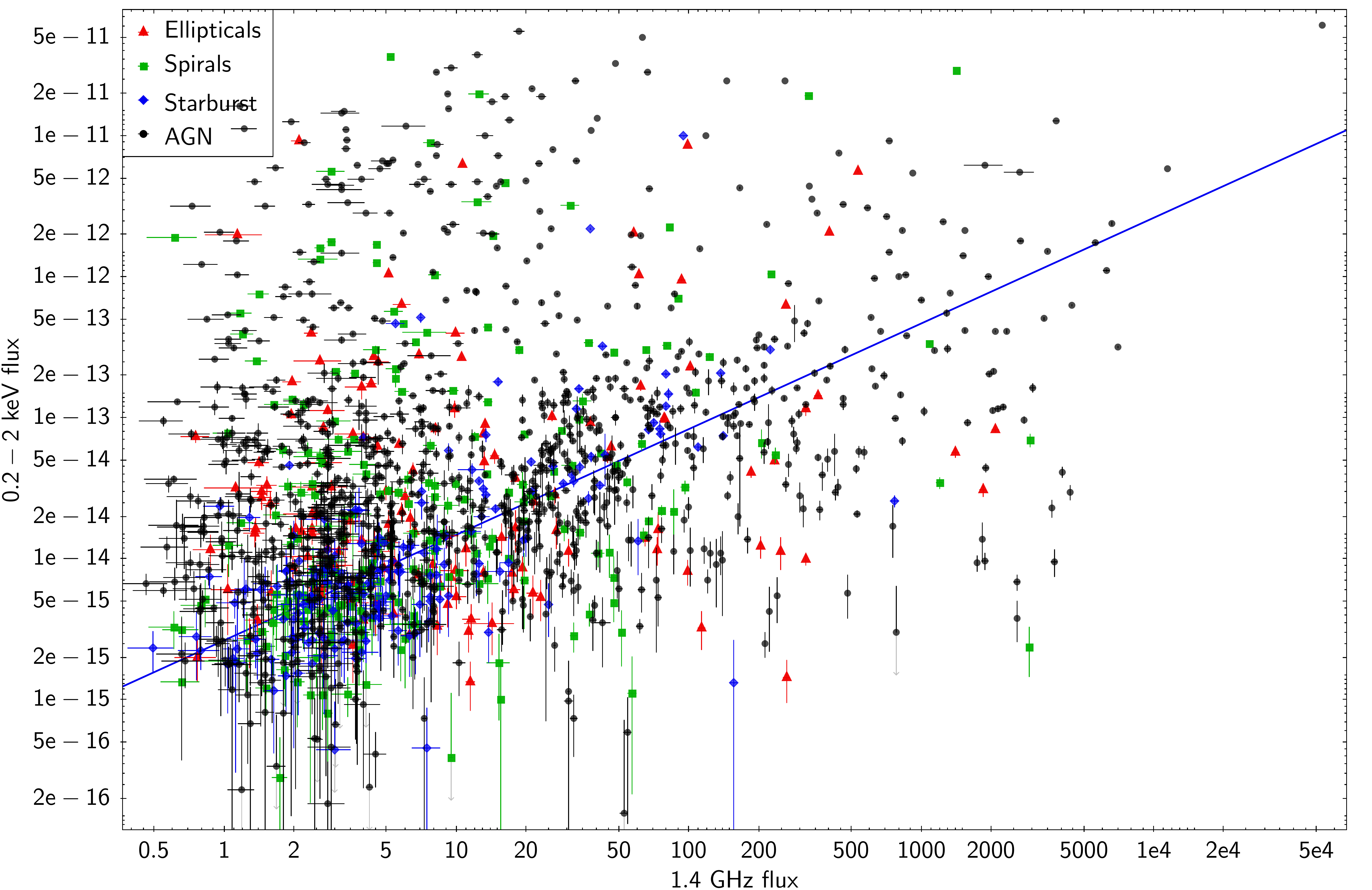}
\caption{Soft (0.2--2 keV) X-ray flux (erg cm$^{-2}$ s$^{-1}$) versus 1.4 GHz radio flux (mJy). The blue line represents the best linear correlation for the starburst sources ($r\sim 0.65$). The colours and symbols follow the same scheme as those in Fig. \ref{C_C_all}. Upper limits (only for the X-rays) are represented with grey arrows.}\label{combFlux_softX}
\end{figure*}

\begin{figure*}
\centering
\includegraphics[width=0.95\textwidth]{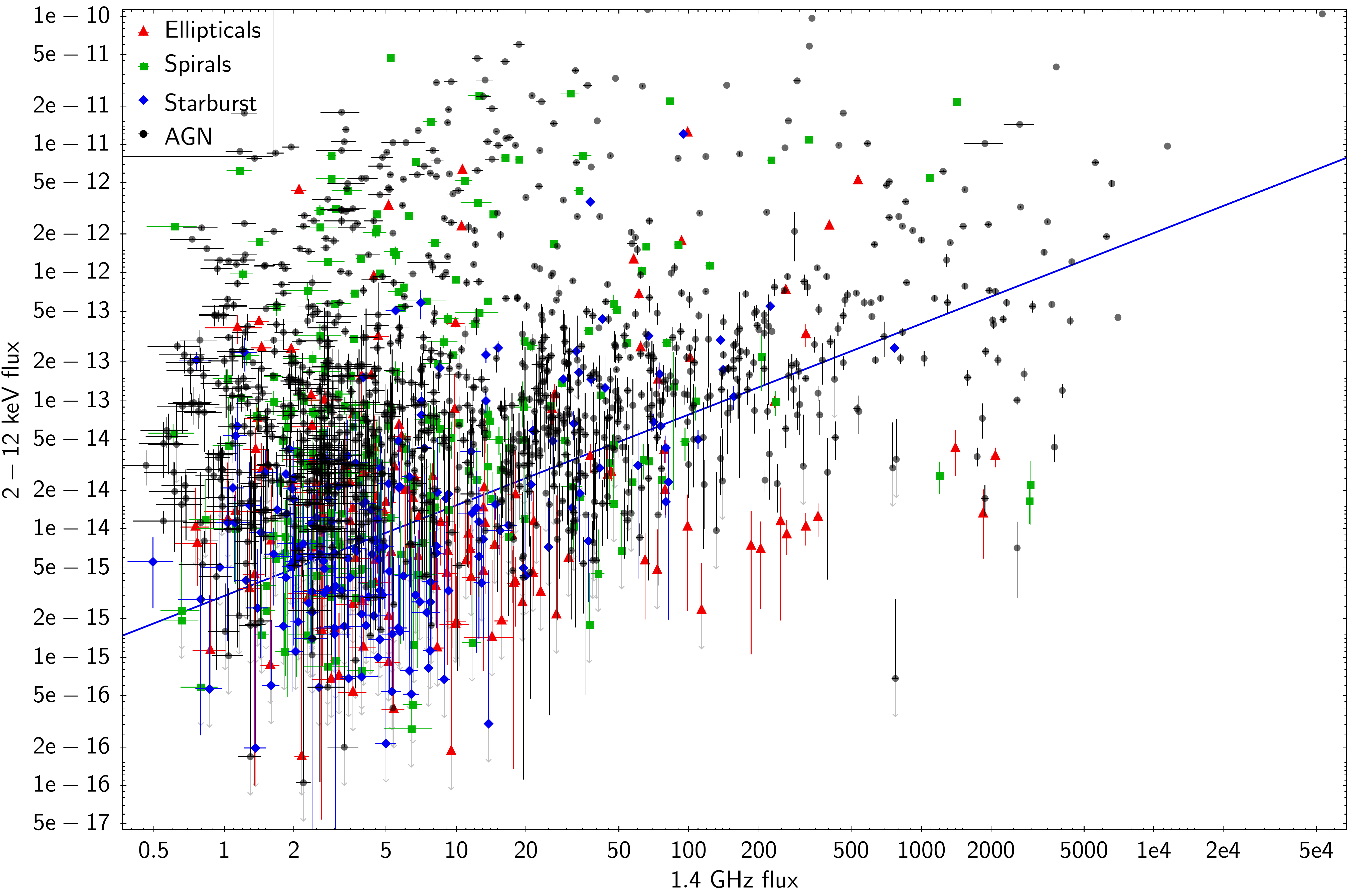}
\caption{Hard X-ray (2--12 keV) flux (erg cm$^{-2}$ s$^{-1}$) versus 1.4 GHz radio flux (mJy). The blue line represents the best attempt at a linear correlation for the starburst sources ($r\sim 0.50$). The colours and symbols follow the same scheme as those in Fig. \ref{C_C_all}. Upper limits (only for the X-rays) are represented with grey arrows.}\label{combFlux_hardX}
\end{figure*}

\begin{figure*}
\centering
\includegraphics[width=0.85\textwidth]{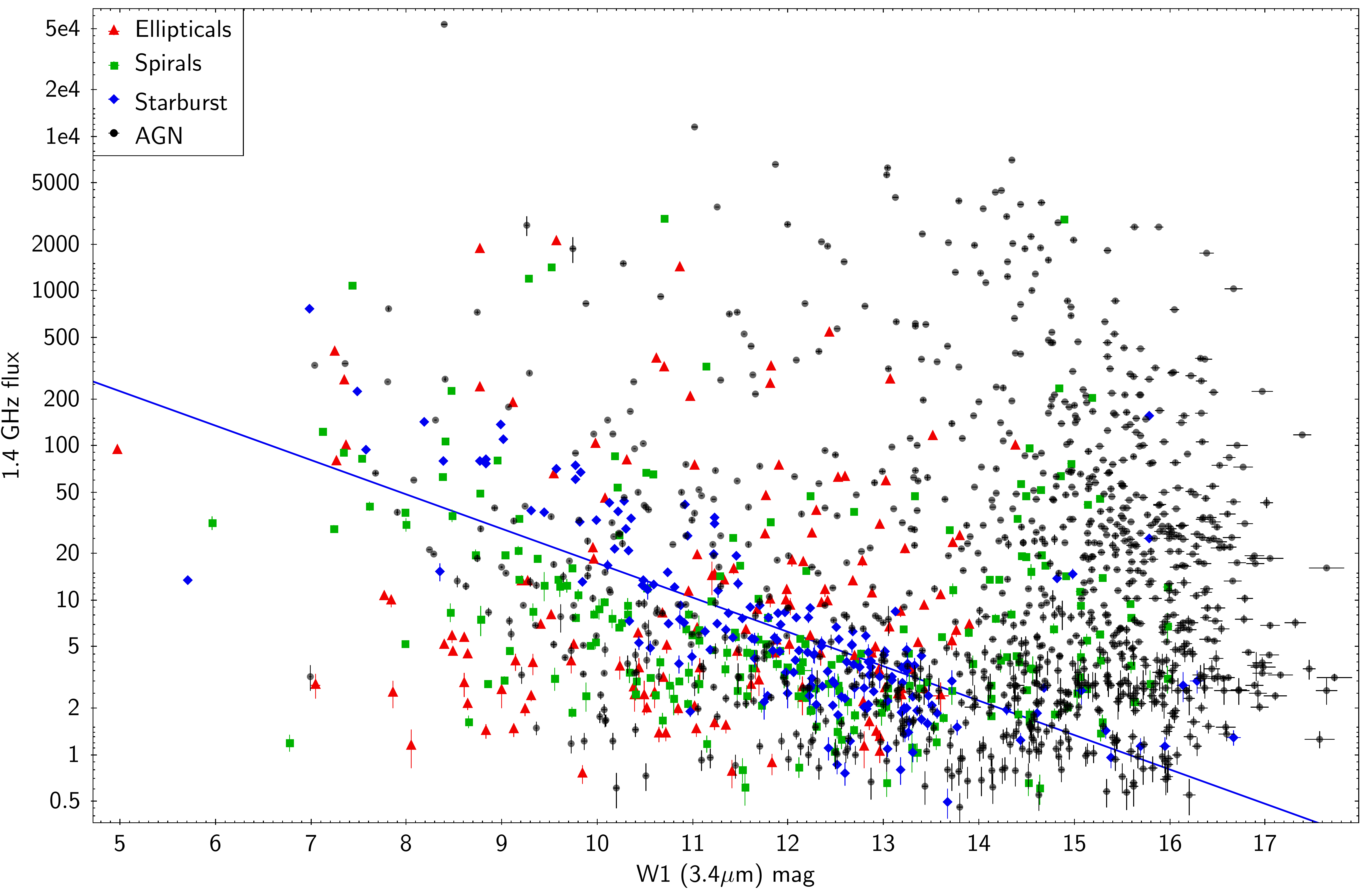}
\caption{1.4 GHz radio flux (mJy) versus W1 (3.4 $\mu$m) magnitude. The blue line represents the best linear correlation for the starburst sources ($r\sim 0.70$). The colours and symbols follow the same scheme as those in Fig. \ref{C_C_all}.}\label{w1_combflux}
\end{figure*}

\begin{figure*}
\centering
\includegraphics[width=0.85\textwidth]{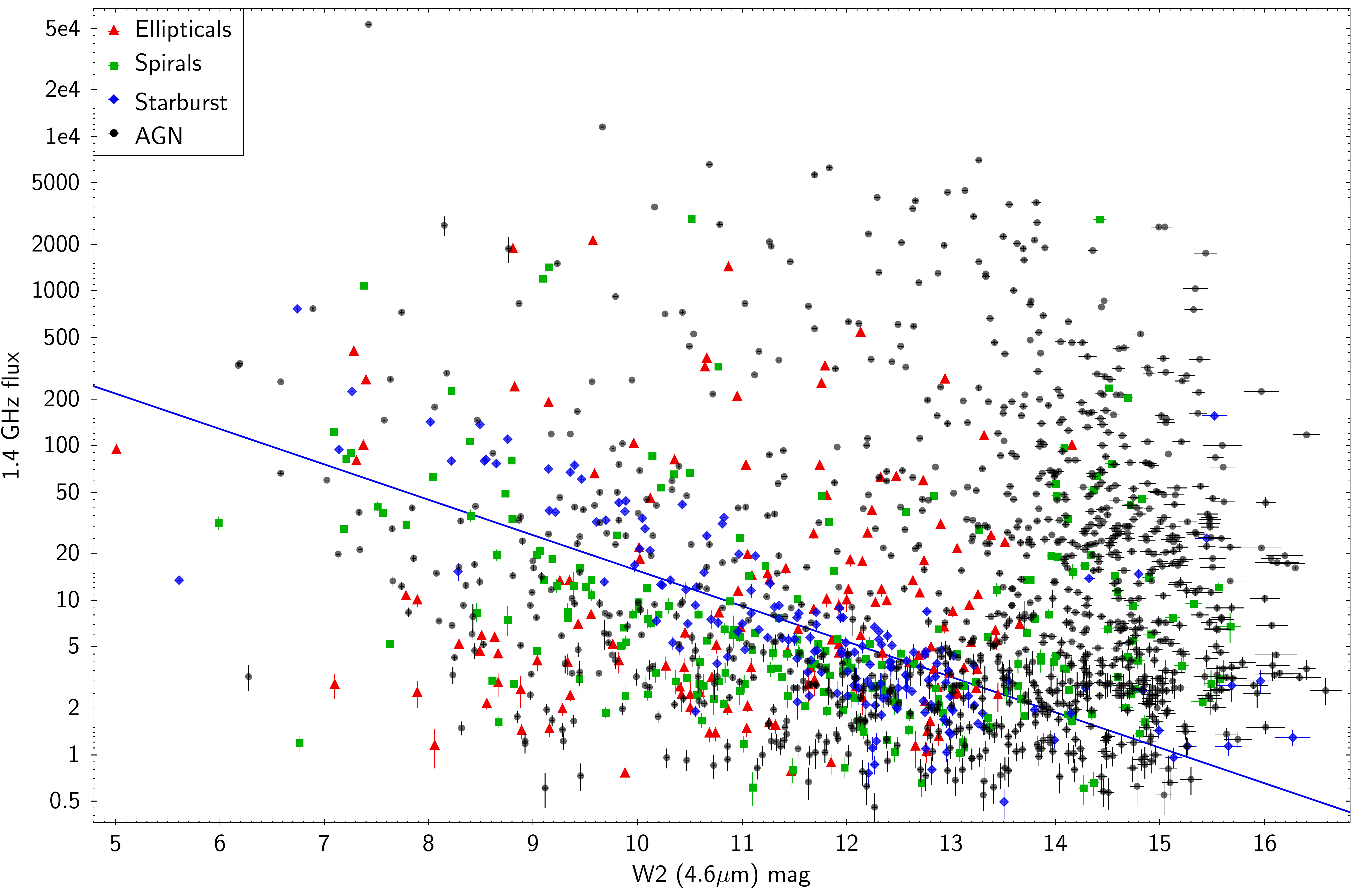}
\caption{1.4 GHz radio flux (mJy) versus W2 (4.6 $\mu$m) magnitude. The blue line represents the best linear correlation for the starburst sources ($r\sim 0.70$). The colours and symbols follow the same scheme as those in Fig. \ref{C_C_all}.}\label{w2_combflux}
\end{figure*}

\begin{figure*}
\centering
\includegraphics[width=0.85\textwidth]{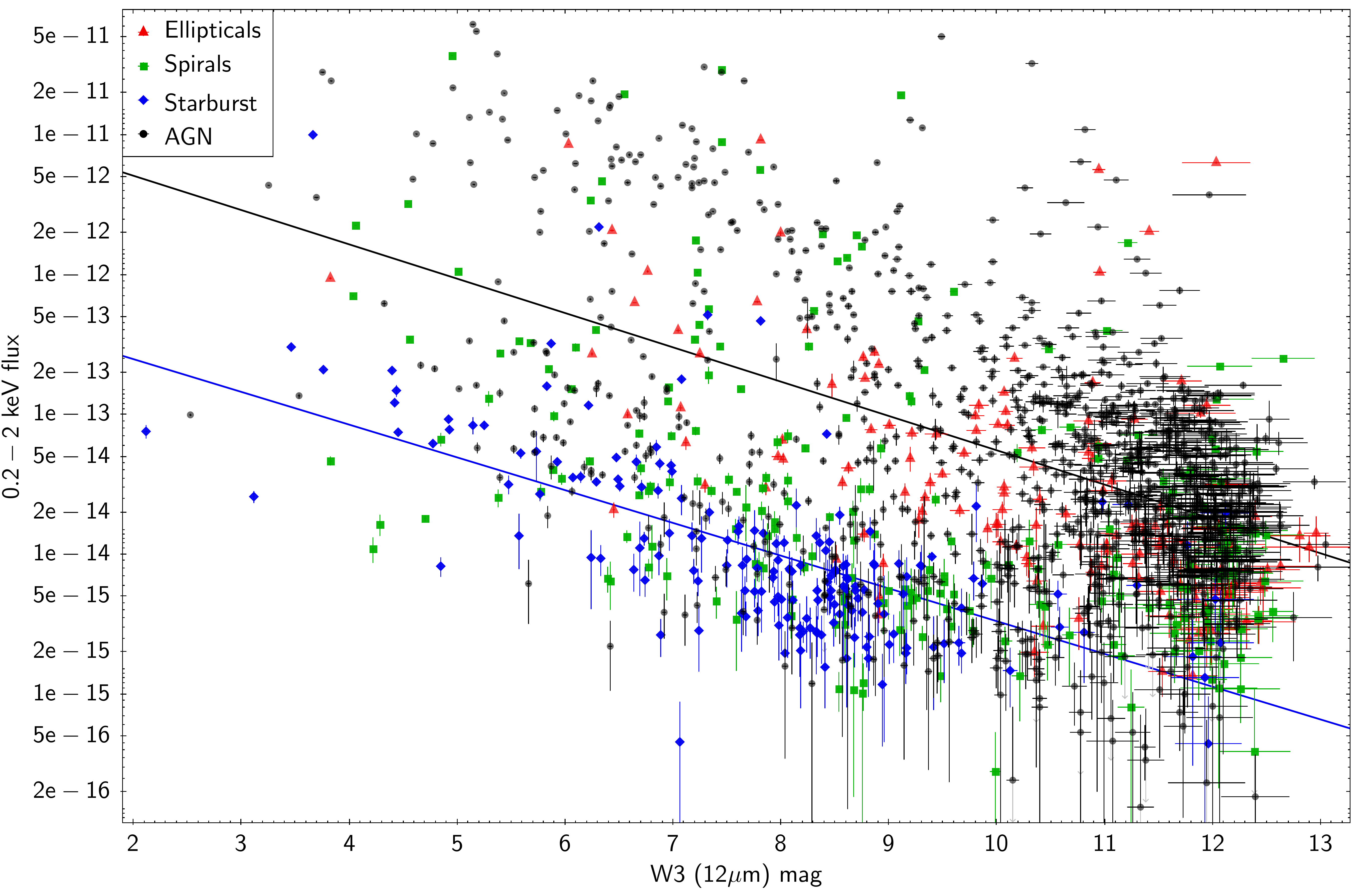}
\caption{Soft (0.2--2 keV) X-ray flux (erg cm$^{-2}$ s$^{-1}$) versus W3 (12 $\mu$m) magnitude. The blue line represents the best linear correlation for the starburst sources ($r\sim 0.66$). The colours and symbols follow the same scheme as those in Fig. \ref{C_C_all}. Upper limits (only for the X-rays) are represented with grey arrows.}\label{w3_softX}
\end{figure*}

Figs. \ref{combFlux_softX} to \ref{w3_softX} supplement the flux diagnostics highlighted in section \ref{FCorr}, adding the soft X-ray emission, and the W1 and W2 WISE bands.

\begin{figure}
\centering
\includegraphics[width=0.46\textwidth]{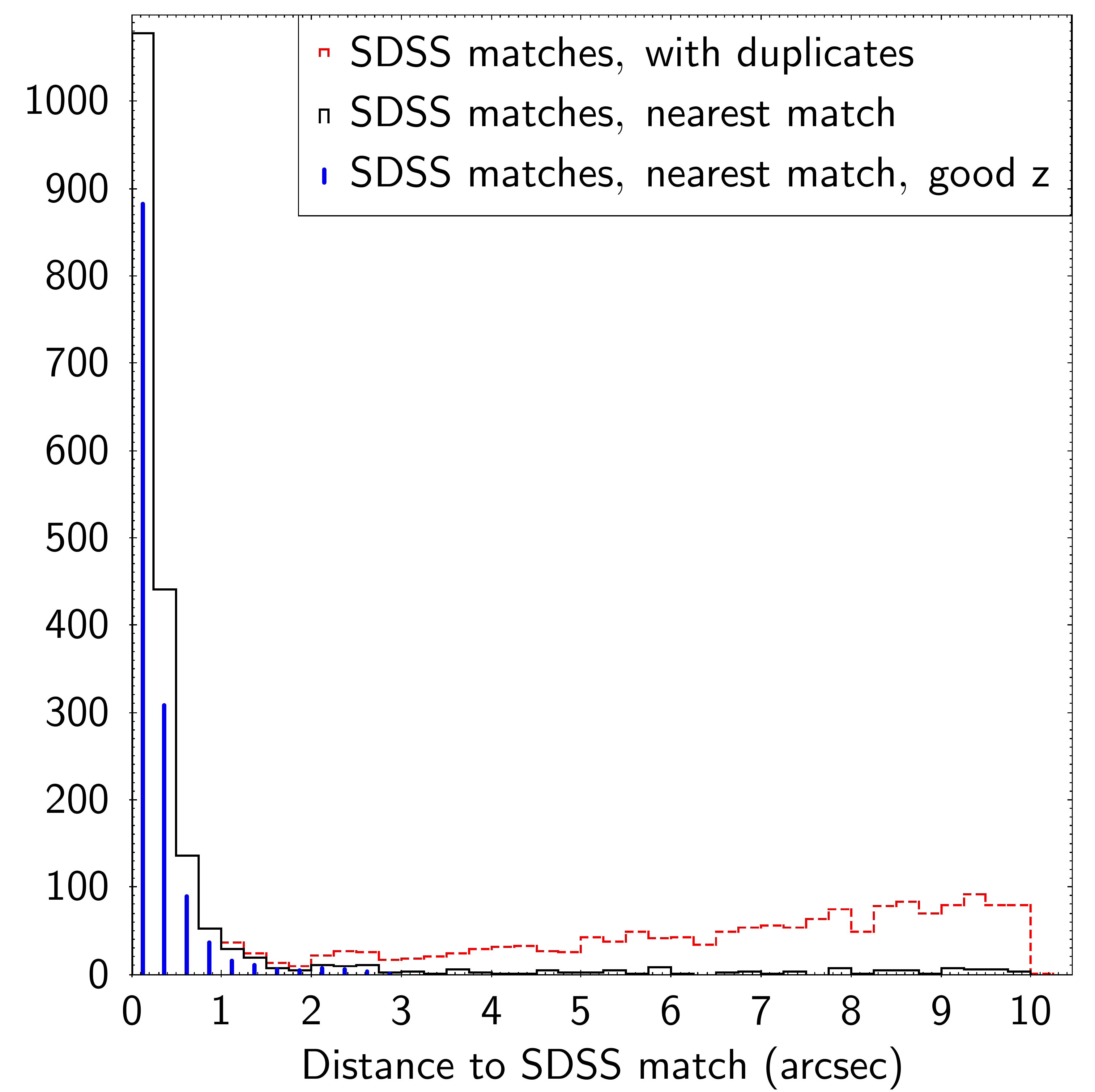}
\caption{Distance histogram for SDSS matches. The distances were calculated between the weighted averaged positions in the MIXR catalogue and the SDSS positions. The red dashed line shows all the matches, including duplicates. The black outline shows the distribution of nearest matches; the distribution is fairly narrow, with very few matches found at distances larger than 2--3 arcsec. The blue vertical lines show the distribution of nearest matches with error-constrained redshift values (not upper limits).}\label{SDSS_dist_histo}
\end{figure}

Fig. \ref{SDSS_dist_histo} shows the distance histogram for the SDSS-MIXR matches, described in section \ref{z}.

\begin{figure}
\centering
\includegraphics[width=0.46\textwidth]{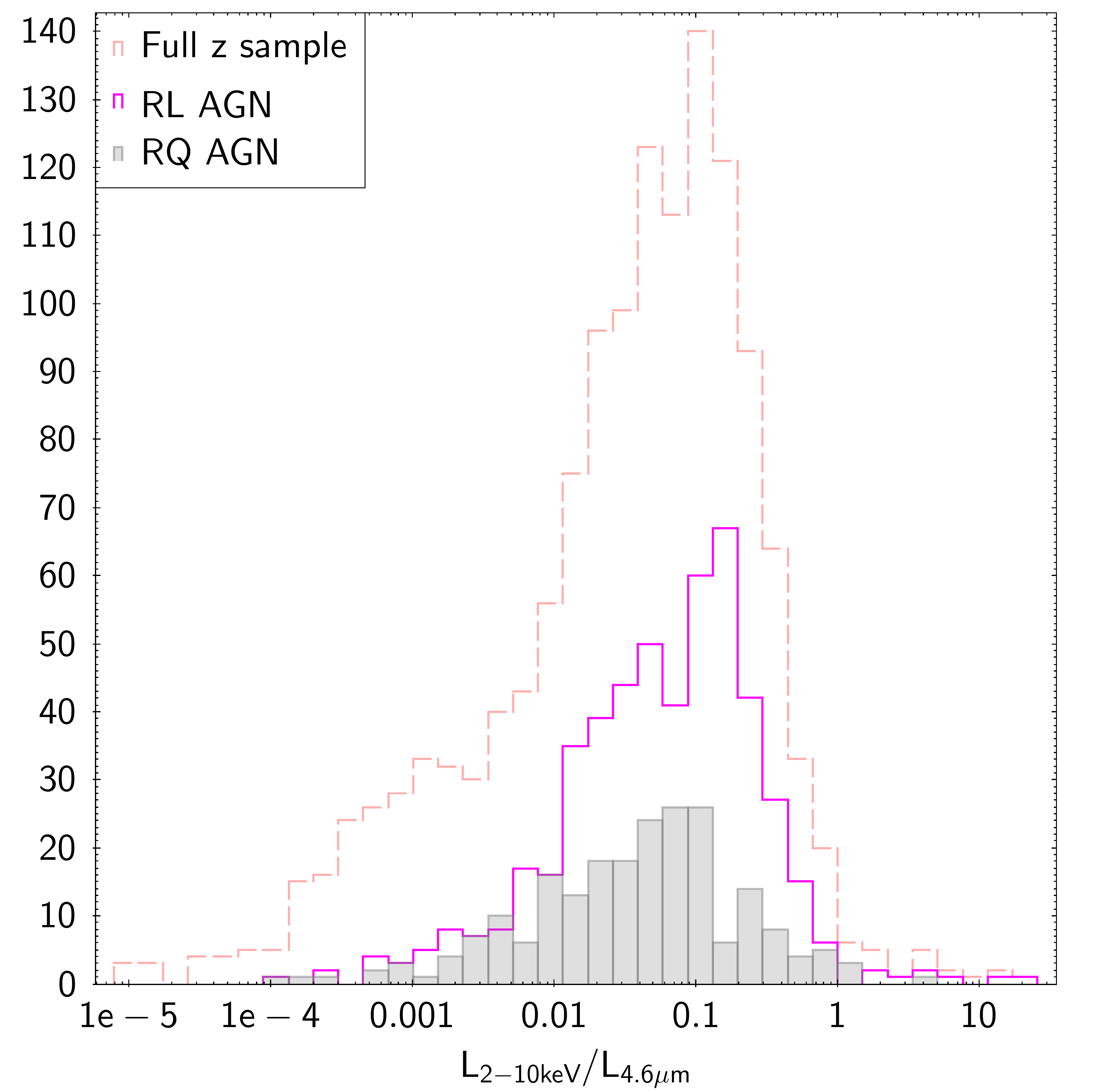}
\caption{$L_{2-10 keV}$/$L_{4.5 \mu m}$ as a proxy for AGN intrinsic obscuration, for the RL and RQ AGN. The distribution for the full z sample (all sources with redshifts, regardless of their W3 S/N, see Table \ref{nSources}) is also plotted as a dashed line, for reference.}\label{CT_RL_RQ_2}
\end{figure}

Fig. \ref{CT_RL_RQ_2} is a counterpart to Fig. \ref{CT_RL_RQ_1}, using the WISE W2 band (4.6 $\mu$m) instead of W3 (12 $\mu$m). 

\begin{figure}
\centering
\includegraphics[width=0.46\textwidth]{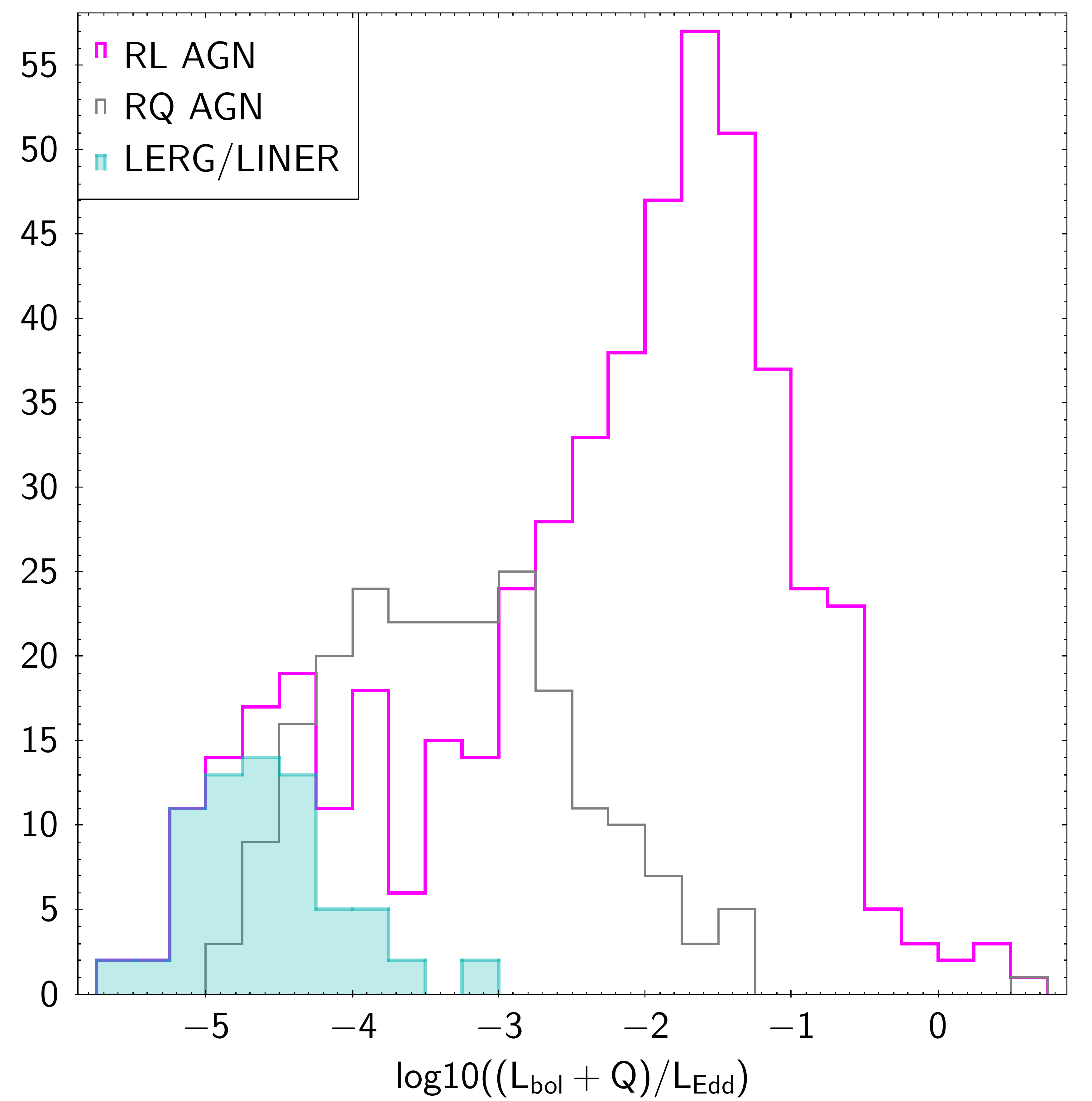}
\caption{Eddington rates for the source subsets defined from Fig. \ref{RL_RQ}, excluding QSOs. Only the sources that pass the W3 S/N cut are considered. Please see section \ref{Power} for the definition of $Q$.}\label{Ledd_withQ_RL_RQ_noQSO}
\end{figure}

\begin{figure}
\centering
\includegraphics[width=0.46\textwidth]{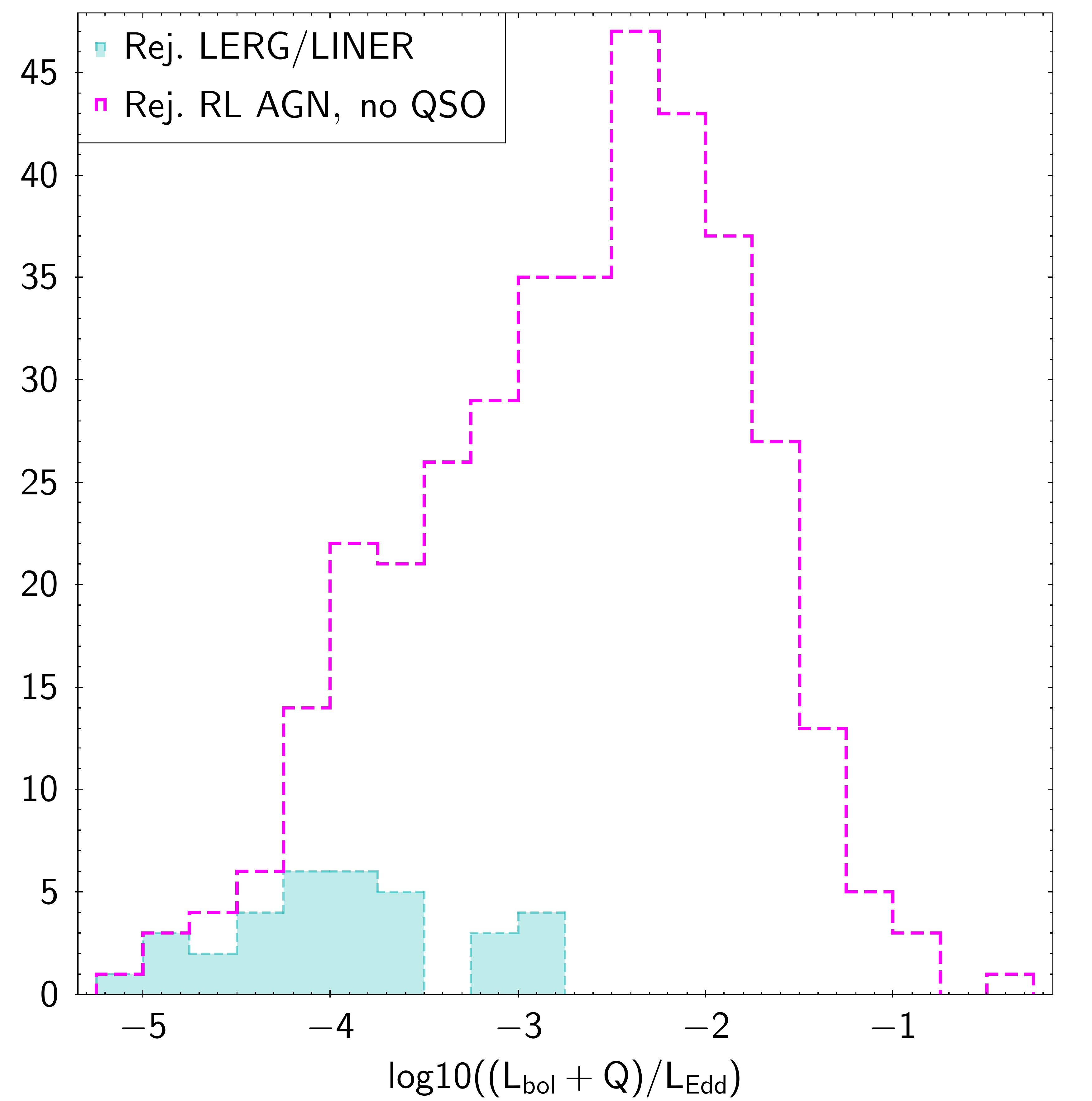}
\caption{Eddington rates for the LERG/LINER and radio-quiet AGN that do not pass the W3 S/N cut. QSOs are excluded. Please see section \ref{Power} for the definition of $Q$.}\label{Ledd_withQ_RL_RQ_noQSO_rej}
\end{figure}

Figs. \ref{Ledd_withQ_RL_RQ_noQSO} and \ref{Ledd_withQ_RL_RQ_noQSO_rej} provide extra background to section \ref{Eddington}, for easier comparison with the earlier results of \citet{Mingo2014}.

\bsp	
\label{lastpage}
\end{document}